\documentclass[12pt,aps,pre,superscriptaddress,showpacs,floatfix,nofootinbib,longbibliography]{revtex4-2}

\usepackage{dcolumn}
\usepackage{array}
\usepackage{amsmath}
\usepackage{amssymb}
\usepackage{amsthm}
\usepackage{bm}
\usepackage{times}
\usepackage{mathtools}
\usepackage{bbm}
\usepackage[T1]{fontenc}
\usepackage[utf8]{inputenc}

\usepackage{graphicx}
\usepackage{color}
\usepackage{url}
\usepackage{rotating}
\usepackage{enumitem}
\usepackage{afterpage} 
\usepackage{booktabs}
\usepackage{multirow}
\usepackage{makecell}
\usepackage{tabularx}
\usepackage{caption}
\usepackage{comment}

\usepackage{algorithm}
\usepackage{algpseudocode}
\usepackage{scalefnt}
\usepackage{booktabs}
\usepackage[table]{xcolor}
\usepackage{hyperref}
\usepackage{siunitx}
\newcommand{\cell}[2]{\parbox[t]{#1}{\raggedright #2}}

\begin{document}

\title{Hypergraph Representations of Single-Cell RNA Sequencing Data for Improved Cell Clustering}

\author{Wan He$^*$}
\affiliation{Network Science Institute, Northeastern University, Boston, MA, USA}

\author{Daniel I. Bolnick}
\affiliation{Department of Ecology and Evolutionary Biology, University of Connecticut Storrs CT, USA}

\author{Samuel V. Scarpino}
\affiliation{Institute for Experiential AI, Northeastern University, Boston, MA, USA}
\affiliation{Bouve College of Health Sciences, Northeastern University, Boston, MA, USA}
\affiliation{Khoury College of Computer Sciences, Northeastern University, Boston, MA, USA}
\affiliation{Network Science Institute, Northeastern University, Boston, MA, USA}
\affiliation{Vermont Complex Systems Institute, University of Vermont, Burlington, VT, USA}
\affiliation{Santa Fe Institute, Santa Fe, NM, USA}

\author{Tina Eliassi-Rad}
\affiliation{Network Science Institute, Northeastern University, Boston, MA, USA}
\affiliation{Khoury College of Computer Sciences, Northeastern University, Boston, MA, USA}
\affiliation{Vermont Complex Systems Institute, University of Vermont, Burlington, VT, USA}
\affiliation{Santa Fe Institute, Santa Fe, NM, USA}

\begin{abstract}

\noindent \textbf{Abstract}\\
\noindent \textbf{Motivation:}
Single-cell RNA sequencing (scRNA-seq) data analysis is often performed using network projections that produce co-expression networks. These network-based algorithms are attractive because regulatory interactions are fundamentally network-based and there are many tools available for downstream analysis. However, most network-based approaches have two major limitations. First, they are typically unipartite and therefore fail to capture higher-order information. Second, scRNA-seq data are often sparse, so most algorithms for constructing unipartite network projections are inefficient and may overestimate co-expression relationships, or may under-utilize the sparsity when clustering (e.g., with cosine distance). To address these limitations, we propose representing scRNA-seq expression data as hypergraphs, which are generalized graphs where a hyperedge can connect more than two nodes. In this context, hypergraph nodes represent cells, and hyperedges represent genes. Each hyperedge connects all cells in which its corresponding gene is actively expressed, indicating the expression of that gene across different cells. The resulting hypergraph can capture higher-order information and appropriately handle varying levels of data sparsity. This representation enables clustering algorithms to leverage higher-order relationships for improved cell-type differentiation.\\
\noindent \textbf{Results:}
To distinguish cell types using hypergraph representations of scRNA-seq data, we introduce two novel clustering algorithms: (1)  Dual-Importance Preference Hypergraph Walk (DIPHW) and (2)  Co-expression and Memory-Integrated Dual-Importance Preference Hypergraph Walk (CoMem-DIPHW). DIPHW is a new hypergraph-based random walk algorithm that computes cell embeddings by considering the relative importance of genes to cells and cells to genes, incorporating a preference exponent to facilitate clustering. CoMem-DIPHW integrates two unipartite projections, the gene co-expression and cell co-expression networks, along with the cell-gene expression hypergraph derived from single-cell abundance count data into the random walk model. The advantage of CoMem-DIPHW is that it accounts for both local information from single-cell gene expression and global information from pairwise similarity in the two co-expression networks. 
 We benchmark the performance of our algorithms against established and state-of-the-art deep learning approaches using both real-world and simulated scRNA-seq data. Real-world datasets include cells from the human pancreas, mouse pancreas, human brain, and mouse brain tissues. We also use a ground-truth labeled cell-type annotation dataset based on human lung adenocarcinoma cell lines. Quantitative evaluation shows that CoMem-DIPHW consistently outperforms established algorithms and state-of-the-art deep learning algorithms for cell-type clustering. Our proposed algorithms show the greatest improvement on scRNA-seq data with weak modularity. Moreover, CoMem-DIPHW successfully annotates clusters with biologically relevant cell types. Our results highlight the utility of hypergraph representations in the analysis of scRNA-seq data.

\textbf{Availability and Implementation:}
Our methods are implemented in Python and are available at \url{https://github.com/wanhe13/CoMem-DIPHW}.

\end{abstract}

\maketitle
 
\section{Introduction}

Network analysis has become a popular tool for studying complex systems in biology, reflecting the combinatorial interactions of biomolecules. Its applications include functional analysis~\cite{szklarczyk2015string,veres2015comppi,vazquez2003global,ashtiani2018systematic} and prediction of interactions in protein-protein interaction (PPI) networks~\cite{kovacs2019network,lei2013novel,murakami2017network}, identification of regulators and analysis of pathways in gene regulatory networks~\cite{coleman2023gene,muzio2021biological,zickenrott2016prediction}, disease modeling, prediction, and intervention in epidemiology~\cite{kraemer2020effect,chinazzi2020effect,davis2021cryptic}, and identification of cell types in co-expression networks~\cite{langfelder2008wgcna,satija2015spatial, butler2018integrating,wolf2018scanpy}.

However, traditional methods represent biological systems using unipartite networks, which capture relationships only between a single type of entity. Such representations are often not the most natural or information-preserving choice, since many complex systems involve interactions between multiple types of entities. Most often, a unipartite network projection is chosen such that the system could be analyzed with the abundant network analysis methods developed based on unipartite networks~\cite{horvat2012one,langfelder2008wgcna}. However, this focus on unipartite networks restricts our attention to biological interactions just between pairs of genes, proteins, or other kinds of nodes.\footnote{We use the terms node and vertex interchangeably.} In reality, biological processes often entail multiway interactions that cannot be represented by unipartite networks. For instance, three-gene epistatic interactions vastly outnumber pairwise epistasis in experimental analyses of yeast genetic networks~\cite{kuzmin2018systematic}.  In protein-protein interaction (PPI) networks, unipartite representations capture relationships between proteins but fail to capture the biological functions associated with groups of proteins. Similarly, unipartite projections of gene regulatory networks connect genes based on regulatory relationships but do not capture relationships among different entities such as transcription factors or their binding sites. Therefore, network analysis methods are needed to more accurately capture the higher-order interactions that are apparently ubiquitous in biological systems.

Single-cell RNA sequencing (scRNA-seq) has enabled the profiling of gene expression at the individual cell level~\cite{yanai2019cel,macosko2015highly,zheng2017massively,picelli2014full}, whereas conventional bulk tissue RNA sequencing measures expression at the tissue level, averaging gene expression over an ensemble of cells. Cell type identification, as one of the most important downstream tasks in single-cell RNA-seq (scRNA-seq) data analysis~\cite{hwang2018single}, has applications in biology and medicine, including tracing the trajectories of different cell lineages in the development of cell differentiation studies~\cite{marioni2017single}, tissue heterogeneity analysis for cancer research~\cite{wong2022single,laviron2022tumor,karaayvaz2018unravelling}, immune cell profiling for therapy development~\cite{landhuis2018single,lyons2017immune,ermann2015immune,chung2017single}, and biomarker discovery for diagnosis. 

However, cell type identification often relies on conventional transcriptomic data analysis pipelines such as WGCNA~\cite{langfelder2008wgcna}, Scanpy~\cite{wolf2018scanpy} and Seurat~\cite{satija2015spatial, butler2018integrating}. These packages embed the cell-gene interactions to a unipartite graph structure such as the co-expression network or the K-nearest neighbor (KNN) graph, which draws edges between the cell pairs based on their similarity, followed by unipartite graph partitioning algorithms to detect closely related cell clusters for cell type identification in the scRNA-seq data. These unipartite projections of the scRNA-seq expression data have two limitations: Firstly, cell and gene co-expression networks capture only pairwise expression similarity among cells or genes. As a result, higher-order information, such as the expression level of a specific gene in a specific cell or the coordinated expression of multiple genes within the same cell, is lost. Secondly, scRNA-seq data are sparse, with non-zero entries often accounting for less than 10$\%$ of the total~\cite{jiang2022statistics, huang2018saver}, compared to 60-90$\%$ in bulk-tissue data. Despite this sparsity, constructing a unipartite co-expression network produces in a fully-connected network, which is an inefficient representation of the originally sparse transcriptomic data. Furthermore, the high sparsity of scRNA-seq data (both dropouts and biological zeros) results in inflated correlations, leading to spurious connections and obscure meaningful biological signals, as will be discussed further in \ref{subsec:sparsity}. In addition, determining whether zero inflation in scRNA-seq data is due to technical dropouts or true biological absence is a challenging task~\cite{qiu2020embracing,jiang2022statistics,kim2020demystifying}.

Given the issues with unipartite projections, it is crucial to explore alternative network representations that can accurately reflect the scRNA-seq data and the underlying biological reality of higher-order interactions among genes. Hypergraphs~\cite{bianconi2021higher,battiston2020networks,battiston2021physics} offer a nice solution by directly representing multiway relationships in scRNA-seq data, without requiring further data projection. Existing work on hypergraph representations has demonstrated the improved performance they can achieve in modeling complex biological systems by incorporating higher-order interactions.
For example, Liu et al.~\cite{liu2022multi} developed a hypergraph-based neural network to predict synergistic drug combinations for cancer treatment. Wang et al.~\cite{wang2024hypertmo} integrated multi-omics data with hypergraph convolutional neural networks to classify patients with diseases such as breast cancer and Alzheimer’s.
Gaudelet et al.~\cite{gaudelet2018higher} used hypergraphs to model interactions across different levels of protein organization (such as protein-protein interactions, complexes, and pathways) to better capture the complexity of biological systems and predict biological functions.
Ma and Ma~\cite{ma2022hypergraph} developed a hypergraph-based logistic matrix factorization method that predicts possible metabolite–disease interactions and uncovers novel disease-related metabolites. These successes highlight the potential of hypergraphs to model multiway relationships inherent in biological systems, motivating their application to scRNA-seq analysis.

\begin{figure*}
\centering
\includegraphics[width=\textwidth]{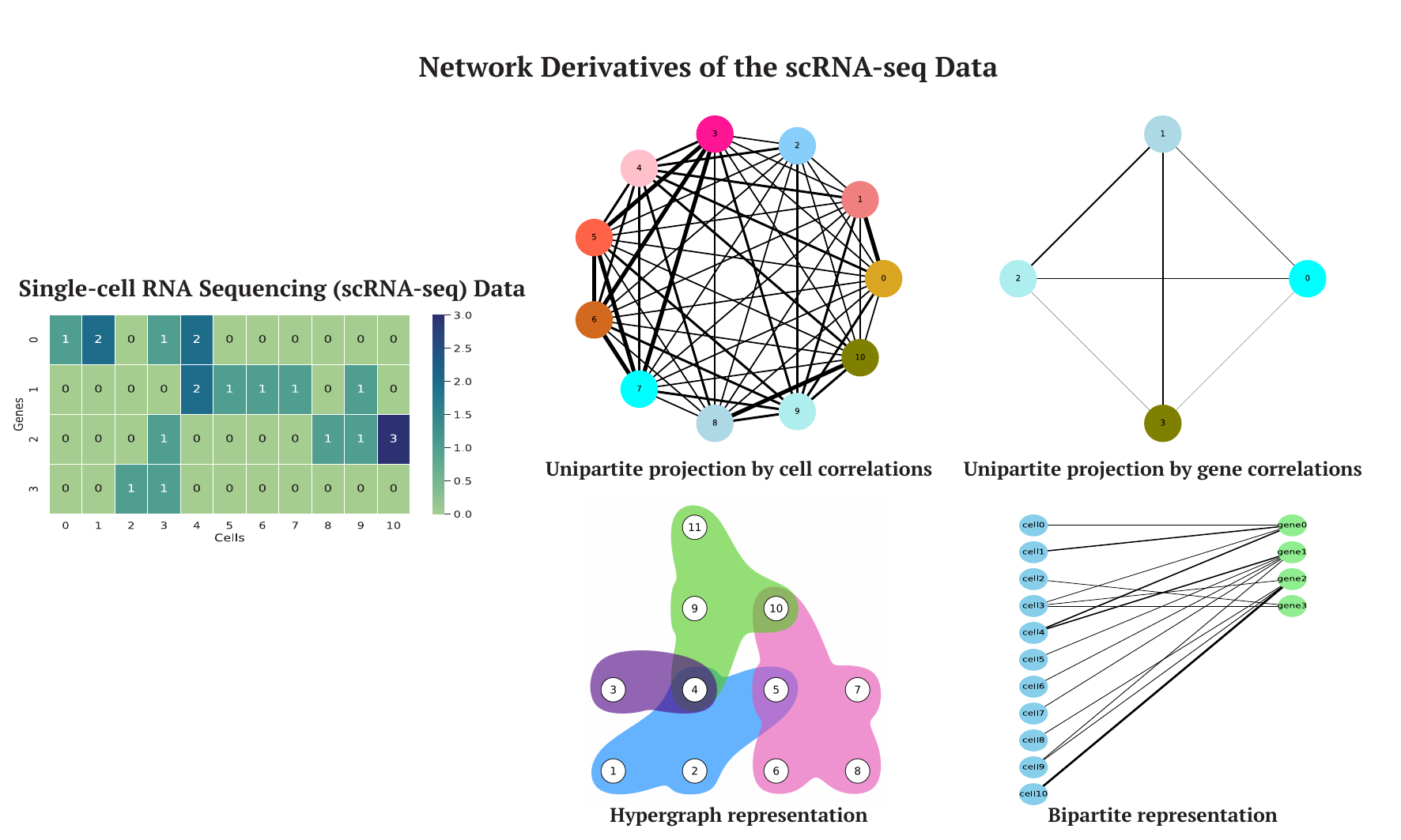}
\caption{Network Representations of scRNA-seq Data. Expression data can be represented by various networks, including unipartite network projections such as cell and gene co-expression networks. In a cell co-expression network, nodes are cells and edge weights represent similarity in cell-pair expression. In gene co-expression networks, nodes represent genes. Hypergraph and bipartite network representations preserve the exact expression level of a gene in a cell. In the hypergraph representation, nodes represent cells and hyperedges represent genes. In the bipartite representation, one set of nodes represents cells and the other set represents genes.}
\label{fig:Network_derivatives_of_the_scRNAseq_data}
\end{figure*}

\begin{figure*}
\centering
\includegraphics[width=\textwidth]{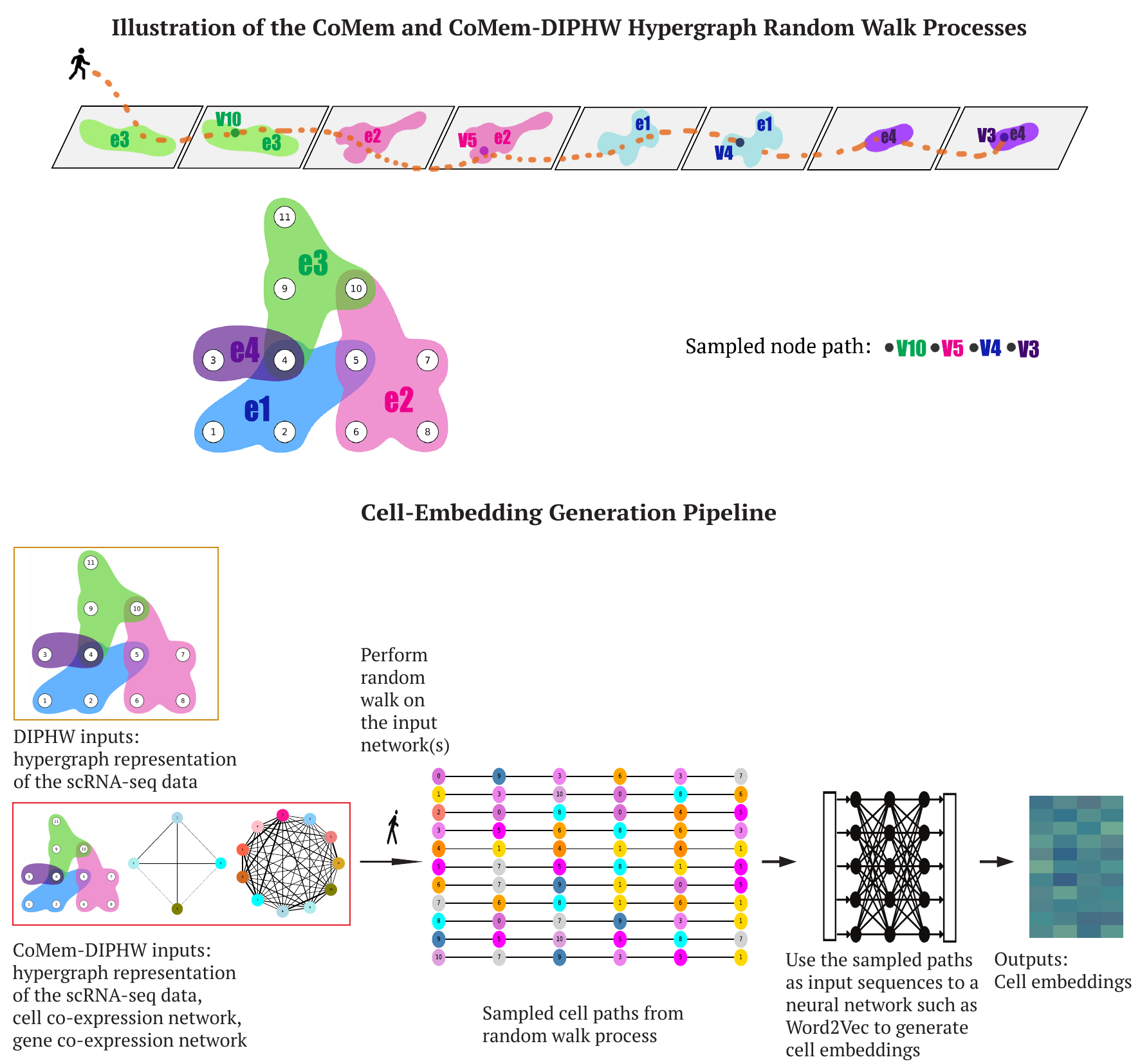}
\caption{Hypergraph Clustering Algorithm Illustration. In the hypergraph representation of scRNA-seq data, nodes represent cells and edges represent genes. The expression level of a gene in a cell is captured by the weight of a node in an edge. Random walks are performed on the hypergraph by alternately considering the probability of the walker choosing a node given an edge and choosing an edge given the current node. For DIPHW, the input is the hypergraph representation of the scRNA-seq data. For CoMem-DIPHW, the input includes the hypergraph, cell co-expression network, and gene co-expression network, which serve as node and edge similarity matrices to incorporate memory during the walk. A small neural network (word2vec) is applied to the sampled node paths from the hypergraph random walk process to compute cell embeddings. Finally, cell clustering is performed by applying K-means to the resulting cell embeddings.}
\label{fig:HypergraphRandomWalkIllustration}
\end{figure*}

\section{Methods}
\label{sec:Methods}
We conceptualize scRNA-seq data as a hypergraph $\mathcal{H}$ and design two hypergraph random walk algorithms that capture cell-cell relationships by sampling cell-to-gene and gene-to-cell interactions in the hypergraph. This hypergraph representation of scRNA-seq data does not require additional data transformation, does not cause information loss, or increase computational cost.

\subsection{Hypergraph representation}
scRNA-seq data can be conceptualized as a hypergraph $\mathcal{H} = (V, E)$, where:
\begin{itemize}
    \item $V$ represents the set of nodes corresponding to the cells profiled in the experiment.
    \item $E$ denotes the set of hyperedges representing genes, with each hyperedge $e \in E$ connecting to all cells $\in V$ in which the gene $e$ is actively expressed. The weight of connection of a hyperedge (gene) to each node (cell) is determined by the gene's expression level in that cell.
\end{itemize}

Thus, each hyperedge captures the expression profile of a gene across all cells, preventing any loss of information. The incidence matrix $\mathbf{I}_\mathcal{H}$ of the hypergraph $\mathcal{H}$ is constructed such that $\mathbf{I}_\mathcal{H}(v,e)$ is given by the expression level, i.e. abundance counts, of gene $e$ in cell $v$.

\begin{table}[htbp]
\small
\centering
\begin{tabular}{|l|l|}
\hline
\textbf{Notation} & \textbf{Description} \\ \hline
$\mathcal{H}$ & Hypergraph \\ \hline
$\mathbf{I}_\mathcal{H}$ & Incidence matrix of the hypergraph $\mathcal{H}$ \\ \hline
$V$ & Set of nodes (cells) \\ \hline
$E$ & Set of hyperedges (genes) in the hypergraph \\ \hline
$G_V$ & Cell co-expression network (node similarity matrix) \\ \hline
$G_E$ & Gene co-expression network (edge similarity matrix) \\ \hline
$\Gamma_e(v)$ &  Weight of node $v$ in hyperedge $e$ (i.e., expression of gene $e$ in cell $v$\strut) \\ \hline
$P(V = v \mid E = e)$ & Probability of selecting node $v$ given hyperedge $e$ \\ \hline
$P(E = e \mid V = v)$ & Probability of selecting hyperedge $e$ given node $v$ \\ \hline
$P(E_{t+1} = e' \mid E_t = e, V_t = v)$ & \parbox[t]{10cm}{\raggedright Probability of transitioning to hyperedge $e'$ at time $t+1$, given current hyperedge $e$ and node $v$ at time $t$\strut} \\ \hline
$P(V_{t+1} = w \mid E_{t+1} = e, V_t = v)$ & \parbox[t]{10cm}{\raggedright Probability of selecting node $w$ at time $t+1$, given hyperedge $e$ at time $t+1$ and previous node $v$\strut} \\ \hline
$P(V_{t+1} = w \mid V_t = v)$ & Unipartite node-to-node transition probability \\ \hline
\end{tabular}
\caption{Notation Table for Our Hypergraph Random Walk Representation.}
\label{tab:hypergraph_notation}
\end{table}

\subsection{Method 1: Dual-Importance Preference (DIP) Hypergraph Walk (DIPHW)}

\label{subsec:DIPHW}
Building on this hypergraph representation, we introduce a novel hypergraph random walk approach for cell embedding computation, the \textit{Dual-Importance Preference (DIP) Hypergraph Walk (DIPHW)}, which accounts for both the relative importance of edges to nodes and nodes to edges, or the importance of genes relative to cells and cells to genes, along with a preference exponent for faster clustering. In a hypergraph random walk process, the walker alternates between transitioning from a hyperedge to a node and from a node to a hyperedge, following the node-to-edge transition probabilities $P_{E|V}(e,v)$ and edge-to-node transition probabilities $P_{V|E}(v,e)$. In the hypergraph random walk with edge-dependent vertex weight (EDVW) by Chitra and Raphael (2019)~\cite{Chitra2019}, the edge-to-node transition probability is edge-dependent, but the node-to-edge transition probability does not consider the importance of the edge to the node. As a natural extension to EDVW, we define the node-to-edge transition probability to be vertex-dependent in DIPHW and use a preference exponent to accelerate the clustering process. The modified random walk process is as follows: 
\begin{enumerate}[leftmargin=0.5cm]

  \item \textit{Node-to-Hyperedge Transition Probability}

   The probability of selecting hyperedge \( e_{t+1} \) given that the walker is currently at vertex \( v_t \) is defined as:

   \[
   P_{E|V}(v_{t} \rightarrow e_{t+1}) = \frac{\omega(e_{t+1})\gamma_{e_{t+1}}(v_t)}{\sum_{e' \in E(v_{t})} \omega(e')\gamma_{e'}(v_t)}
   \]

In scRNA-seq cell clustering, this represents the probability of selecting a gene based on its expression level relative to other genes in the same cell.   
  \item \textit{Hyperedge-to-Node Transition Probability}
 
 Upon the walker's arrival at hyperedge \( e_{t+1} \), the probability of moving to vertex \( v_{t+1} \in e_{t+1} \) is given by:

   \[
   P_{V|E}(e_{t+1} \rightarrow v_{t+1}) = \frac{\gamma_{e_{t+1}}^{\alpha}(v_{t+1})}{\sum_{v' \in e_{t+1}} \gamma_{e_{t+1}}^{\alpha}(v')}
   \]

    In scRNA-seq cell clustering, the transition from a gene to a cell is based on the relative expression level of a gene in a cell, amplified by a preference exponent to improve clustering convergence.

  \item \textit{Node-to-Node Transition Probability}

The unipartite node-to-node transition probability of DIPHW is calculated by summing the products of the node-to-edge and edge-to-node transition probabilities across all hyperedges. This probability is used to sample random walk node paths, which are then input into a neural network to compute cell embeddings.

\[P(v_{t} \rightarrow v_{t+1}) = \sum_{e \in E} P_{E|V}(e | v_t) P_{V|E}(v_{t+1} | e).\]

Substituting the expressions for $P_{E|V}(e_{t+1} | v_t)$ and $P_{V|E}(v_{t+1} | e_{t+1})$, we get:

\[P(v_{t} \rightarrow v_{t+1}) = \sum_{e_{t+1} \in E} \frac{\omega(e_{t+1})\gamma_{e_{t+1}}(v_t)}{\sum_{e' \in E} \omega(e')\gamma_{e'}(v_t)}\frac{\gamma_{e_{t+1}}^{\alpha}(v_{t+1})}{\sum_{v' \in e_{t+1}} \gamma_{e_{t+1}}^{\alpha}(v')}
\]

The expression of node-to-node transition probability could be simplified by removing reference to any particular time points as,

\[P(u \rightarrow v) = \sum_{e \in E(u)} \frac{\omega(e)\gamma_e(u)}{\sum_{e' \in E} \omega(e')\gamma_{e'}(u)}\frac{\gamma_e^{\alpha}(v)}{\sum_{u' \in e} \gamma_e^{\alpha}(u')}\]

  \item \textit{Node-to-Node Transition Probability in Vectorized Format}

The $|V|\times|V|$ node-to-node transition probability matrix $P_{V|V}$ could be written in the matrix form as
\[P=D_{E|V}^{-1}W_{E|V}D_{V|E}^{-1}W_{V|E}\]

The matrix formulation eliminates explicit iteration and reduces computation time. Here, \( W_{E|V}=\mathbf{I}_\mathcal{H}^T W_E\) is the $|V| \times |E|$ node-to-edge transition weight matrix, where column $e$ of $W_{E|V}$ contains the vertex weights in hyperedge $e$ multiplied by the weight of hyperedge $e$. Similarly, $W_{V|E}$ represents the edge-to-node transition weight matrix before normalization with dimensions \( |E| \times |V| \), given by
$W_{V|E} = \mathbf{I}_\mathcal{H}^{\alpha}, \quad w_{ve} = \left[\mathbf{I}_\mathcal{H}(e,v)\right]^{\alpha}$. Finally, $D_{E|V}$ and $D_{V|E}$ are diagonal matrices that normalize vertex-to-edge and edge-to-vertex transitions to ensure the probability distribution sums to 1.

\label{eq:DIPHW}

\end{enumerate}

\subsection{Method 2: co-expression and Memory-Integrated Dual-Importance Preference Hypergraph Walk (CoMem-DIPHW)}

CoMem-DIPHW is an extension of DIPHW. It incorporates co-expression networks and single-cell transcriptomic profiles. The memory mechanism of CoMem-DIPHW leverages cell and gene co-expression networks to account for previously visited nodes and edges during transitions to capture both local expression relationships and global co-expression patterns.

\begin{enumerate}[leftmargin=0.5cm]

  \item \textit{Hypergraph Representation of the scRNA-seq data}
  
We note that a matrix can be directly interpreted as the incidence matrix of a hypergraph to represent the relationship between two variables, without additional computation. 

 \item \textit{The Node and Edge Similarity Networks: Cell and Gene co-expression Networks and their construction}

Using scRNA-seq data, we construct two co-expression networks: the cell co-expression network (\(G_V\)) and the gene co-expression network (\(G_E\)). In the cell co-expression network, cells are nodes, and edges represent correlation coefficients between cell pairs, indicating the similarity of their expression profiles. Similarly, in the gene co-expression network, genes are nodes, and edges represent correlation coefficients between gene pairs. We use Spearman's rank coefficient to measure pairwise similarity of expression profiles, as it captures non-linearity better than Pearson’s correlation. These unipartite networks serve as node and edge similarity matrices, embedding global-level information that summarizes associations between cell and gene pairs.

 \item \textit{Memory-Integrated Random Walk}

CoMem-DIPHW incorporates a memory component by considering the similarity between consecutively visited nodes and edges. This memory mechanism adjusts the transition probabilities so that the node-to-edge transition probability depends on the previously visited edge, and the edge-to-node transition probability depends on the previously visited node. Node-node and edge-edge dependencies are incorporated through the node and edge similarity networks, which are two correlation networks derived from transcriptomic expression data. The memory mechanism constrains the hypergraph walker to focus on cell-gene interactions among similar nodes (cells) and edges (genes), thus better revealing modularity within the data.

While the co-expression networks provide global perspectives through unipartite projections, the specific expression levels of genes in individual cells (local information) are preserved in the hypergraph structure. CoMem-DIPHW integrates global and local information for more accurate cell type identification.

\item \textit{Formulation}
\end{enumerate}

The random walk process in CoMem-DIPHW is formulated as follows. Let $G_V$ be the vertex similarity matrix, where $G_V(v, w)$ quantifies the similarity between nodes $v$ and $w$. Similarly, let $G_E$ be the edge similarity matrix, where each element $G_E(e, f)$ represents the similarity between edge $e$ and $f$.

\subsubsection*{Method 2.0: CoMem}
\label{subsubsec:CoMem}

In CoMem, we calculate the unipartite node-to-node transition probability of the hypergraph random walker. To do this, we first calculate the following four probabilities.

\begin{enumerate}[leftmargin=0.5cm]

  \item The probability of selecting vertex $v$ in edge $e$ is:
  \[ P_{V|E}(e \rightarrow v) = \frac{\Gamma_e(v)}{\sum_{u \in e} \Gamma_e(u)} \]
 
  \item The probability of selecting edge $e$ given vertex $v$ is:
  \[ P_{E|V}(v \rightarrow e) = \frac{\Gamma_e(v)}{\sum_{e \in E} \Gamma_e(v)} \]

  \item The probability of transitioning from node $v$ in edge $e_1$ to edge $e_2$, considering edge similarity, is given by:
  \[ P_{E_{t+1}|V,E_t}(e_2|v,e_1) = \frac{G_E(e_1, e_2) P(e_2|v)}{\sum_{e' \in E} 
  G_E(e_1, e') P(e'|v) }\]

  \item Similarly, the probability of selecting node $w$ from edge $e$ after transitioning from node $v$, 
  incorporating node similarity and the node-edge connection, is formulated as:
  \[ P_{V_{t+1}|E,V_t}(w|e, v) = \frac{G_V(v, w)P(w|e)}{\sum_{w' \in e} G_V(v, w') P(w'|e)} \]
\end{enumerate}

Finally, the unipartite node-to-node transition probability of the hypergraph random walk process is:
\begin{align*}
 &P_{V_{t} \rightarrow V_{t+1}(v_1,v_2)} \\
 \propto &
 \sum_{e_1, e_2 \in E} P_{V|E}(v_1|e_1) P_{E_{2}|V,E_1}(e_2|v_1,e_1) P_{V_{t+1}|E,V_t}(v_2|e_2 , v_1) 
\end{align*}

To encourage exploration and ensure non-lazy walks, we modify the node and edge similarity graphs such that the diagonal elements are set to zero. This ensures that during the random walk, the random walker does not remain at the same node or edge in consecutive steps.

\subsubsection*{Method 2.1: CoMem based on DIPHW}
For CoMem-DIPHW, we integrate DIPHW into the CoMem walk framework from section \ref{subsubsec:CoMem}. Specifically, the vertex selection probability \( P_{V|E}(e \rightarrow v) \) and the edge selection probability \( P_{E|V}(v \rightarrow e) \) are both defined as in DIPHW (section \ref{subsec:DIPHW}), while all other probability distributions remain as specified in section \ref{subsubsec:CoMem}.

\subsection{Cell Embedding and Clustering}

We perform random walks using the defined hypergraph node-to-edge and edge-to-node transition probabilities to sample sequences of visited cells. These sampled cell paths, which capture the geometry of the hypergraph representing the scRNA-seq data, are then used as input to word2vec~\cite{mikolov2013distributed} to learn low-dimensional vector representations of cells. Cells with similar expression profiles frequently co-occur in the sampled random walk paths and are closer in the embedded space. Finally, K-means clustering is applied to cell embeddings to obtain cluster assignments.

\subsection{scRNA-seq Data Simulation}
\label{subsec:Simulation}
In our study, we develop an algorithm to simulate data that mimics the characteristics of scRNA-seq expression data (see Algorithm~\ref{alg:simulation_sc}). In addition to the underlying assumption of within-cell-type homogeneous expression that governs most cell clustering methods, we incorporate between-cell-type crosstalk~\cite{armingol2021deciphering,zhang2023defining,zepp2019cellular,bayik2021cancer,lehuen2010immune} by integrating configurable intermodular covariance, density, and signal strength into the simulation model. The algorithm generates a sparse matrix representing the simulated scRNA-seq expression data, with numerous user-configurable parameters, including number of modules, module density, shape, background signal strength, modular signal strength, intermodular signal strength, intermodular covariance, and noise level (see Table~\ref{tab:hyperparameters}). This method is implemented as a function in our codebase and is available for use by other researchers \cite{wan2025comem}.

Our simulation model is highly flexible to accommodate various types of scRNA-seq data. For example, the model can simulate both sequenced datasets with high density (up to 50\%) and sequenced datasets with low density (as low as 1\%)~\cite{andrews2021tutorial}. The proportion of non-housekeeping genes can be adjusted by changing the number of differentially expressed genes (DEGs) in each module. In addition, the variance in module sizes and the number of co-expressed genes in each cell type can be configured by setting different variances for the Poisson distribution used in the embedded module simulation. The modules embedded in the simulated data can be configured with higher density and average signal strength to model the expression profiles of biomarker genes that are highly expressed in certain cell types. This enables us to mimic the sparsity and modularity of real scRNA-seq data, where a small number of genes are highly expressed in specific cell types. Alternatively, the embedding modules can be user-specified without using the embedded module simulation.

    \begin{algorithm}[H]
        \caption{Simulation Algorithm for scRNA-seq Data}
        \label{alg:simulation_sc}
        \begin{algorithmic}[1]
            \State \textbf{Input:} Number of genes $E$, number of cells $V$, number of cell types $K$; average genes per cell type $\bar{g}$ and average cells per cell type $\bar{c}$; average expression levels for within-cell-type $\lambda_{\text{ct}}$, cross-cell-type $\lambda_{\text{cross}}$, and background $\lambda_{\text{bg}}$; densities within-cell-type $\rho_{\text{ct}}$, cross-cell-type $\rho_{\text{cross}}$, and background $\rho_{\text{bg}}$; and crosstalk probability $p_{\text{cross}}$.
            \State \textbf{Output:} Sparse expression matrix $\mathbf{X} \in \mathbb{R}^{E \times V}$ with ground-truth labels.
            \State Initialize $\mathbf{X} \gets \mathbf{0}_{E \times V}$.
            \State Sample cell type sizes: $g_k \sim \text{Poisson}(\bar{g})$, $c_k \sim \text{Poisson}(\bar{c})$ for $k = 1, \ldots, K$.  
            \State \textit{Phase 1: Within-Cell-Type Signal}
            \For{each cell type $k = 1, \ldots, K$}
                \State For each gene $e \in E_k$ and cell $v \in V_k$: with probability $\rho_{\text{ct}}$, set $\mathbf{X}[e, v] \gets \text{Poisson}(\lambda_{\text{ct}})$.
            \EndFor            
            \State \textit{Phase 2: Background Noise}
            \State Randomly fill $\rho_{\text{bg}}$ fraction of remaining zero entries with $\text{Poisson}(\lambda_{\text{bg}})$ values. 
            \State \textit{Phase 3: Cross-Cell-Type Crosstalk}
            \For{each cell type pair $(k, l)$ where $k \neq l$}
                \State With probability $p_{\text{cross}}$: randomly fill $\rho_{\text{cross}}$ fraction of $\mathbf{X}[E_k, V_l]$ with $\text{Poisson}(\lambda_{\text{cross}})$ values.
            \EndFor
            \State \Return $\mathbf{X}$, ground-truth labels.
        \end{algorithmic}
    \end{algorithm}

\section{Experiments and Results}

\subsection{Baseline and Competing Methods}

To evaluate the performance of DIPHW and CoMem-DIPHW, we compared them with 13 cell-clustering methods from various categories, including community detection, embedding-based methods, and recent deep learning methods that cluster scRNA-seq data. 
 Table~\ref{tab:clustering_methods} summarizes these methods. For a fair comparison under consistent conditions, we focus on core clustering algorithms rather than specific implementations in packages such as Seurat~\cite{satija2015spatial}, Scanpy~\cite{wolf2018scanpy}, and WGCNA~\cite{langfelder2008wgcna}. This approach allows us to evaluate the clustering capabilities of each method independently of any additional preprocessing or optimization steps. The hyperparameters used for each experiment are detailed in Table~\ref{tab:hyperparameters} in the supplementary materials.

\begin{table*}[!h]
\centering
\scriptsize
\begin{tabular}{|l|l|l|l|}
    \hline
    \textbf{Method} & \textbf{Type} & \textbf{Input} & \textbf{Packages} \\ \hline
    Greedy Modularity \cite{newman2004fast} & Community detection & Cell co-expression network & igraph \cite{csardi2006igraph} \\ \hline
    Louvain \cite{blondel2008fast} & Community detection & Cell co-expression network & \parbox[t]{5.5cm}{\raggedright Seurat \cite{satija2015spatial}, Scanpy \cite{wolf2018scanpy}, WGCNA \cite{langfelder2008wgcna}\strut} \\ \hline
    Infomap \cite{rosvall2008maps} & Community detection & Cell co-expression network & igraph \cite{csardi2006igraph} \\ \hline
    Leiden \cite{traag2019louvain} & Community detection & Cell co-expression network & igraph \cite{csardi2006igraph} \\ \hline
    PCA \cite{pearson1901liii} & Linear embedding & Gene $\times$ cell matrix & \parbox[t]{5.5cm}{\raggedright Seurat \cite{satija2015spatial}, Scanpy \cite{wolf2018scanpy}, SC3 \cite{kiselev2017sc3}, Cell Ranger \cite{zheng2017massively}\strut} \\ \hline
    t-SNE \cite{van2008visualizing} & Manifold embedding & Gene $\times$ cell matrix & \parbox[t]{5.5cm}{\raggedright Seurat~\cite{satija2015spatial}, Scanpy \cite{wolf2018scanpy}\strut} \\ \hline
    UMAP \cite{mcinnes2018umap} & Manifold embedding & Gene $\times$ cell matrix & \parbox[t]{5.5cm}{\raggedright Seurat~\cite{satija2015spatial}, Scanpy \cite{wolf2018scanpy}\strut} \\ \hline
    Node2Vec \cite{grover2016Node2Vec} & Graph embedding & Cell co-expression network & \parbox[t]{5.5cm}{\raggedright Multiple implementations available~\cite{grover_Node2Vec_2016,snap_Node2Vec_nodate}\strut} \\ \hline
    graph-sc \cite{ciortan2022gnn} & Deep graph embedding & Cell co-expression network & GitHub \cite{graphsc_github} \\ \hline
    tsImpute \cite{zheng2023tsimpute} & Imputation & Gene $\times$ cell matrix & GitHub \cite{tsimpute_github} \\ \hline
    CAKE \cite{liu2024cake} & Deep contrastive clustering & Gene $\times$ cell matrix & GitHub \cite{cake_github} \\ \hline
    scASDC \cite{min2024scasdc} & Deep clustering & Gene $\times$ cell matrix & GitHub \cite{scasdc_github} \\ \hline
    \parbox[t]{2cm}{\raggedright EDVW \cite{Chitra2019} + word2vec \cite{mikolov2013efficient}\strut} & Hypergraph embedding & Gene $\times$ cell matrix & \parbox[t]{5.5cm}{\raggedright Implemented based on \cite{Chitra2019}; GitHub \cite{wan2025comem}\strut} \\ \hline
    \parbox[t]{2cm}{\raggedright DIPHW + word2vec \cite{mikolov2013efficient}\strut} & Hypergraph embedding & Gene $\times$ cell matrix & \parbox[t]{5.5cm}{\raggedright Our proposed method; GitHub \cite{wan2025comem}\strut} \\ \hline
    \parbox[t]{2cm}{\raggedright CoMem + word2vec \cite{mikolov2013efficient}\strut} & Hypergraph embedding & \parbox[t]{2.8cm}{\raggedright Gene $\times$ cell matrix, plus gene and cell co-expression networks\strut} & \parbox[t]{5.5cm}{\raggedright Our proposed method; GitHub \cite{wan2025comem}\strut} \\ \hline
\end{tabular}
\caption{List of Clustering Methods Evaluated in Our Study. K-means was used to cluster the output of all embedding-based methods that do not include a clustering algorithm. Since tsImpute is an imputation method, we evaluated its impact on clustering performance by applying PCA followed by K-means to its imputed data.}
\label{tab:clustering_methods}
\end{table*}

\subsection{Data and Preprocessing}
\label{subsec:dataandpreprocessing}

We use simulated and publicly available scRNA-seq datasets for our study. The simulated data is generated using our simulation algorithm described in Section~\ref{subsec:Simulation}, with pseudocode in Algorithm~\ref{alg:simulation_sc} and code at~\cite{wan2025comem}.

For both simulated and real scRNA-seq data, preprocessing includes Counts Per Million normalization~\cite{vallejos2017normalizing}  to account for differences in sequencing depth, log transformation to reduce skewness, and removal of genes and cells with zero total expression. For real scRNA-seq datasets, we additionally retain the top $n$ highly variable genes using Scanpy's implementation of the Seurat~\cite{satija2015spatial} method. The values of $n$ used for each dataset are provided in Table~\ref{tab:hyperparameters}. 

We use six publicly available scRNA-seq datasets: (1) human brain~\cite{camp2015human}, (2) human pancreas~\cite{muraro2016single}, (3) mouse brain~\cite{zeisel2015cell}, (4) mouse pancreas~\cite{baron2016single}, and (5) scMixology benchmark with 3 classes and (6) scMixology benchmark with 5 classes. scMixology benchmark which is a curated benchmark dataset with ground-truth cell line labels~\cite{tian2019benchmarking}. Table~\ref{tab:clustering_methods} provides a summary of these datasets.

\begin{table}[!h]
\centering
\begin{tabular}{|c|c|c|c|c|}
\hline
\textbf{Datasets} & ~\# of Cells~ & ~\# of Genes~ & ~\textbf{Sparsity}~ & \textbf{Results in} \\
\hline
scMixology Benchmark (3-class)~\cite{tian2019benchmarking} & 902 & 16,468 & 45.02\% & Table~\ref{tab:scMixology} \\
\hline
scMixology Benchmark (5-class)~\cite{tian2019benchmarking} & 3,918 & 11,786 & 63.01\% & Table~\ref{tab:scMixology} \\
\hline
Human Pancreas ~\cite{muraro2016single} & 3,072 & 18,348 & 78.74\% & Section~\ref{subsec:human_pancreas_results} \\
\hline
Human Brain ~\cite{camp2015human} & 735 & 18,929 & 80.11\% & Section~\ref{subsec:human_brain_results}\\
\hline
Mouse Brain ~\cite{zeisel2015cell} & 3,006 & 19,973 & 81.21\% & Section~\ref{subsec:mouse_pancreas_brain_results} \\
\hline
Mouse Pancreas ~\cite{baron2016single} & 1,065 & 14,881 & 87.79\% & Section~\ref{subsec:mouse_pancreas_brain_results} \\
\hline
\end{tabular}
\caption{List of Publicly Available Datasets Used in Our Study. Sparsity measures the percentage of zero entries in a cell $\times$ gene matrix.}
\label{tab:data_summary}
\end{table}

\subsection{Sparsity-Induced Inflated Correlation in scRNA-seq Co-expression Networks}
\label{subsec:sparsity}
The zero inflation problem in scRNA-seq data analysis refers to the excessive number of zero counts (or "dropouts") observed in the data. These zeros may arise from technical limitations in detecting low-abundance transcripts (false negatives) or from the true absence of transcripts (true negatives)~\cite{jiang2022statistics,qiu2020embracing,kim2020demystifying}. Integrating individual cell expression profiles into a unified scRNA-seq dataset requires constructing a matrix that includes every gene detected in any cell, which introduces zero entries wherever genes are not expressed in those cells. Because cells typically express different sets of genes, this process creates significant sparsity in the merged dataset. In co-expression network construction, edge weights are inflated due to these artificially introduced common zeros, as the common zeros are interpreted as evidence of expression homogeneity. This issue is especially pronounced at the single-cell level compared to bulk tissue RNA-seq data, since the proportion of expressed genes per cell is much smaller, resulting in greater data sparsity.

\begin{figure}[!h]
\centering
\includegraphics[scale=0.4]{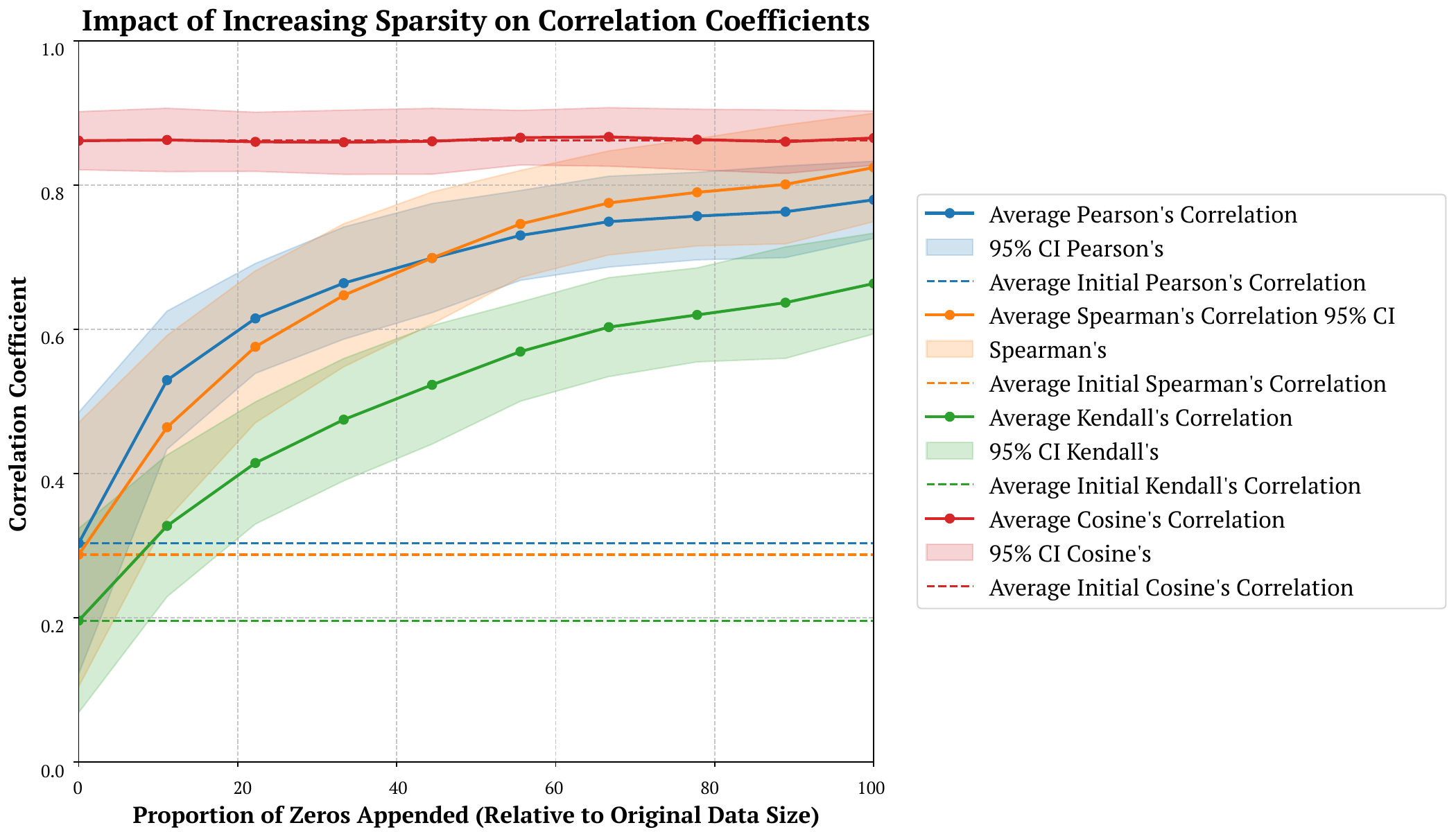}
\caption{Impact of Increased Sparsity on Correlation Coefficients. This figure shows the relationship between the proportion of zeros added to the expression profile (representing increased sparsity) and the resulting correlation coefficients. This process was repeated 100 times. Shaded areas indicate the 95\% confidence intervals around the average correlation values. The dashed lines represent the average initial correlation without added sparsity. Induced sparsity in scRNA-seq data can inflate correlations in co-expression networks, highlighting the need for alternative scRNA-seq data representations.}
\label{fig:zeroInflation}
\end{figure}

In Fig.~\ref{fig:zeroInflation}, we examine how induced sparsity affects correlation coefficients and the interpretation of gene and cell connections in co-expression networks. To simulate this effect, we generate gene expression data with a fixed base size and introduce varying proportions of zeros to represent different sparsity levels. Our results indicate that as induced sparsity increases, correlation coefficients between expression profiles become artificially inflated, leading to an overestimation of similarity between cell expression profiles. These findings highlight the importance of addressing zero inflation in scRNA-seq data analysis and the need for accurate representation of underlying biological relationships.

Table~\ref{tab:CorrelationClusteringComparison} (in the Supplementary Section~\ref{subsec:cosine_similarity}) shows that ignoring shared zeros using cosine similarity results in poorer clustering performance, as these inactive genes still have biological significance.

\subsection{Results on the Simulated Data}
\label{subsubsec:modularity_clustering}

To evaluate performance of the clustering methods on datasets with known ground-truth, we employ five widely used measures to compute coherence between the identified clusters and the ground-truth clusters: Adjusted Rand Index (ARI), Normalized Mutual Information (NMI), Adjusted Mutual Information (AMI), Clustering Accuracy (ACC), and F1 Score. All evaluation measures are invariant to permutations, with a score of 0 representing random labeling and a score of 1 indicating identical clusters. K-means is used to cluster the output of all embedding-based methods that do not directly assign cluster membership.

\begin{figure*}[!h]
\centering
\includegraphics[width=\textwidth]{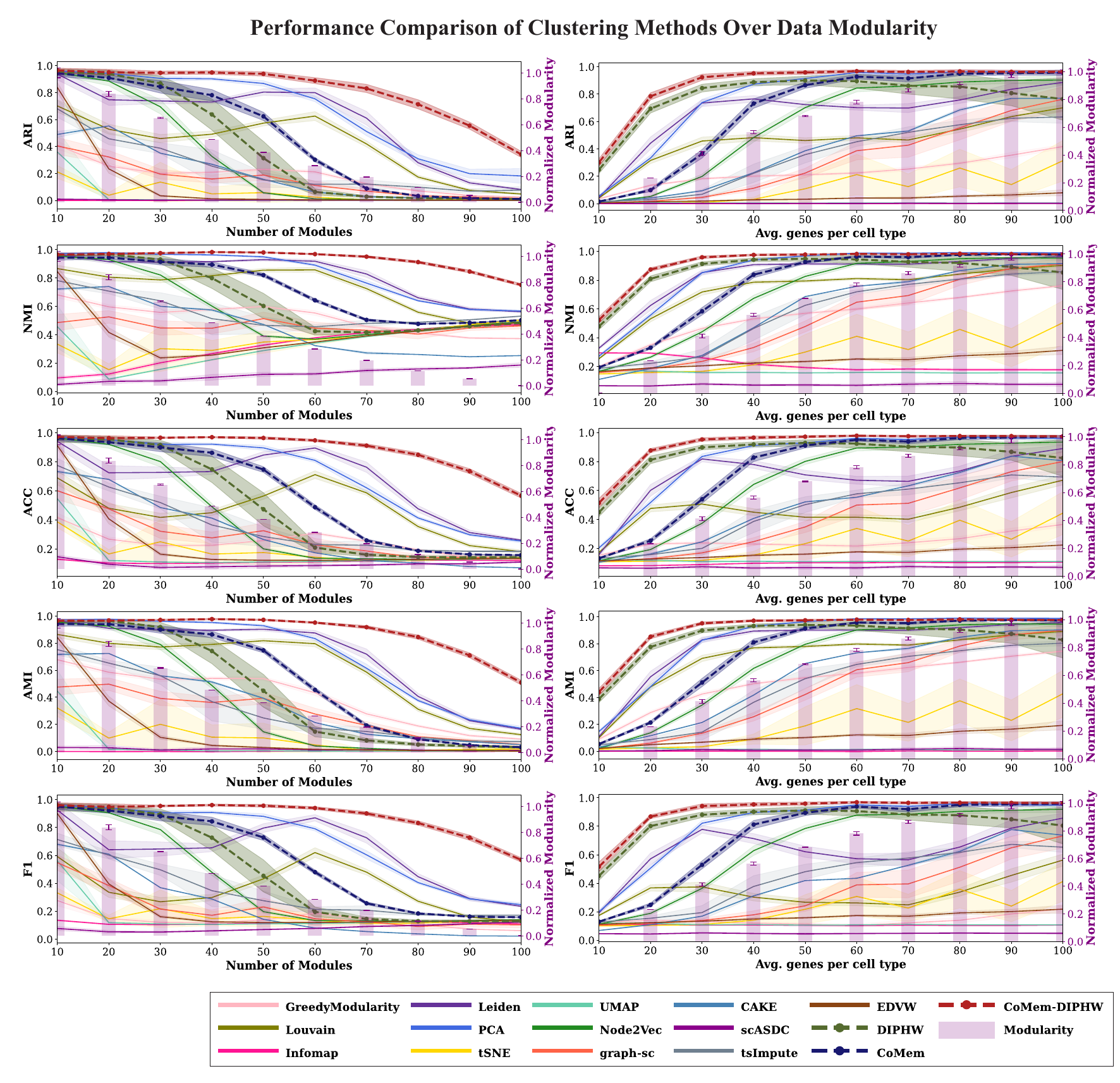}
\caption{Comparing Performances of Clustering Methods on Simulated scRNA-seq Data Across Modularity. The performance of clustering methods is measured by ARI, NMI, ACC, AMI, and F1 across different levels of data modularity. The x-axis represents parameters determining data modularity, including the average number of co-expressed genes per cell type and the number of embedded modules. Our proposed methods, DIPHW, CoMem, and CoMem-DIPHW, are highlighted with dashed lines. CoMem-DIPHW, in particular, maintains strong performance and is least affected by decreases in modularity, performing significantly better even under conditions of extremely weak modularity. This behavior is observed consistently across all evaluation measures. Experiments are repeated 10 times for each parameter setting, and the 95\% confidence intervals (CIs) are shown. The bar plots represent Barber's bipartite modularity, normalized so that the most modular graphs have modularity = 1. K-means is used to cluster the output of all embedding-based methods that do not include an inherent clustering mechanism.}
\label{fig:PerformanceByModularity}
\end{figure*}

\begin{figure*}[!h]
\centering
\includegraphics[width=\textwidth]{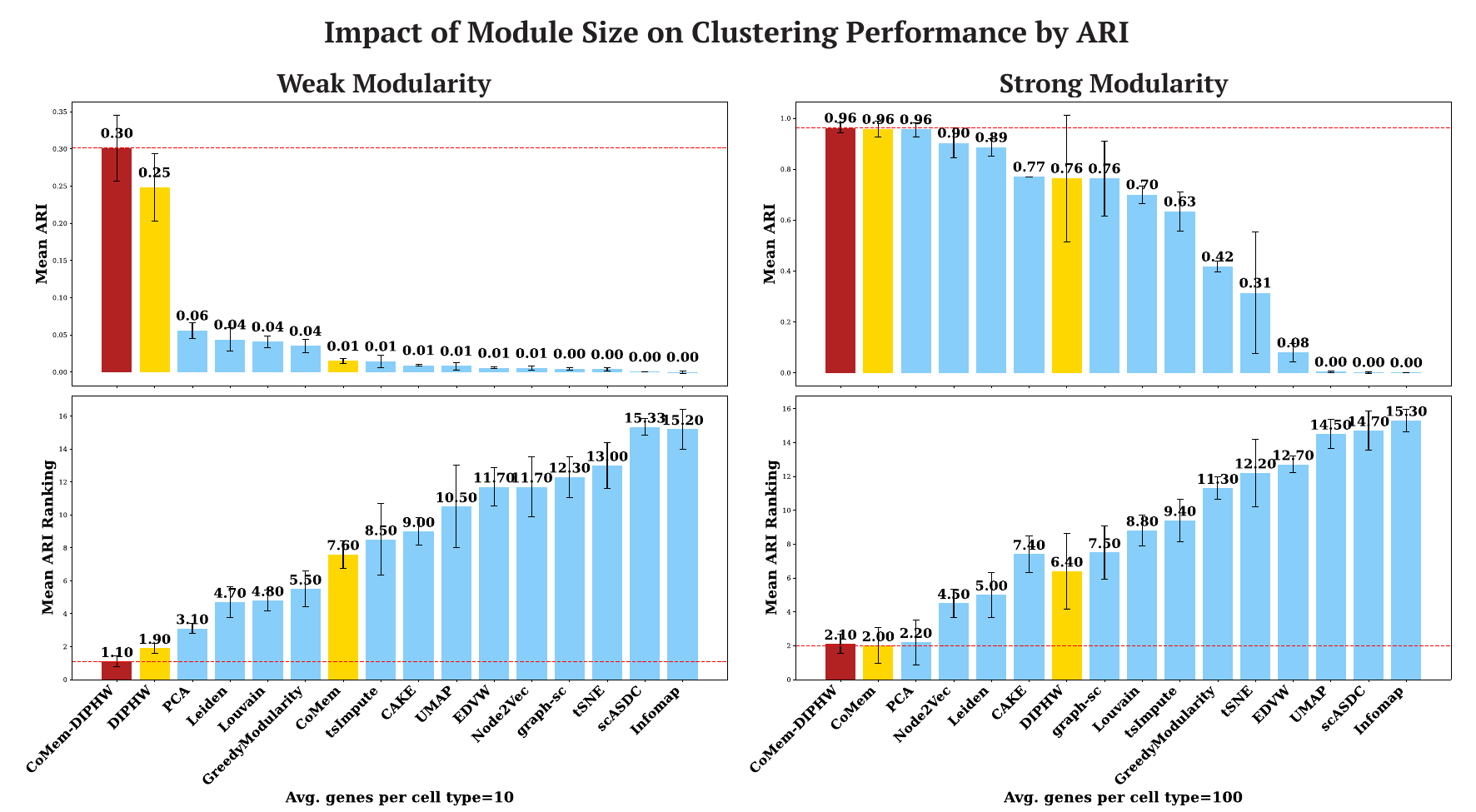}
\caption{Impact of Module Size on Clustering Performance by ARI. The bar plots show the mean ARI (top) and average ranking (bottom) for each method, ordered by their average ARI. Each experiment was repeated 10 times per parameter setting, with error bars representing the 95\% confidence interval. Red dashed lines indicate the highest ARI values or best ARI rankings. Our proposed hypergraph-based methods, highlighted in red and yellow, show significant advantages when data modularity is weak, that is, when the average number of co-expressed genes per cell type module is low (10 genes per module). Under strong modularity conditions, where the average number of genes per cell type module is high (100 genes per module), many methods (e.g., PCA, CoMem-DIPHW, CoMem, and Node2Vec) perform well and achieve average ARI scores above 0.9. These observations are consistent across NMI, ACC, AMI, and F1, as shown in Supplementary Section~\ref{subsec:additionalresults}. K-means is used to cluster the output of all embedding-based methods that do not directly assign cluster membership.}
\label{fig:Ngenes_ARI}
\end{figure*}

\begin{figure*}[!h]
  \centering
  \includegraphics[width=1\textwidth]{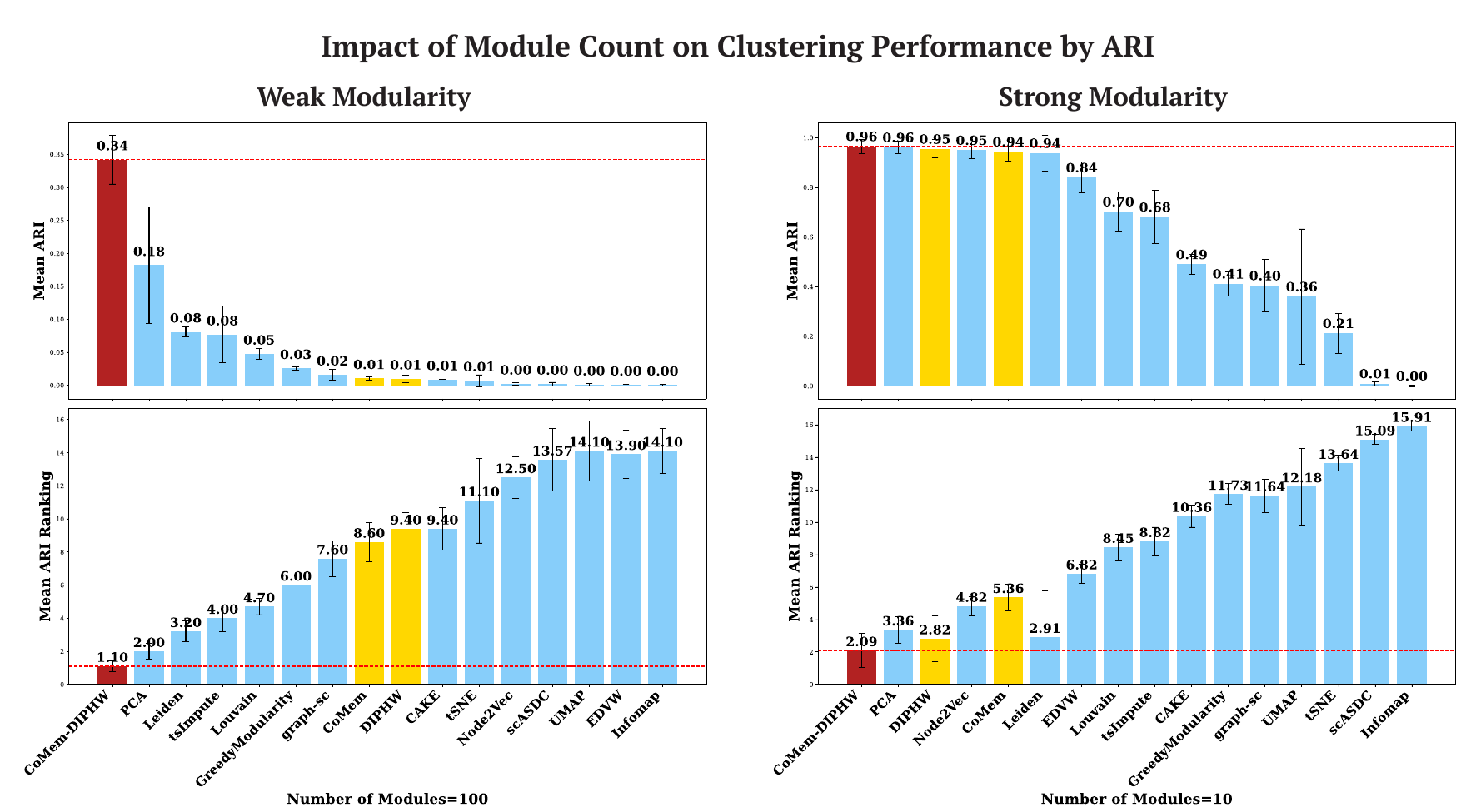} 
  \caption{Impact of Module Count on Clustering Performance by ARI. The bar plots show the mean ARI (top) and average ranking (bottom) of each method, ordered by their average ARI. Our proposed hypergraph-based methods, highlighted in red and yellow, demonstrate significant advantages in weak modularity regimes, where the number of embedded modules (representing the number of cell types) in the simulated data is high (100 modules). Each experiment is repeated 10 times per parameter setting, with error bars representing the 95\% confidence interval. Red dashed lines indicate the highest ARI values or best ARI rankings. These observations are consistent across NMI, ACC, AMI, and F1, as shown in Supplementary Section~\ref{subsec:additionalresults}. K-means is used to cluster the output of all embedding-based methods that do not directly assign cluster membership.}
\label{fig:Nmodules_ARI}
\end{figure*}

We evaluate the performance of clustering methods under varying data modularity conditions. Modularity is controlled by two key parameters: the average number of co-expressed genes per cell type and the number of embedded cell modules. The number of modules corresponds to the number of cell types in the simulated scRNA-seq data. Weak modularity refers to a small number of co-expressed genes per module or a large number of embedded modules, which makes data structures more difficult to detect. The modularity bar plots in Fig.~\ref{fig:PerformanceByModularity} and the heatmap visualization in Fig.~\ref{fig:ModularityVisual} show that the simulated scRNA-seq data exhibit stronger modularity when the average number of co-expressed genes per cell type is higher or when the number of modules is smaller. We implemented Barber's bipartite modularity~\cite{barber2007modularity}, as shown in the bar plots in Fig.~\ref{fig:PerformanceByModularity}, to quantify the modularity of the underlying data for clustering and to analyze the association between modularity and clustering performance. Clustering performance generally improves as data modularity increases. Our proposed methods, DIPHW, CoMem, and CoMem-DIPHW, demonstrate strong and highly competitive performance across varying modularity. Specifically, CoMem-DIPHW consistently remains competitive with the best-performing methods evaluated, with its advantage particularly pronounced under weak data modularity conditions. For clarity, we present the most contrasting cases from our analysis: (i) an average of 10 and 100 co-expressed genes per module in Fig.~\ref{fig:Ngenes_ARI} and (ii) 10 and 100 embedded modules in Fig.~\ref{fig:Nmodules_ARI}.

In Figures~\ref{fig:Ngenes_ARI} and~\ref{fig:Nmodules_ARI}, we compare the performance of our proposed hypergraph-based clustering methods (highlighted in red and yellow) with five embedding-based approaches, four community detection algorithms, and four state-of-the-art scRNA-seq clustering methods. K-means is used to cluster the output of all embedding-based methods that do not directly assign cluster membership. In weak modularity regimes (average number of co-expressed genes per module = 10 or number of modules = 100), CoMem-DIPHW and DIPHW show a clear performance advantage. For example, when the average number of co-expressed genes is 10, CoMem-DIPHW achieves an ARI of 0.30 and DIPHW achieves an ARI of 0.25, compared to 0.05 for the best baseline method (PCA), representing a performance improvement of nearly 450\%. Under strong modularity conditions (average number of genes per module = 100 or number of modules = 10), multiple methods (PCA, CoMem-DIPHW, CoMem, and Node2Vec) performed well, achieving average ARI scores above 0.9.

Additional results from intermediate parameter settings (average number of co-expressed genes per module: 20, 30, 40, 70, 80, 90; number of embedded modules: 20, 30, 40, 70, 80, 90), as measured by NMI, ACC, AMI, and F1, support the same observation. Across the 16 parameter settings under different modularity conditions, CoMem-DIPHW ranks first in all 16 settings by ACC and F1, 15 by ARI, and 10 each by NMI and AMI. DIPHW and CoMem also perform strongly. Either DIPHW or CoMem rank among the top three methods in at least 10 of 16 settings across all evaluation measures. Our hypergraph-based methods show significant advantages in weak modularity regimes, where data structures are harder to detect. Under comparatively strong modularity conditions, several methods (PCA, CoMem-DIPHW, CoMem, Node2Vec, DIPHW, and Leiden) perform well, with average NMI scores above 0.9. Notably, PCA slightly outperforms CoMem-DIPHW by ARI, NMI, and AMI under strong modularity, with differences in scores ranging from 0.01 to 0.02. These additional results are in Supplementary Section~\ref{subsec:additionalresults}.


\subsection{Results on the scMixology Benchmark Datasets}
\label{subsec:scMixology}

A major challenge in scRNA-seq analysis is the lack of reliable ground truth for cell type annotation, which complicates the evaluation of clustering algorithms. In most scRNA-seq datasets, cell type labels are assigned using a combination of clustering results, differential expression (DE) analysis, and canonical marker genes. First, cells are grouped into clusters based on transcriptomic similarity using clustering algorithms such as K-means. Next, DE analysis identifies genes that are differentially expressed across clusters. Finally, these clusters are annotated based on known canonical marker genes associated with specific cell types. Because this annotation process relies on clustering results, using these labels to evaluate clustering performance introduces circular logic and bias toward the method originally used for annotation. To address this issue, we use the mixture control benchmarking dataset scMixology~\cite{tian2019benchmarking}, where ground truth is established based on cell line identity.

We evaluate the performance of our proposed method, CoMem-DIPHW, using two benchmarking datasets from scMixology~\cite{tian2019benchmarking}. These datasets provide ground truth labels for cell type annotation based on human lung adenocarcinoma cell lines. As shown in Table~\ref{tab:scMixology}, tsImpute achieved the highest scores across all five evaluation measures (ARI, NMI, AMI, ACC, and F1) on the 3-class dataset, with CoMem-DIPHW, Louvain, and Leiden tied for second. On the 5-class dataset, CoMem-DIPHW achieved the best performance across all evaluation measures, followed by tsImpute and graph-sc. This result is consistent with the pattern observed in our simulation analysis (see Section~\ref{subsubsec:modularity_clustering}), where CoMem-DIPHW maintains strong performance under low modularity conditions.

\begin{table}[!h]
\centering
\resizebox{\textwidth}{!}{%
\begin{tabular}{|l|c|c|c|c|c|c|c|c|}
\toprule
\multicolumn9{c}{\textbf{Traditional Embedding and Community Detection Methods}} \\
\midrule
\textbf{Dataset / Evaluation Measures} & \textbf{GreedyModularity} & \textbf{Louvain} & \textbf{Infomap} & \textbf{Leiden} & \textbf{PCA} & \textbf{t-SNE} & \textbf{Node2Vec} & \textbf{UMAP} \\
\midrule
\rowcolor{gray!10}
\multicolumn9{l}{\textit{scMixology 3 class}} \\ \hline

ARI & \num{0.5836} $\pm$ 0.000 & \underline{\num{0.9933} $\pm$ 0.000} & \num{0.2327} $\pm$ 0.000 & \underline{\num{0.9933} $\pm$ 0.000} & \num{0.9900} $\pm$ 0.000 & \num{0.9898} $\pm$ 0.000 & \num{0.9913} $\pm$ \num{0.0037} & \num{0.9898} $\pm$ 0.000 \\

NMI & \num{0.7193} $\pm$ 0.000 & \underline{\num{0.9864} $\pm$ 0.000} & \num{0.3843} $\pm$ 0.000 & \underline{\num{0.9864} $\pm$ 0.000} & \num{0.9797} $\pm$ 0.000 & \num{0.9810} $\pm$ 0.000 & \num{0.9829} $\pm$ \num{0.0075} & \num{0.9810} $\pm$ 0.000 \\

ACC & \num{0.6929} $\pm$ 0.000 & \underline{\num{0.9978} $\pm$ 0.000} & \num{0.3370} $\pm$ 0.000 & \underline{\num{0.9978} $\pm$ 0.000} & \num{0.996} $\pm$ 0.000 & \num{0.9967} $\pm$ 0.000 & \num{0.9971} $\pm$ \num{0.0013} & \num{0.9967} $\pm$ 0.000 \\

AMI & \num{0.7189} $\pm$ 0.000 & \underline{\num{0.9864} $\pm$ 0.000} & \num{0.2335} $\pm$ 0.000 & \underline{\num{0.9864} $\pm$ 0.000} & \num{0.9796} $\pm$ 0.000 & \num{0.9810} $\pm$ 0.000 & \num{0.9829} $\pm$ \num{0.0075} & \num{0.9810} $\pm$ 0.000 \\

F1 & \num{0.5868} $\pm$ 0.000 & \underline{\num{0.9978} $\pm$ 0.000} & \num{0.4588} $\pm$ 0.000 & \underline{\num{0.9978} $\pm$ 0.000} & \num{0.9967} $\pm$ 0.000 & \num{0.9967} $\pm$ 0.000 & \num{0.9971} $\pm$ \num{0.0013} & \num{0.9967} $\pm$ 0.000 \\ \hline

\rowcolor{gray!10}
\multicolumn9{l}{\textit{scMixology 5 class}} \\ \hline

ARI & \num{0.6645} $\pm$ 0.000 & \num{0.6746} $\pm$ 0.000 & \num{0.3002} $\pm$ 0.0008 & \num{0.6746} $\pm$ 0.000 & \num{0.9873} $\pm$ 0.000 & \num{0.9861} $\pm$ 0.000 & \num{0.9451} $\pm$ \num{0.0073} & \num{0.9838} $\pm$ 0.000 \\

NMI & \num{0.7852} $\pm$ 0.000 & \num{0.7953} $\pm$ 0.000 & \num{0.4355} $\pm$ \num{0.0003} & \num{0.7917} $\pm$ 0.000 & \num{0.9747} $\pm$ 0.000 & \num{0.9739} $\pm$ 0.000 & \num{0.9150} $\pm$ \num{0.0086} & \num{0.9707} $\pm$ 0.000 \\

ACC & \num{0.7325} $\pm$ 0.000 & \num{0.7361} $\pm$ 0.000 & \num{0.4150} $\pm$ \num{0.0018} & \num{0.7358} $\pm$ 0.000 & \num{0.9941} $\pm$ 0.000 & \num{0.9939} $\pm$ 0.000 & \num{0.9621} $\pm$ \num{0.0059} & \num{0.9929} $\pm$ 0.000 \\

AMI & \num{0.7851} $\pm$ 0.000 & \num{0.7951} $\pm$ 0.000 & \num{0.2735} $\pm$ \num{0.0003} & \num{0.7916} $\pm$ 0.000 & \num{0.9747} $\pm$ 0.000 & \num{0.9738} $\pm$ 0.000 & \num{0.9149} $\pm$ \num{0.0086} & \num{0.9707} $\pm$ 0.000 \\

F1 & \num{0.6556} $\pm$ 0.000 & \num{0.6576} $\pm$ 0.000 & \num{0.5250} $\pm$ \num{0.0028} & \num{0.6571} $\pm$ 0.000 & \num{0.9941} $\pm$ 0.000 & \num{0.9939} $\pm$ 0.000 & \num{0.9618} $\pm$ \num{0.0062} & \num{0.9929} $\pm$ 0.000 \\ \hline

\addlinespace[2mm]
\multicolumn9{c}{\textbf{Advanced ScRNA-seq and Hypergraph-Based Methods}} \\
\midrule
\textbf{Dataset / Evaluation Measure} & \textbf{EDVW} & \textbf{graph-sc} & \textbf{tsImpute} & \textbf{CAKE} & \textbf{scASDC} & \textbf{DIPHW} & \textbf{CoMem} & \textbf{CoMem-DIPHW} \\ 
\midrule
\rowcolor{gray!10}
\multicolumn9{l}{\textit{scMixology 3 class}} \\ \hline

ARI & \num{0.5960} $\pm$ \num{0.0454} & \num{0.7020} $\pm$ \num{0.2650} &  \textbf{0.997 $\pm$ 0.000} & \num{0.8318} $\pm$ \num{0.0196} & \num{0.4227} $\pm$ \num{0.4342} & \num{0.9900} $\pm$ \num{0.0052} & \num{0.9912} $\pm$ \num{0.0019} & \underline{\num{0.9933} $\pm$ 0.000} \\

NMI & \num{0.6140} $\pm$ \num{0.0266} & \num{0.7654} $\pm$ \num{0.1883} & \textbf{0.993 $\pm$ 0.000} & \num{0.8609} $\pm$ \num{0.0205} & \num{0.4500} $\pm$ \num{0.4397} & \num{0.9803} $\pm$ \num{0.0097} & \num{0.9829} $\pm$ \num{0.0033} & \underline{\num{0.9864} $\pm$ 0.000} \\

ACC & \num{0.7996} $\pm$ \num{0.0557} & \num{0.8612} $\pm$ \num{0.1275} & \textbf{0.999 $\pm$ 0.000} & \num{0.9958} $\pm$ \num{0.0011} & \num{0.6435} $\pm$ \num{0.2956} & \num{0.9967} $\pm$ \num{0.0018} & \num{0.9971} $\pm$ \num{0.0006} & \underline{\num{0.9978} $\pm$ 0.000} \\

AMI & \num{0.6132} $\pm$ \num{0.0266} & \num{0.7649} $\pm$ \num{0.1887} & \textbf{0.993 $\pm$ 0.000} & \num{0.8605} $\pm$ \num{0.0206} & \num{0.4479} $\pm$ \num{0.4418} & \num{0.9803} $\pm$ \num{0.0097} & \num{0.9829} $\pm$ \num{0.0033} & \underline{\num{0.9864} $\pm$ 0.000} \\

F1 & \num{0.7991} $\pm$ \num{0.0561} & \num{0.8513} $\pm$ \num{0.1380} & \textbf{0.999 $\pm$ 0.000} & \num{0.9957} $\pm$ \num{0.0011} & \num{0.5477} $\pm$ \num{0.3810} & \num{0.9967} $\pm$ \num{0.0018} & \num{0.9971} $\pm$ \num{0.0006} & \underline{\num{0.9978} $\pm$ 0.000} \\ \hline

\rowcolor{gray!10}
\multicolumn9{l}{\textit{scMixology 5 class}} \\ \hline

ARI & \num{0.6572} $\pm$ \num{0.0221} & \num{0.9884} $\pm$ \num{0.0015} & \underline{\num{0.9892} $\pm$ 0.000} & \num{0.7333} $\pm$ \num{0.0746} & \num{0.3919} $\pm$ \num{0.3602} & \num{0.9440} $\pm$ \num{0.0031} & \num{0.9879} $\pm$ \num{0.0013} & \textbf{0.990 $\pm$ 0.002} \\

NMI & \num{0.7421} $\pm$ \num{0.0197} & \num{0.9782} $\pm$ \num{0.0019} & \underline{\num{0.9796} $\pm$ 0.000} & \num{0.8435} $\pm$ \num{0.0217} & \num{0.4575} $\pm$ \num{0.4160} & \num{0.9092} $\pm$ \num{0.0043} & \num{0.9771} $\pm$ \num{0.0026} & \textbf{0.980 $\pm$ 0.003} \\

ACC & \num{0.6985} $\pm$ \num{0.0161} & \num{0.9950} $\pm$ \num{0.0006} & \underline{\num{0.9954} $\pm$ 0.000} & \num{0.8568} $\pm$ \num{0.0192} & \num{0.5553} $\pm$ \num{0.2156} & \num{0.9628} $\pm$ \num{0.0020} & \num{0.9946} $\pm$ \num{0.0007} & \textbf{0.996 $\pm$ 0.001} \\

AMI & \num{0.7418} $\pm$ \num{0.0197} & \num{0.9782} $\pm$ \num{0.0019} & \underline{\num{0.9796} $\pm$ 0.000} & \num{0.8431} $\pm$ \num{0.0218} & \num{0.4565} $\pm$ \num{0.4169} & \num{0.9091} $\pm$ \num{0.0043} & \num{0.9771} $\pm$ \num{0.0026} & \textbf{0.980 $\pm$ 0.003} \\

F1 & \num{0.7113} $\pm$ \num{0.0143} & \underline{\num{0.9950} $\pm$ \num{0.0006}} & \num{0.9949} $\pm$ \num{0.000} & \num{0.8579} $\pm$ \num{0.0175} & \num{0.3443} $\pm$ \num{0.2246} & \num{0.9628} $\pm$ \num{0.0018} & \num{0.9946} $\pm$ \num{0.0007} & \textbf{0.996 $\pm$ 0.001} \\

\bottomrule
\end{tabular}}
\caption{Performance of Clustering Methods on the scMixology Benchmark Datasets~\cite{tian2019benchmarking} across Five Evaluation Measures (ARI, NMI, ACC, AMI, and F1). The scMixology benchmark datasets include a 3-class and a 5-class dataset. Results for each are shown in the top and bottom sections, respectively. Best values are in bold, and second best are underlined. Results are reported as mean $\pm$ standard deviation over five independent runs. On the 3-class dataset, tsImpute achieved the highest scores across all evaluation measures, with CoMem-DIPHW, Louvain, and Leiden tied for second. On the more challenging 5-class dataset, CoMem-DIPHW achieved the best performance across all evaluation measures, outperforming tsImpute, which ranked second.}
\label{tab:scMixology}
\end{table}

\subsection{Results on the Human Pancreas Dataset}
\label{subsec:human_pancreas_results}

For datasets without ground-truth annotations (specifically, human pancreas, mouse pancreas, human brain, and mouse brain), we qualitatively evaluate clustering performance using differential expression (DE) analysis. We present detailed DE analysis results for the human pancreas dataset using CoMem-DIPHW, PCA, graph-sc, tsImpute, CAKE, and scASDC in this section and provide results for the human brain, mouse brain, and mouse pancreas in the Supplementary Sections~\ref{subsec:human_brain_results} and~\ref{subsec:mouse_pancreas_brain_results}.

\begin{figure*}[htbp]
  \centering
  \vspace{-10pt}
  \includegraphics[width=0.88\textwidth]{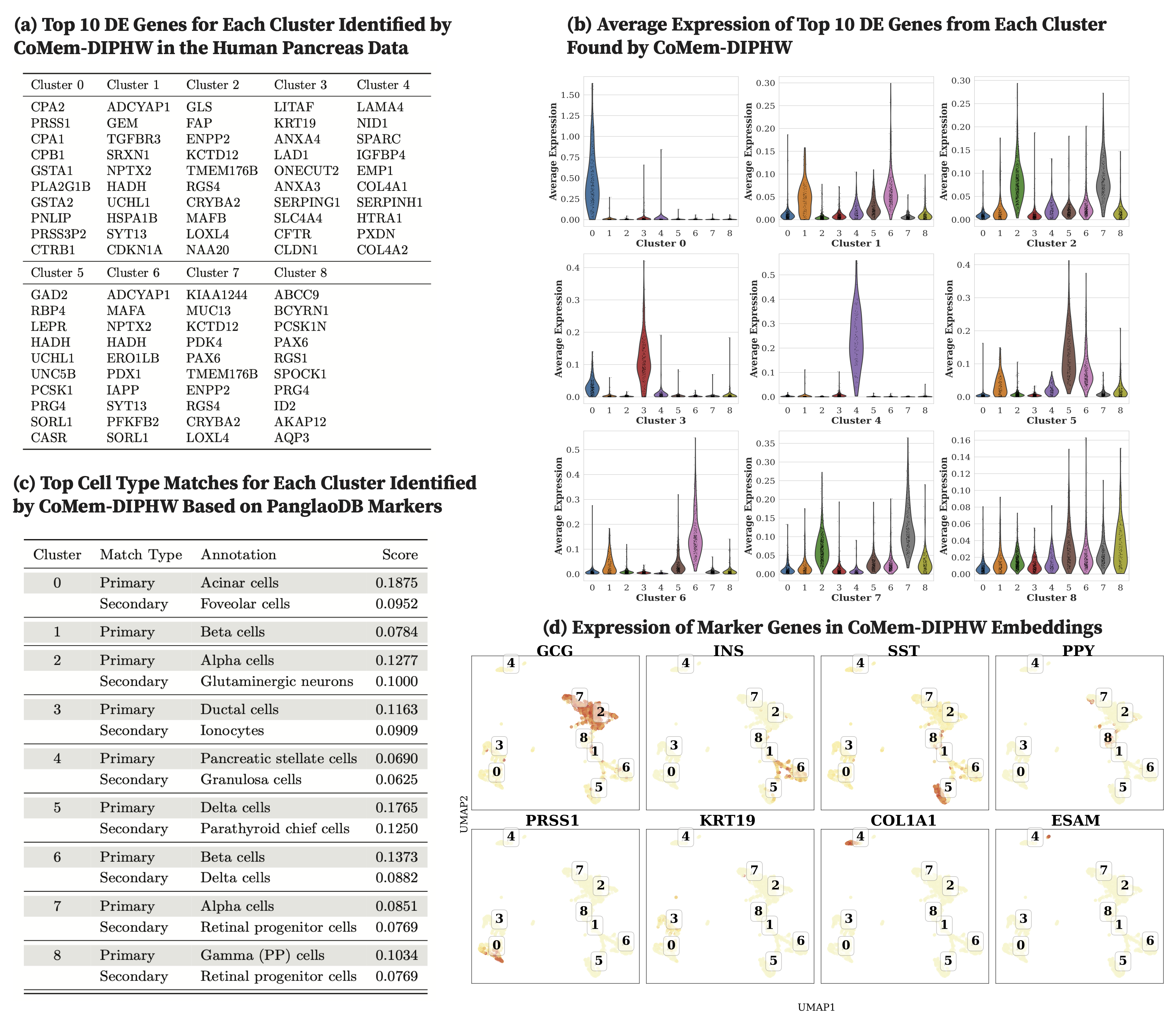} 
  \vspace{-5pt}
  \caption{Clustering Performance of CoMem-DIPHW on the Human Pancreas Dataset and Cell Type Annotation Using Differentially Expressed Genes (DEGs) and Canonical Markers. (a) Top 10 DEGs per cluster identified by CoMem-DIPHW. For example, Cluster 0 (acinar cells) includes genes associated with digestive enzyme production, a primary function of pancreatic acinar cells, including CPA1, CPA2~\cite{tamura2018mutations}, PLA2G1B~\cite{hui2019group}, CTRB1~\cite{pinho2011adult}, and PRSS1~\cite{masson2013conservative}. Clusters 2 and 7 (alpha cells) include markers GLS, CRYBA2, FAP, LOXL4, MAFB, and RGS4 (cluster 2), and LOXL4, PAX6, RGS4, and CRYBA2 (cluster 7). Cluster 3 (ductal cells) includes SLC4A4, ANXA4, KRT19, CFTR, CLDN1, and ONECUT2. Cluster 4, likely mesenchymal or endothelial cells, aligns with 9 of the top 10 DEGs found by CoMem-DIPHW, including COL4A2, COL4A1, SPARC, EMP1, IGFBP4, HTRA1, PXDN, LAMA4, and NID1. Cluster 5 (delta cells) includes GAD2, PCSK1, LEPR, and CASR, while Cluster 6 (beta cells) includes MAFA, ERO1LB, PDX1, IAPP, SYT13, HADH, and ADCYAP1. Cell type assignments are based on canonical marker expression patterns shown in (d). (b) Across-cluster average expression of cluster-specific DEGs. Violin plots show the distribution of average expression levels of these DEGs across all clusters. Strong clustering performance is indicated by high within-cluster expression of DEGs and low expression in other clusters. (c) Cell type annotation using the PanglaoDB marker database. Cell types are determined by the overlap between each cluster's DEGs and PanglaoDB cell type-specific markers, with match scores computed based on the proportion of matched markers. (d) Expression of human pancreatic canonical markers in CoMem-DIPHW embeddings. UMAP visualizations show CoMem-DIPHW cell embeddings, with cluster IDs assigned by K-means clustering. Each subplot displays the expression of a canonical marker gene for a specific human pancreatic cell type. The cell-type-specific markers considered include GCG (alpha cells), INS (beta cells), SST (delta cells), PPY (PP cells), PRSS1 (acinar cells), KRT19 (ductal cells), COL1A1 (mesenchymal cells), and ESAM (endothelial cells). Color intensity represents gene expression levels, with red hues indicating higher expression. Clusters showing high expression of cell-type-specific canonical markers are assigned to the corresponding cell types. Overall, clustering quality is demonstrated by (b) high within-cluster DEG expression, (c) consistency between PanglaoDB-based annotation and clustering, and (d) distinct separation of cell types.}
\label{fig:CoMem_HumanPancreas}
\end{figure*}

\begin{figure*}[htbp]
  \centering
  \includegraphics[width=\textwidth]{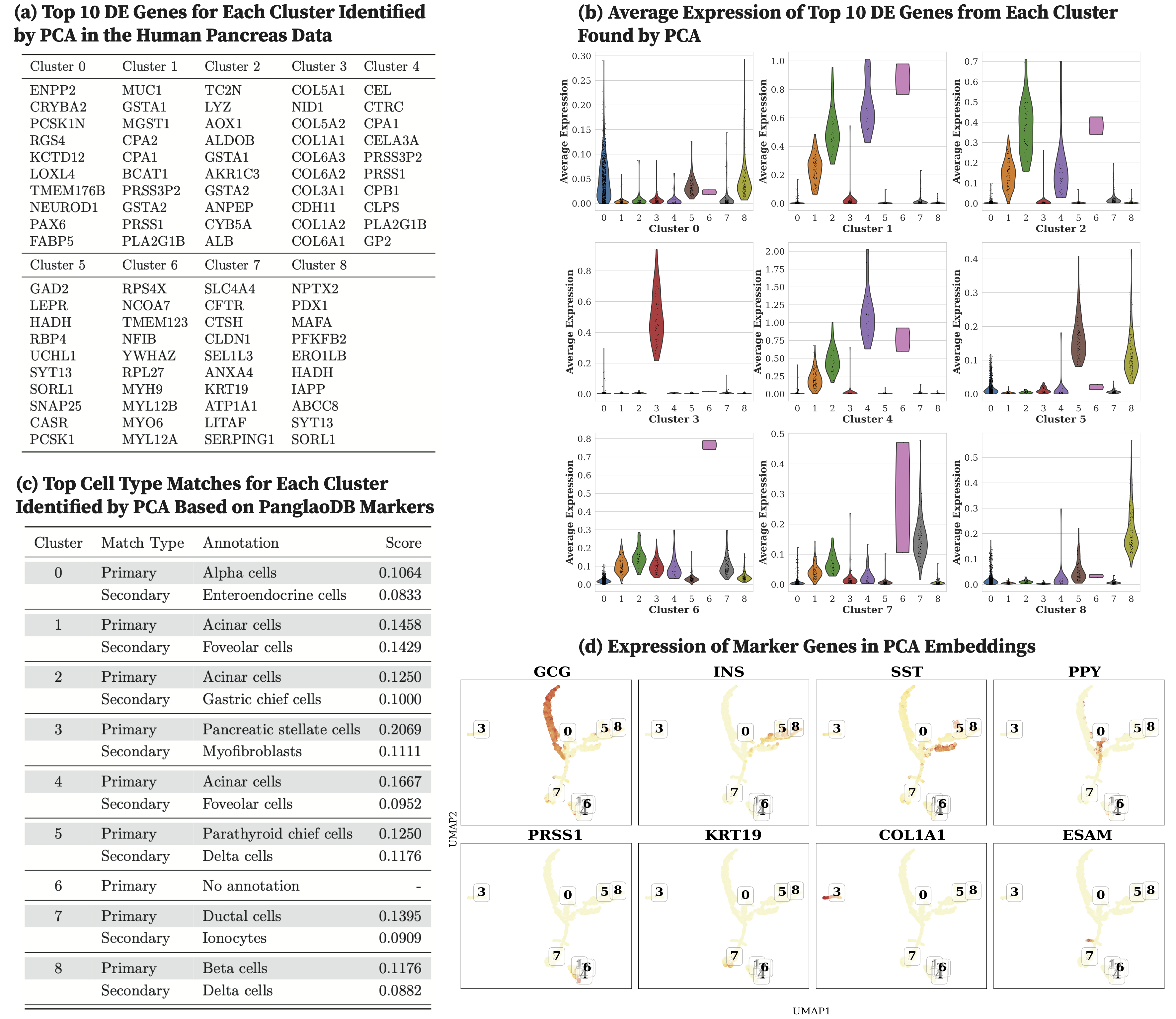} 
  \caption{Clustering Performance of PCA on the Human Pancreas Dataset and Cell Type Annotation Using Differentially Expressed Genes (DEGs) and Canonical Markers. (a) Top 10 DEGs per cluster identified by PCA. (b) Average expression of cluster-specific DEGs across clusters. Strong clustering performance is indicated by high expression of cluster-specific DEGs within their respective clusters and low expression in other clusters. (c) Cell type annotation using the PanglaoDB marker database. (d) Expression of human pancreatic canonical markers in PCA embeddings. UMAP visualizations show PCA cell embeddings, with cluster IDs assigned by K-means clustering. Overall, the clustering of cell types is limited compared to CoMem-DIPHW clustering, as shown by the overlapping beta cells, delta cells, and PP cells in (d) and the absence of distinct high expression patterns of the DEGs in (b).}
\label{fig:PCA_HumanPancreas}

\end{figure*}

\begin{figure*}[htbp]
  \centering
  \includegraphics[width=\textwidth]{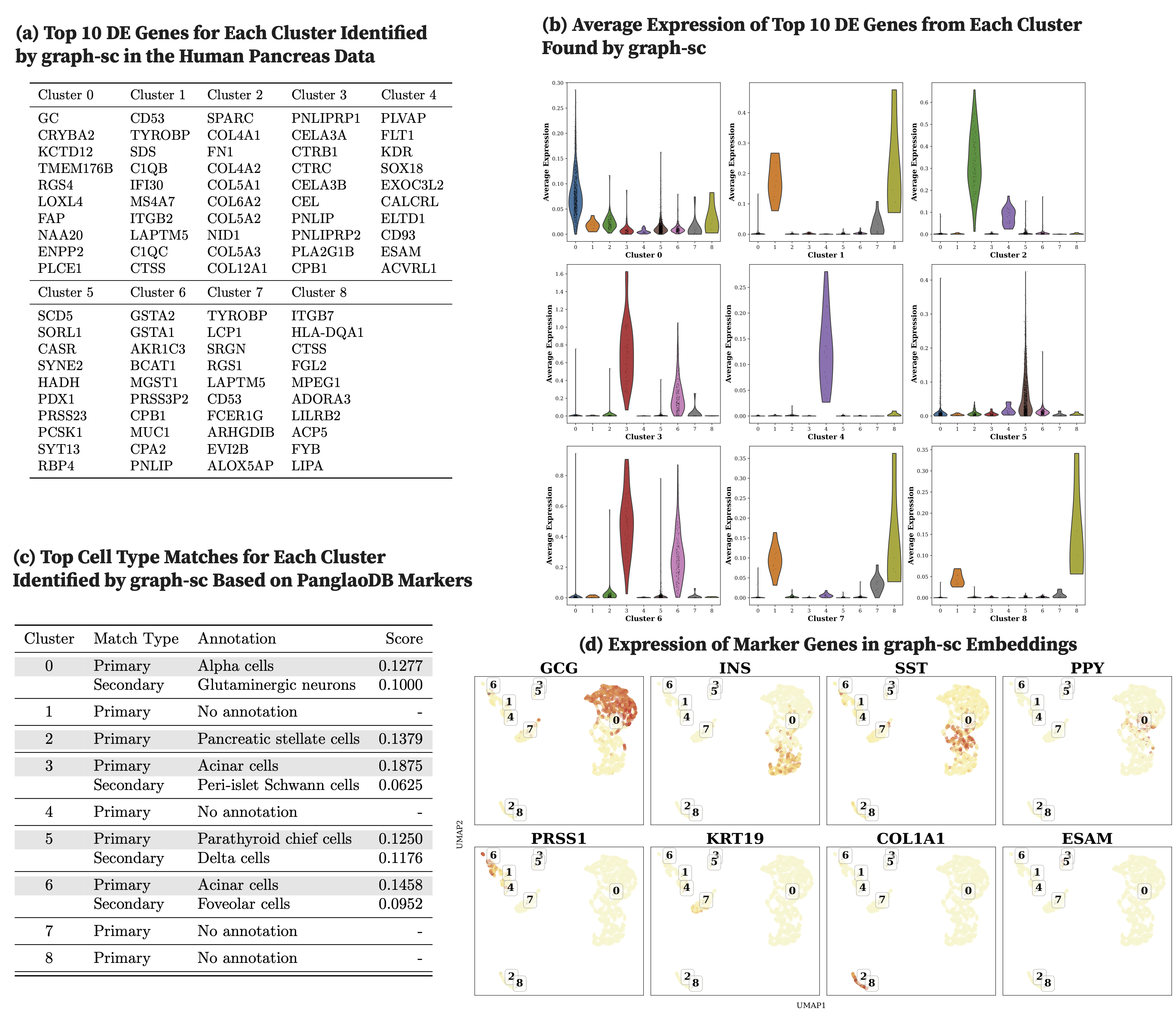} 
  \caption{Clustering Performance of graph-sc on the Human Pancreas Dataset and Cell Type Annotation Using Differentially Expressed Genes (DEGs) and Canonical Markers. (a) Top 10 DEGs per cluster identified by graph-sc. (b) Across expression of cluster-specific DEGs across clusters. Strong clustering performance is indicated by high expression of cluster-specific DEGs within their respective clusters and low expression in other clusters. Graph-sc clusters generally show moderate specificity with substantial cross-cluster overlap. (c) Cell type annotation using the PanglaoDB marker database. Four clusters (1, 4, 7, and 8) yield no annotation, indicating that their DEGs did not match known cell type markers. (d) Expression of human pancreatic canonical markers in graph-sc embeddings. UMAP visualizations show graph-sc cell embeddings, with cluster IDs assigned by K-means clustering. Overall, cell cluster separation is limited.}
\label{fig:graph-sc_HumanPancreas}
\end{figure*}

\begin{figure*}[htbp]
  \centering
  \includegraphics[width=\textwidth]{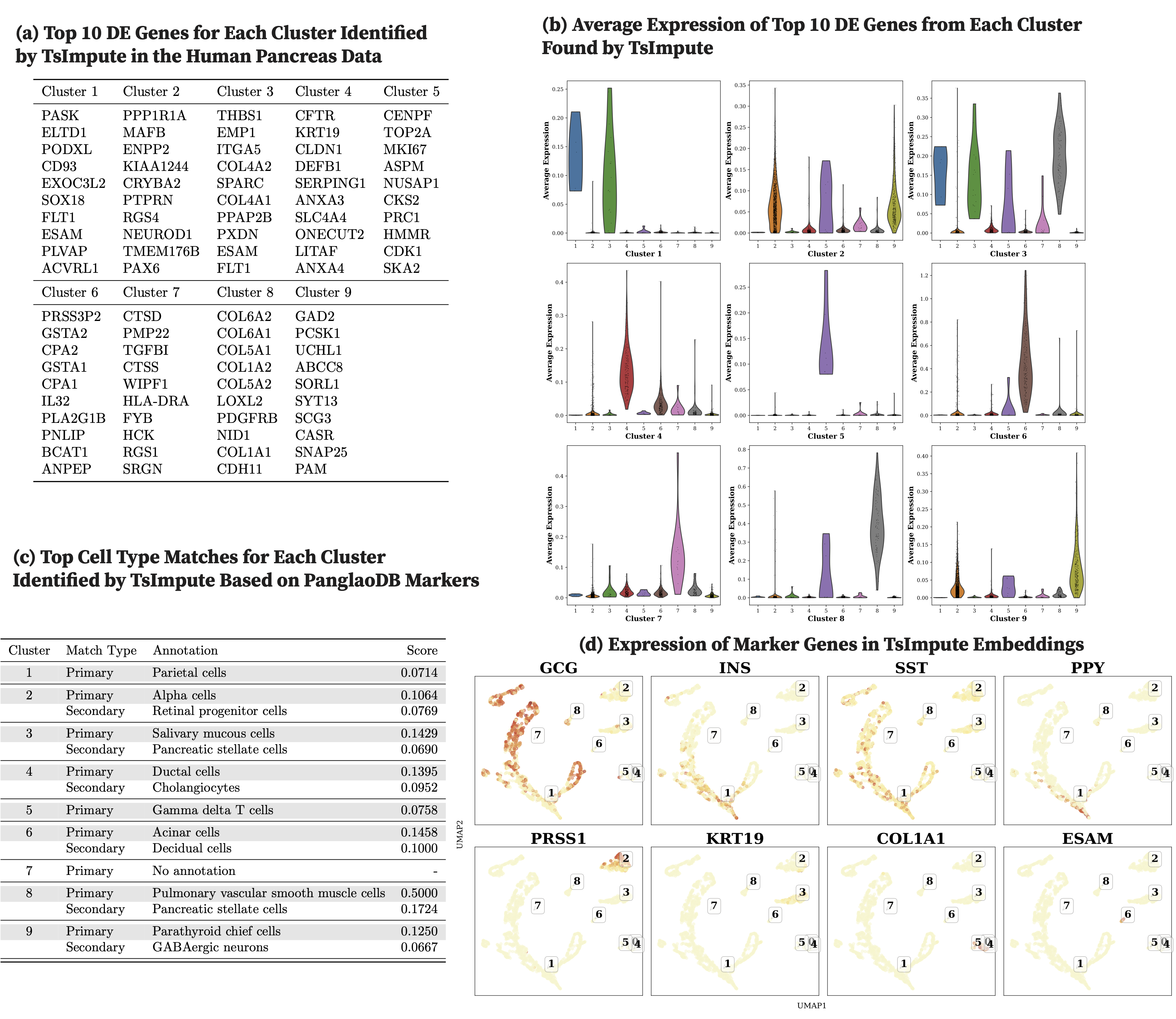} 
  \caption{Clustering Performance of tsImpute on the Human Pancreas Dataset and Cell Type Annotation Using Differentially Expressed Genes (DEGs) and Canonical Markers. (a) Top 10 DEGs per cluster identified by tsImpute combined with PCA embeddings and K-means. (b) Average expression of cluster-specific DEGs across clusters. Strong clustering performance is indicated by high expression of cluster-specific DEGs within their respective clusters and low expression in other clusters. tsImpute DEGs show high specificity in Clusters 2, 4, 5, 6, 7, 8, and 9. (c) Cell type annotation using the PanglaoDB marker database. Several annotations are biologically implausible for pancreas tissue, including parietal cells (Cluster 1), salivary mucous cells (Cluster 3), gamma delta T cells (Cluster 5), and pulmonary vascular smooth muscle cells (Cluster 8). These unexpected annotations suggest that the imputation process may introduce artifacts that confound cell type identification. (d) Expression of human pancreatic canonical markers in tsImpute combined with PCA embeddings. UMAP visualizations show tsImpute plus PCA cell embeddings, with cluster IDs assigned by K-means clustering. tsImpute improves cluster separation compared to PCA without imputation.}
\label{fig:tsImpute_HumanPancreas}

\end{figure*}

\begin{figure*}[htbp]
  \centering
  \includegraphics[width=\textwidth]{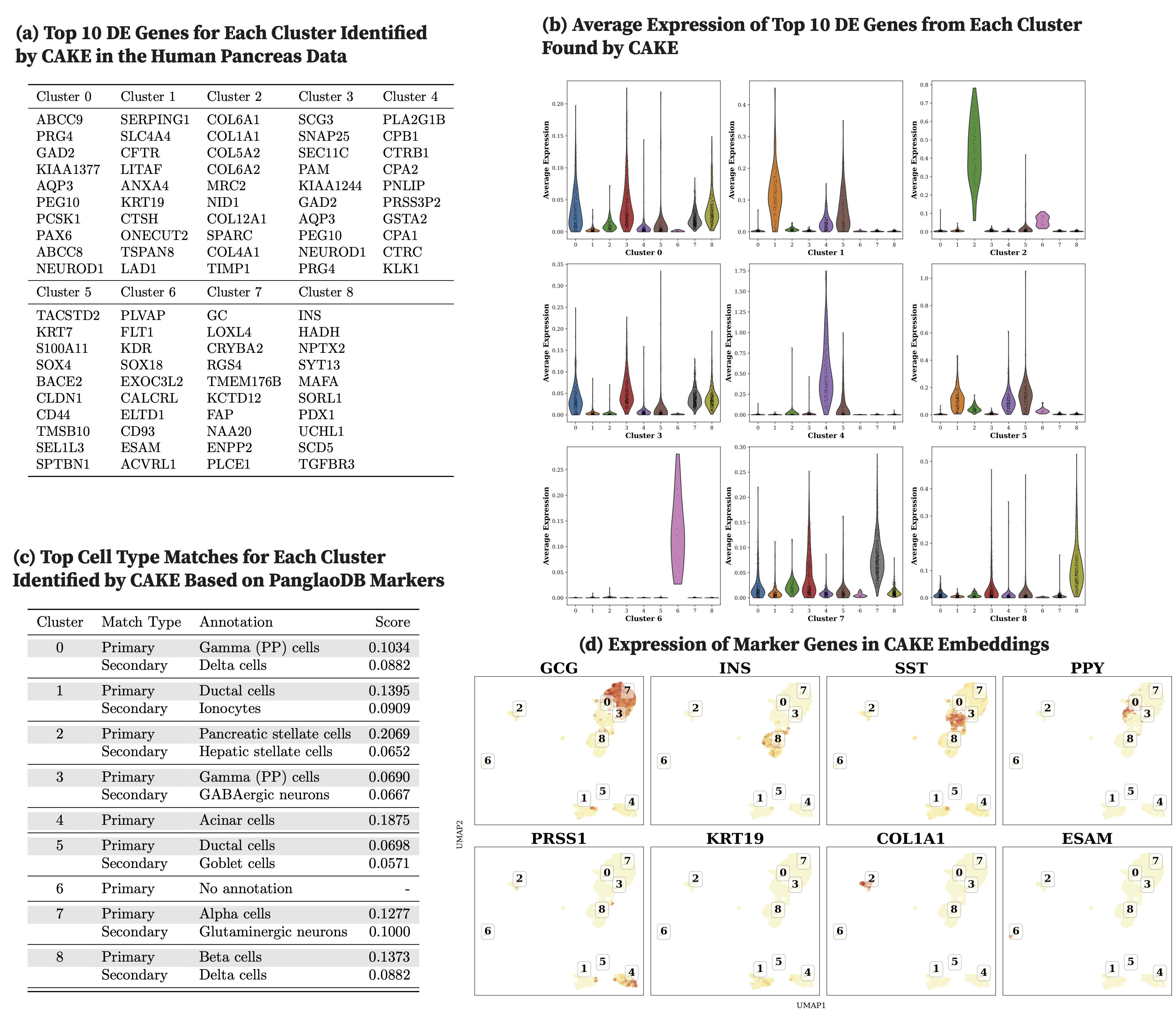} 
  \caption{Clustering Performance of CAKE on the Human Pancreas Dataset and Cell Type Annotation Using Differentially Expressed Genes (DEGs) and Canonical Markers. (a) Top 10 DEGs per cluster identified by CAKE. (b) Average expression of cluster-specific DEGs across clusters. Strong clustering performance is indicated by high expression of cluster-specific DEGs within their respective clusters and low expression in other clusters. CAKE DEGs show good cluster specificity. (c) Cell type annotation using the PanglaoDB marker database. Most CAKE clusters correspond to biologically relevant cell types: gamma/PP cells (Clusters 0 and 3), ductal cells (Clusters 1 and 5), pancreatic stellate cells (Cluster 2), acinar cells (Cluster 4), alpha cells (Cluster 7), and beta cells (Cluster 8). (d) Expression of human pancreatic canonical markers in CAKE embeddings. UMAP visualizations show CAKE cell embeddings, with cluster IDs assigned by K-means clustering. Marker genes show distinct concentration in the well-separated CAKE embeddings: GCG in Cluster 7, INS in Cluster 8, SST in Clusters 0 and 3, PRSS1 in Cluster 4, and KRT19 in Cluster 1.}
\label{fig:CAKE_HumanPancreas}
\end{figure*}

\begin{figure*}[htbp]
  \centering
  \includegraphics[width=\textwidth]{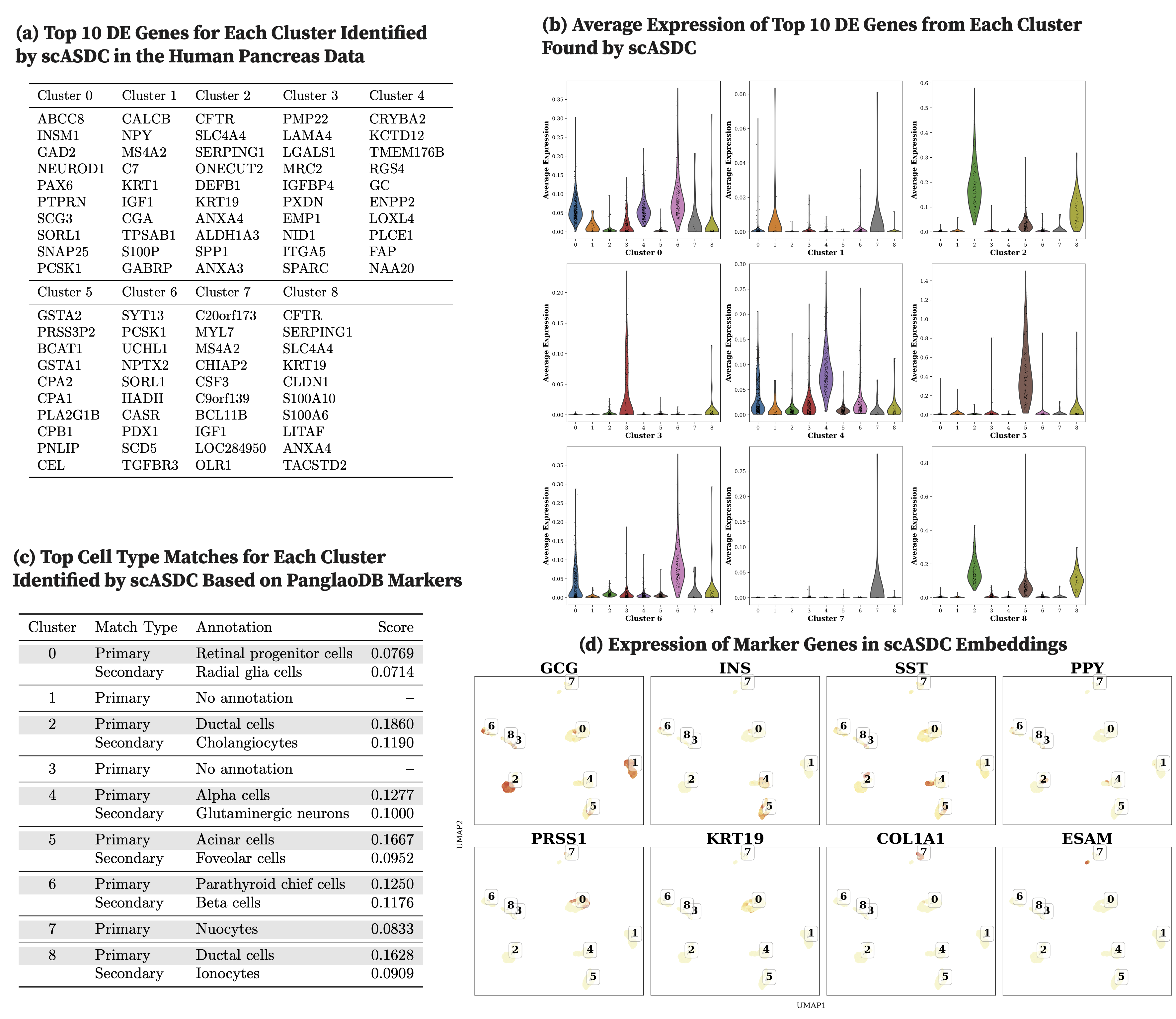} 
  \caption{Clustering Performance of scASDC on the Human Pancreas Dataset and Cell Type Annotation Using Differentially Expressed Genes (DEGs) and Canonical Markers. (a) Top 10 DEGs per cluster identified by scASDC. (b) Average expression of cluster-specific DEGs across clusters. Strong clustering performance is indicated by high expression of cluster-specific DEGs within their respective clusters and low expression in other clusters. scASDC DEGs show good cluster specificity. (c) Cell type annotation using the PanglaoDB marker database. While most clusters are matched to relevant pancreatic cell types, some are annotated with unexpected cell types, including retinal progenitor cells (Cluster 0) and nuocytes (Cluster 7). Clusters 1 and 3 received no annotation. (d) Expression of human pancreatic canonical markers in scASDC embeddings. scASDC embeddings show the best internal cohesion and separability. The marker genes show distinct concentration in the well-separated scASDC embeddings. However, the cell type annotations suggest that the embeddings may not necessarily correspond to biologically meaningful clusters.}
\label{fig:scASDC_HumanPancreas}
\end{figure*}

We report the top 10 differentially expressed genes (DEGs) from each human pancreas cell cluster identified by CoMem-DIPHW (Fig.~\ref{fig:CoMem_HumanPancreas}a), PCA (Fig.~\ref{fig:PCA_HumanPancreas}a), graph-sc (Fig.~\ref{fig:graph-sc_HumanPancreas}a), tsImpute (Fig.~\ref{fig:tsImpute_HumanPancreas}a), CAKE (Fig.~\ref{fig:CAKE_HumanPancreas}a), and scASDC (Fig.~\ref{fig:scASDC_HumanPancreas}a). To evaluate cluster purity, we examine the expression of these DEGs across clusters (Fig.~\ref{fig:CoMem_HumanPancreas}b, Fig.~\ref{fig:PCA_HumanPancreas}b, Fig.~\ref{fig:graph-sc_HumanPancreas}b, Fig.~\ref{fig:tsImpute_HumanPancreas}b, Fig.~\ref{fig:CAKE_HumanPancreas}b, Fig.~\ref{fig:scASDC_HumanPancreas}b). An effective clustering method should yield DEGs that uniquely characterize each cluster while showing minimal expression in other clusters.

Violin plots show that the DEGs identified by CoMem-DIPHW exhibit more cluster-specific expression patterns than those derived from PCA, which tend to show higher expression in multiple clusters. For example, the average expression of the DEGs from PCA Cluster 1 is higher in Clusters 2, 5, and 6. Similarly, the DEGs from Cluster 7 show higher average expression in Cluster 6. Among the comparison methods, CAKE and scASDC produced highly distinctive DEGs, with scASDC yielding clusters that showed the strongest intra-cluster cohesion and inter-cluster separability. Graph-sc and PCA show less cluster-specific profiles, with DEGs frequently expressed across multiple clusters.

To assess the biological relevance of the clusters, we cross-reference the cluster-specific DEGs (see Fig.~\ref{fig:CoMem_HumanPancreas}a) with markers from the PanglaoDB cell type marker database (version 27 March 2020)\cite{franzen2019panglaodb}. Cell-type assignment is based on the overlap between the DEGs of each cluster and the cell type-specific markers in the PanglaoDB marker database, with a match score calculated as the proportion of matched markers (Table \ref{fig:CoMem_HumanPancreas}c). CoMem-DIPHW (Fig.~\ref{fig:CoMem_HumanPancreas}c) successfully annotates all nine clusters with biologically relevant pancreatic cell types. CAKE produces similarly meaningful annotations, while graph-sc has several clusters with no annotation or biologically implausible labels.

To further validate the clustering results, we examine the expression of canonical pancreatic marker genes~\cite{muraro2016single}, including GCG (alpha cells), INS (beta cells), SST (delta cells), PPY (PP cells), PRSS1 (acinar cells), KRT19 (ductal cells), COL1A1 (mesenchymal cells), and ESAM (endothelial cells) in the embeddings (Fig.~\ref{fig:CoMem_HumanPancreas}d, Fig.~\ref{fig:PCA_HumanPancreas}d, Fig.~\ref{fig:graph-sc_HumanPancreas}d, Fig.~\ref{fig:tsImpute_HumanPancreas}d, Fig.~\ref{fig:CAKE_HumanPancreas}d, Fig.~\ref{fig:scASDC_HumanPancreas}d). Each marker is distinctly expressed in the embedding generated by CoMem-DIPHW, as shown in Fig. \ref{fig:CoMem_HumanPancreas}d, confirming that CoMem-DIPHW embeddings effectively distinguish different cell types. CAKE and scASDC also produce well-separated embeddings. In contrast, PCA embeddings show poor separation. The results from CoMem-DIPHW align well with the cell-type annotation by the PanglaoDB database. The only discrepancy is in Cluster 4. Matching the DEGs with PanglaoDB cell type markers suggests Pancreatic stellate cells (PSCs) or Granulosa cells, while the COL1A1 and ESAM expression plots indicate mesenchymal or endothelial cells. However, this apparent discrepancy can be explained by the fact that PSCs can differentiate into cells with endothelial-like properties~\cite{cheng2021generation}, and granulosa cells can exhibit mesenchymal-like behaviors~\cite{jozkowiak2020stemness}.

In contrast, PCA clustering results (Fig. \ref{fig:PCA_HumanPancreas}) showed limited cluster separation. While some PCA clusters could be annotated consistently using both PanglaoDB markers and canonical markers, significant overlap remained among clusters. For example, PCA Cluster 8, identified as beta cells by both the PanglaoDB markers and the canonical marker INS, is not well separated from Cluster 5 (potentially delta cells, marked by SST expression). Similarly, Cluster 7, identified as ductal cells by both the PanglaoDB markers and the canonical marker KRT19, is closely embedded with endothelial cells (ESAM). These results highlight CoMem-DIPHW’s improved ability to distinguish pancreatic cell types compared to PCA.

In addition to the human pancreas results presented here, we also analyze mouse pancreas, human brain, and mouse brain datasets. For these datasets, we include visualizations in the supplementary material in Figures \ref{fig:CoMem_HumanBrain}b through \ref{fig:scASDC_MouseBrain}b that show the average expression of the top DEGs across clusters identified by different methods. These results show our proposed algorithm's ability in effectively distinguishing cell clusters by their expression profiles. The supplementary material (Figures \ref{fig:CoMem_HumanBrain}a through \ref{fig:scASDC_MouseBrain}a) also contain tables listing the top DEGs for clusters identified by CoMem-DIPHW, PCA, graph-sc, tsImpute, CAKE, and scASDC in each dataset for reference.

To complement the qualitative biological validation presented above, we further assess clustering performance using quantitative cluster quality measures, including the Silhouette score, Calinski-Harabasz index, Davies-Bouldin index, and Coefficient of Variation. These evaluations are conducted in two settings: (1) directly in the learned embedding space and (2) by evaluating the same cluster assignments on the preprocessed expression data used as input to the embedding algorithms. We refer the reader to Supplementary Tables~\ref{tab:evaluate_clustering_embeddings} and~\ref{tab:evaluate_clustering_preprocesseddata}. Evaluation in the embedding space captures the geometric separability and cohesion of clusters, whereas evaluation on the preprocessed data assesses whether the resulting cluster assignments capture meaningful structure in the input gene expression data. As demonstrated by differential expression and cell-type annotation analyzes in the human pancreas data set, strong embedding-space cohesion and separation alone do not necessarily guarantee biological interpretability.

\section{Conclusion}

In this study, we develop and analyze two hypergraph-based clustering algorithms, DIPHW and CoMem-DIPHW, to improve cell clustering performance on scRNA-seq data. First, the Dual-Importance Preference (DIP) Hypergraph Walk (DIPHW) extends the edge-dependent vertex weight (EDVW) hypergraph random walk~\cite{Chitra2019} and consistently outperforms it in our experiments. The key idea of DIPHW is its symmetrized treatment of importance, considering both the relative importance of edges to nodes and nodes to edges, or the importance of genes relative to cells and cells to genes, along with a preference exponent for faster clustering. Second, the CoMem and CoMem-DIPHW clustering methods incorporate a memory mechanism that considers both the abundance counts data (conceptualized as hypergraphs) and the two co-expression networks to integrate local and global information when computing embeddings for clustering. Third, we identify and demonstrate through experiments (see Fig.~\ref{fig:zeroInflation}) the issue of inflated signals in co-expression networks caused by the sparsity of scRNA-seq data, which further motivates the use of hypergraph representations. Fourth, we compare our proposed methods with 13 cell-clustering methods from various categories, including community detection, embedding-based methods, and recent deep learning methods used in clustering scRNA-seq data. Our evaluation covers a broad range of modularity conditions in the data, where CoMem-DIPHW and DIPHW consistently show superior performance, particularly in weak modularity regimes. When clusters are harder to detect, other methods experience a much more pronounced decline in performance compared to CoMem-DIPHW. Fifth, we test our proposed methods on four real scRNA-seq datasets and two benchmark datasets. In these experiments, CoMem-DIPHW consistently demonstrates superior performance in differentiating and classifying distinct cell types compared to the other methods. Sixth, we design a versatile simulation algorithm for scRNA-seq data that is highly customizable through many user-specified parameters. This algorithm models a wide range of expression scenarios and modular structures. Notably, it incorporates intermodular co-expressions, which mimic cell-type cross-talk observed in real biological systems.

\subsection{Future Work} 

We plan to optimize the implementation of CoMem-DIPHW to reduce its memory complexity, enabling it to scale to larger datasets. We will also explore alternative methods for incorporating the memory mechanism into the hypergraph random walk. Currently, CoMem uses a simple product approach; moving beyond this could reveal more effective clustering strategies. Although our current evaluation focuses on the biological relevance of clusters through cell-type annotation using PanglaoDB markers, future application-focused studies could include Gene Ontology enrichment analysis to characterize the functions of identified clusters.

With the increasing availability of data and studies in spatial transcriptomics~\cite{li2024spadit, rao2021exploring, longo2021integrating, kleshchevnikov2022cell2location}, extending the CoMem-DIPHW formulation for spatial transcriptomics analysis is a promising direction. While the current CoMem-DIPHW formulation focuses on cell clustering based on expression profile similarity, it could be extended to incorporate spatial proximity by replacing the cell expression similarity matrix $G_V$ with functions that characterize joint cell similarity based on both expression profiles and spatial information.

\bibliographystyle{apsrev}
\bibliography{main.bib}

\newpage
\appendix

\renewcommand{\thesubsection}{S\arabic{subsection}}
\renewcommand{\thetable}{S\arabic{table}}
\renewcommand{\thefigure}{S\arabic{figure}}

\setcounter{table}{0}
\setcounter{figure}{0}

\section*{Supplementary Materials}
\label{sec:supplementary}

\subsection{Evaluating the Impact of Correlation Methods on Network-Based Clustering Performance}
\label{subsec:cosine_similarity}
There are two primary strategies for handling shared zeros in cell expression profiles of scRNA-seq data during correlation computation. The first strategy, exemplified by cosine similarity, is to ignore the shared zeros. In this approach, inactive expression (zeros) does not contribute to the similarity between cells. However, our experiments indicate that this strategy leads to significantly worse clustering performance. The absence of expression in the same genes between cell pairs provides information about cellular similarity and should not be disregarded.


We evaluate how co-expression networks generated by different correlation computation methods affect the performance of various network-based clustering methods. See Table~\ref{tab:CorrelationClusteringComparison}. Our results show that cosine similarity often produces the worst clustering results among Pearson, Spearman, and Kendall Tau.

\begin{table}[htbp] 
    \centering 
    \begin{tabular}{@{}l|cccc|cccc@{}} 
        \toprule 
        \textbf{Method} & \multicolumn{4}{c|}{\textbf{ARI}} & \multicolumn{4}{c}{\textbf{NMI}} \\ 
        \cmidrule(lr){2-5} \cmidrule(lr){6-9}  
        & \textbf{Pearson} & \textbf{Cosine} & \textbf{Spearman} & \textbf{Kendall}  
        & \textbf{Pearson} & \textbf{Cosine} & \textbf{Spearman} & \textbf{Kendall} \\ 
        \midrule 
        GreedyModularity & 0.205 & \textbf{0.135} & 0.194 & 0.197  
        & 0.534 & \textbf{0.429} & 0.496 & 0.502 \\ 
        Louvain & 0.493 & \textbf{0.322} & 0.445 & 0.473  
        & 0.784 & \textbf{0.666} & 0.768 & 0.781 \\ 
        Infomap & \textbf{0.000} & \textbf{0.000} & \textbf{0.000} & \textbf{0.000}  
        & \textbf{0.000} & \textbf{0.000} & \textbf{0.000} & \textbf{0.000} \\ 
        Leiden & \textbf{-0.001} & \textbf{-0.001} & -0.000 & -0.000  
        & 0.062 & \textbf{0.040} & 0.082 & 0.073 \\ 
        Multilevel & \textbf{0.032} & \textbf{0.034} & 0.034 & 0.035
        & 0.305 & \textbf{0.289} & 0.325 & 0.319 \\ 
        Eigenvector & 0.002 & \textbf{0.000} & 0.003 & 0.005
        & 0.097 & \textbf{0.060} & 0.082 & 0.091 \\ 
        \bottomrule 
    \end{tabular} 
\caption{Impact of Correlation Methods on Network-Based Clustering Performance. The table shows how different correlation measures (specifically, Pearson, Spearman, Kendall Tau, and cosine similarity) affect the performance of various network-based clustering methods. The correlation computation method that results in the worst clustering performance is marked in bold. Cosine similarity, which ignores shared zeros in cell expression profiles, often produces the worst clustering performance. These results indicate that while correlation computation methods that treat zeros the same as non-zero signals may inflate correlations, ignoring these zeros, as in cosine similarity, can lead to poorer performance.}
\label{tab:CorrelationClusteringComparison}
\end{table}

The main implication of this experiment is that both strategies have limitations when handling sparse scRNA-seq data. Cosine similarity’s omission of shared zeros results in the poorest performance. In contrast, methods that treat all shared zeros in expression profiles as indicators of cell homogeneity (such as Pearson, Spearman, and Kendall Tau) overlook the fact that many shared zeros arise simply from the high dimensionality of scRNA-seq data. When sequencing results from multiple cells are combined, many zero entries appear because different cells often express different sets of genes. Our results indicate that alternative approaches, such as hypergraph-based methods, offer a better solution by avoiding the need for unipartite network projections and the complexity of handling zeros in correlation calculations.

\subsection{Computational and Memory Complexities of DIPHW and CoMem-DIPHW}

\paragraph{DIPHW.}
The computational cost of each element $P(u \rightarrow v)$ in the $|V|\times|V|$ node-to-node transition probability matrix is $O(|E|)$ when the intermediate matrices for the node-to-edge and edge-to-node transition probabilities $P_{E|V}(e|u)$ and $P_{V|E}(v|e)$ are precomputed. The construction of the entire $|V|\times|V|$ node-to-node transition probability matrix involves computing these probabilities for all node pairs, resulting in an overall time complexity of $O(|V|^2\cdot|E|)$.

For efficient computation, the $|V|\times|V|$ node-to-node transition probability matrix $P_{V|V}$ in martix form is given by:
\[
P_{V|V} = D_{E|V}^{-1} W_{E|V} D_{V|E}^{-1} W_{V|E}
\]

The theoretical complexity remains $O(|E|^2 \cdot |V| + |E|\cdot |V|^2)$. However, the matrix representation utilizes parallel computation and is significantly faster, making it more suitable for large datasets. 

DIPHW's memory complexity is $O(|E|\cdot |V|)$ for the node-to-edge and edge-to-node transition probability matrices.

\paragraph{CoMem-DIPHW.}

The computational complexity of CoMem-DIPHW is $O(|E|^2 \cdot |V|^2)$. For each pair of nodes, the memory-incorporated transition traverses all possible edges that could have reached the first node and all possible edges that could then connect the first node with the second node. 

CoMem-DIPHW's memory complexity is $O(|E|^2 \cdot |V|+|E|\cdot |V|^2)$ for the memory-incorporated node-to-edge and edge-to-node transition probability matrices.

\subsection{Bipartite representation}
We choose the hypergraph conceptualization over the bipartite representation for consistency, as the DIPHW algorithm we introduce is an extension of the edge-dependent vertex weight hypergraph random walk (EDVW)~\cite{Chitra2019}. DIPHW extends the EDVW hypergraph random walk by incorporating a vertex-dependent edge selection probability and a preference exponent to accelerate clustering.

A bipartite conceptualization could also be applied to our proposed methods by interpreting the incidence matrix $\mathbf{I}_\mathcal{H}$ of the hypergraph $\mathcal{H}$ as a bipartite graph. In this representation, the two sets of nodes are the set of cells $V$ and the set of genes $E$. Each gene is connected to all cells in which it is actively expressed, with weights corresponding to the expression level. This alternative representation enables the use of bipartite graph algorithms. For example, we use Barber's bipartite modularity~\cite{barber2007modularity} to assess how the clustering performance of different algorithms varies with the modularity of the underlying scRNA-seq dataset.

\subsection{Quantitative Evaluation of Clustering Quality on Tissue Datasets}

The quantitative evaluations in Tables~\ref{tab:evaluate_clustering_embeddings} and~\ref{tab:evaluate_clustering_preprocesseddata} offer complementary perspectives on clustering performance for tissue scRNA-seq datasets. The evaluation in the learned embedding space (Table~\ref{tab:evaluate_clustering_embeddings}) reflects the geometric separability and compactness of the clusters produced by each method, while the evaluation of the preprocessed expression data (Table~\ref{tab:evaluate_clustering_preprocesseddata}) assesses whether the same cluster assignments capture the structure of the preprocessed gene expression data used as input to the embedding methods. In the embedding space, graph-sc and scASDC achieve strong geometric separation and cohesion across datasets. However, they show reduced clustering performance when evaluated on the preprocessed data. This aligns with the results in Fig.~\ref{fig:scASDC_HumanPancreas}, where scASDC produces embeddings with strong cluster separation and cohesion, but some identified DEGs correspond to biologically implausible cell types. In particular, PCA performs best in the preprocessed data evaluation. This is consistent with the findings of Ciortan et al.~\cite{ciortan2022gnn}, where PCA outperformed many state-of-the-art methods, including graph-sc, when evaluated by the Silhouette score and the Calinski-Harabasz index on real and simulated datasets.

\begin{table}[!h]
\centering
\resizebox{\textwidth}{!}{%
\begin{tabular}{l | cccc | cccc | cccc | cccc}
\toprule
 & \multicolumn{4}{c}{\textbf{Human Brain}} & \multicolumn{4}{c}{\textbf{Human Pancreas}} & \multicolumn{4}{c}{\textbf{Mouse Brain}} & \multicolumn{4}{c}{\textbf{Mouse Pancreas}} \\
Method  & SC & CH & DB & CV & SC & CH & DB & CV & SC & CH & DB & CV & SC & CH & DB & CV \\
\midrule
scASDC & \textbf{0.421} & 244.710 & \underline{0.975} & 2.373 & \textbf{0.763} & \textbf{7005.583} & \textbf{0.549} & \underline{0.821} & \textbf{0.816} & \underline{5230.004} & \textbf{0.640} & 1.845 & \textbf{0.874} & \textbf{2843.512} & \textbf{0.462} & \underline{0.563} \\
graph-sc & \underline{0.352} & \textbf{705.099} & \textbf{0.844} & \textbf{0.560} & \underline{0.529} & \underline{4164.981} & \underline{0.584} & \textbf{0.816} & 0.418 & 2431.617 & 0.982 & \textbf{0.905} & 0.394 & \underline{976.784} & \underline{0.768} & \textbf{0.495} \\
CoMem\_DIPHW & 0.285 & 224.443 & 1.207 & 1.287 & 0.384 & 1048.921 & 1.119 & 1.058 & 0.396 & 1222.851 & 0.936 & 1.403 & 0.274 & 313.957 & 1.183 & 1.677 \\
CoMem & 0.265 & 199.820 & 1.285 & 1.922 & 0.474 & 1447.208 & 0.852 & 3.575 & 0.455 & 1432.727 & 0.817 & \underline{1.086} & 0.291 & 287.810 & 1.179 & 4.768 \\
tsImpute & 0.263 & \underline{703.410} & 1.163 & 8.200 & 0.457 & 715.814 & 0.789 & 5.051 & 0.409 & 867.564 & 1.095 & 8.499 & \underline{0.424} & 666.677 & 0.902 & 10.673 \\
PCA & 0.258 & 236.768 & 1.412 & 6.802 & 0.294 & 542.832 & 1.270 & 89.515 & 0.229 & 512.127 & 1.652 & 83.396 & 0.107 & 120.386 & 1.804 & 125.705 \\
CAKE & 0.156 & 164.736 & 1.679 & 2.849 & 0.417 & 3999.076 & 0.794 & 1.944 & \underline{0.461} & \textbf{12194.111} & \underline{0.800} & 2.684 & 0.237 & 373.750 & 1.391 & 1.825 \\
node2vec & 0.133 & 172.629 & 2.118 & 2.307 & 0.409 & 1405.125 & 1.034 & 1.359 & 0.332 & 1108.268 & 1.188 & 27.464 & 0.198 & 265.955 & 1.467 & 1.014 \\
DIPHW & 0.122 & 110.724 & 2.138 & \underline{0.903} & 0.204 & 480.360 & 1.328 & 1.060 & 0.127 & 264.797 & 2.444 & 5.145 & 0.055 & 35.369 & 3.380 & 3.147 \\
EDVW & 0.109 & 97.634 & 2.315 & 1.720 & 0.152 & 402.944 & 2.039 & 6.156 & 0.126 & 246.265 & 2.616 & 5.840 & 0.060 & 40.552 & 3.482 & 5.804 \\
\bottomrule
\end{tabular}
}

\caption{Clustering Quality Evaluated in the Learned Embedding Space on Tissue scRNA-seq Datasets. Clustering performance is assessed using the Silhouette score (SC), Calinski–Harabasz index (CH), Davies–Bouldin index (DB), and Coefficient of Variation (CV) on the learned embeddings produced by each method. Higher values indicate better performance for the Silhouette score and Calinski–Harabasz index, while lower values indicate better performance for the Davies–Bouldin index and Coefficient of Variation. This evaluation reflects the geometric separability and cohesion of clusters in the embedding space. The best values within each dataset/evaluation measure column are highlighted in bold, and the second-best values are underlined.}

\label{tab:evaluate_clustering_embeddings}

\end{table}

\begin{table}[!h]
\centering
\resizebox{\textwidth}{!}{%
\begin{tabular}{l | cccc | cccc | cccc | cccc}
\toprule
 & \multicolumn{4}{c}{\textbf{Human Brain}} & \multicolumn{4}{c}{\textbf{Human Pancreas}} & \multicolumn{4}{c}{\textbf{Mouse Brain}} & \multicolumn{4}{c}{\textbf{Mouse Pancreas}} \\
Method  & SC & CH & DB & CV & SC & CH & DB & CV & SC & CH & DB & CV & SC & CH & DB & CV \\
\midrule
CoMem & \textbf{0.027} & 17.529 & \underline{4.654} & 3.951 & 0.071 & 230.969 & 2.048 & 4.559 & 0.041 & 173.331 & \underline{2.838} & 7.603 & -0.002 & 37.954 & 4.182 & 5.064 \\
CoMem\_DIPHW & \underline{0.027} & \underline{17.529} & \textbf{4.575} & \underline{3.951} & 0.021 & 235.958 & 2.364 & 4.834 & 0.018 & 169.640 & 3.363 & 7.795 & 0.002 & \underline{40.049} & 3.891 & 5.008 \\
PCA & 0.025 & \textbf{19.080} & 4.991 & 4.012 & \textbf{0.276} & \textbf{395.673} & \textbf{1.461} & \underline{4.012} & \textbf{0.087} & \textbf{183.428} & \textbf{2.829} & 6.995 & \underline{0.016} & \textbf{46.001} & \textbf{3.051} & \textbf{4.699} \\
tsImpute & 0.009 & 12.901 & 7.122 & 3.958 & \underline{0.114} & 171.384 & 2.715 & \textbf{3.734} & -0.005 & 67.857 & 6.082 & \underline{6.721} & -0.025 & 22.239 & 6.907 & 5.127 \\
graph-sc & 0.008 & 10.692 & 6.286 & 4.043 & 0.042 & 201.025 & 2.218 & 4.137 & 0.047 & 127.852 & 5.446 & 6.905 & \textbf{0.052} & 28.514 & \underline{3.696} & \underline{4.749} \\
node2vec & 0.007 & 15.731 & 8.386 & 4.010 & 0.060 & \underline{259.873} & \underline{2.019} & 4.609 & 0.040 & \underline{174.822} & 3.240 & 7.726 & -0.001 & 37.939 & 4.459 & 5.027 \\
DIPHW & 0.006 & 14.937 & 7.794 & 4.018 & 0.015 & 195.839 & 3.152 & 4.328 & 0.012 & 153.568 & 4.081 & 7.782 & -0.006 & 36.516 & 5.353 & 5.115 \\
EDVW & 0.002 & 13.243 & 8.843 & 4.042 & -0.075 & 177.732 & 7.734 & 4.956 & 0.034 & 172.331 & 3.403 & 7.805 & -0.001 & 39.774 & 4.790 & 4.991 \\
scASDC & -0.009 & 7.670 & 7.899 & \textbf{3.947} & -0.128 & 139.358 & 3.605 & 4.655 & \underline{0.070} & 166.681 & 3.032 & 7.177 & -0.027 & 26.217 & 6.188 & 5.148 \\
CAKE & -0.012 & 7.713 & 8.723 & 4.002 & -0.025 & 184.071 & 3.063 & 4.444 & 0.068 & 75.122 & 6.077 & \textbf{5.624} & 0.006 & 33.657 & 6.005 & 5.042 \\
\bottomrule
\end{tabular}
}
\caption{Clustering Quality Evaluated on Preprocessed Expression Data for Tissue scRNA-seq Datasets. Using the same cluster assignments as in Supplementary Table~\ref{tab:evaluate_clustering_embeddings}, we evaluate cluster quality on the preprocessed gene expression data used as input to each embedding method. We measure cluster quality using the Silhouette score (SC), Calinski–Harabasz index (CH), Davies–Bouldin index (DB), and Coefficient of Variation (CV). Higher values indicate better performance for SC and CH, while lower values indicate better performance for DB and CV. This evaluation assesses whether the cluster assignments by each method capture meaningful structure in the input gene expression data. The best values within each dataset/evaluation measure column are highlighted in bold, and the second-best values are underlined.}

\label{tab:evaluate_clustering_preprocesseddata}
\end{table}

\subsection{Additional Results for Clustering Performance Across Modularity Levels}
\label{subsec:additionalresults}

\begin{figure}[H]
\centering
\includegraphics[width=\textwidth]{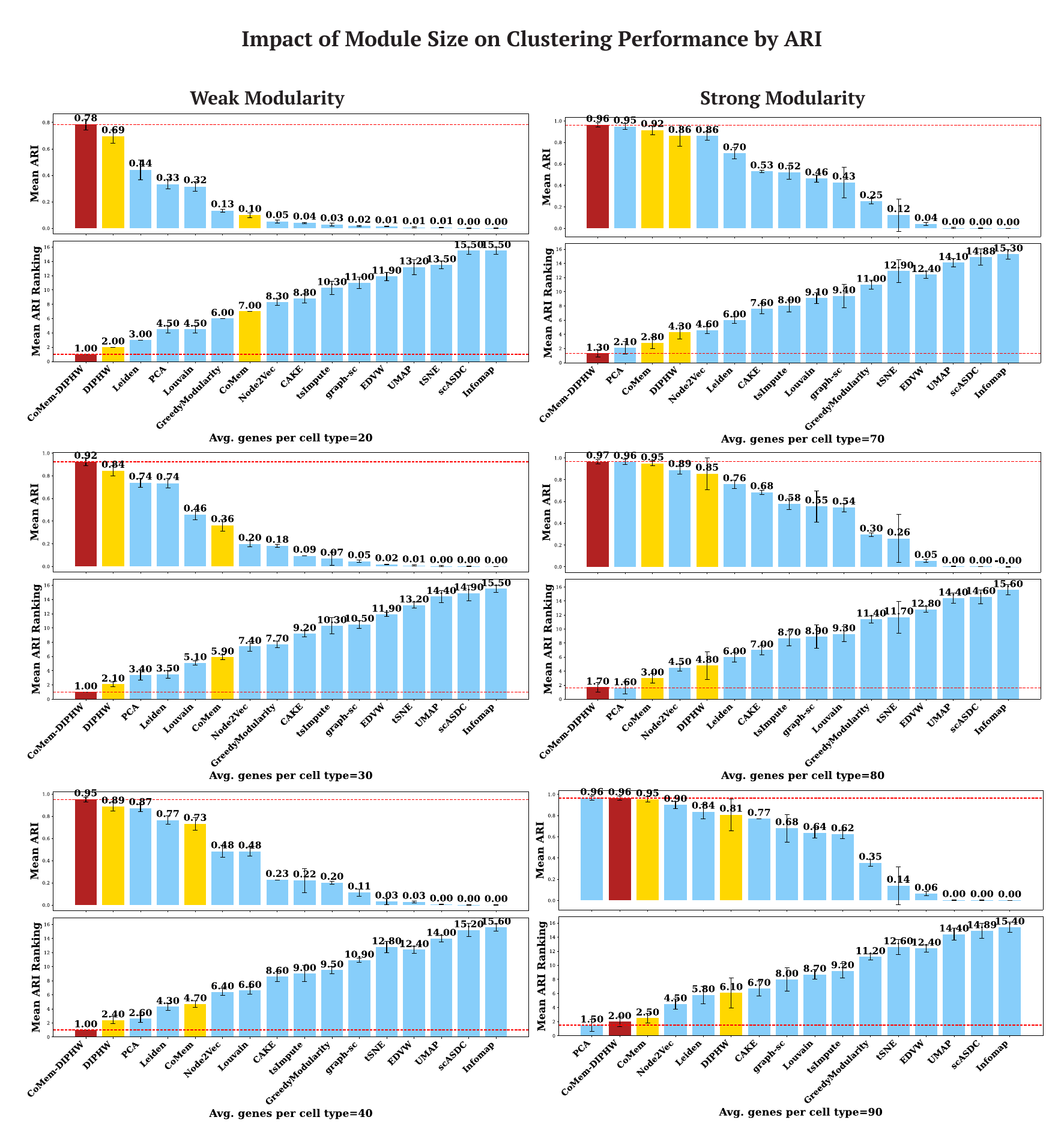}
\caption{Clustering Performance Comparison by ARI Across Varying Module Sizes. Simulated scRNA-seq data were used for this evaluation. Here, we present the complementary results across intermediate parameter settings, which show consistent patterns: our proposed methods (highlighted in red and yellow) consistently ranked first by ARI in all scenarios, with a stronger advantage when data modularity is weak, i.e., when the average number of co-expressed genes per module is small. Each experiment was repeated 10 times per parameter setting, with error bars representing the 95\% confidence interval. Red dashed lines indicate the highest ARI values or best ARI rankings. K-means was used to cluster the output of all embedding-based methods that do not directly assign cluster membership.}
\label{fig:Ngenes_ARI_supplement}
\end{figure}

\begin{figure}[H]
\centering
\includegraphics[width=\textwidth]{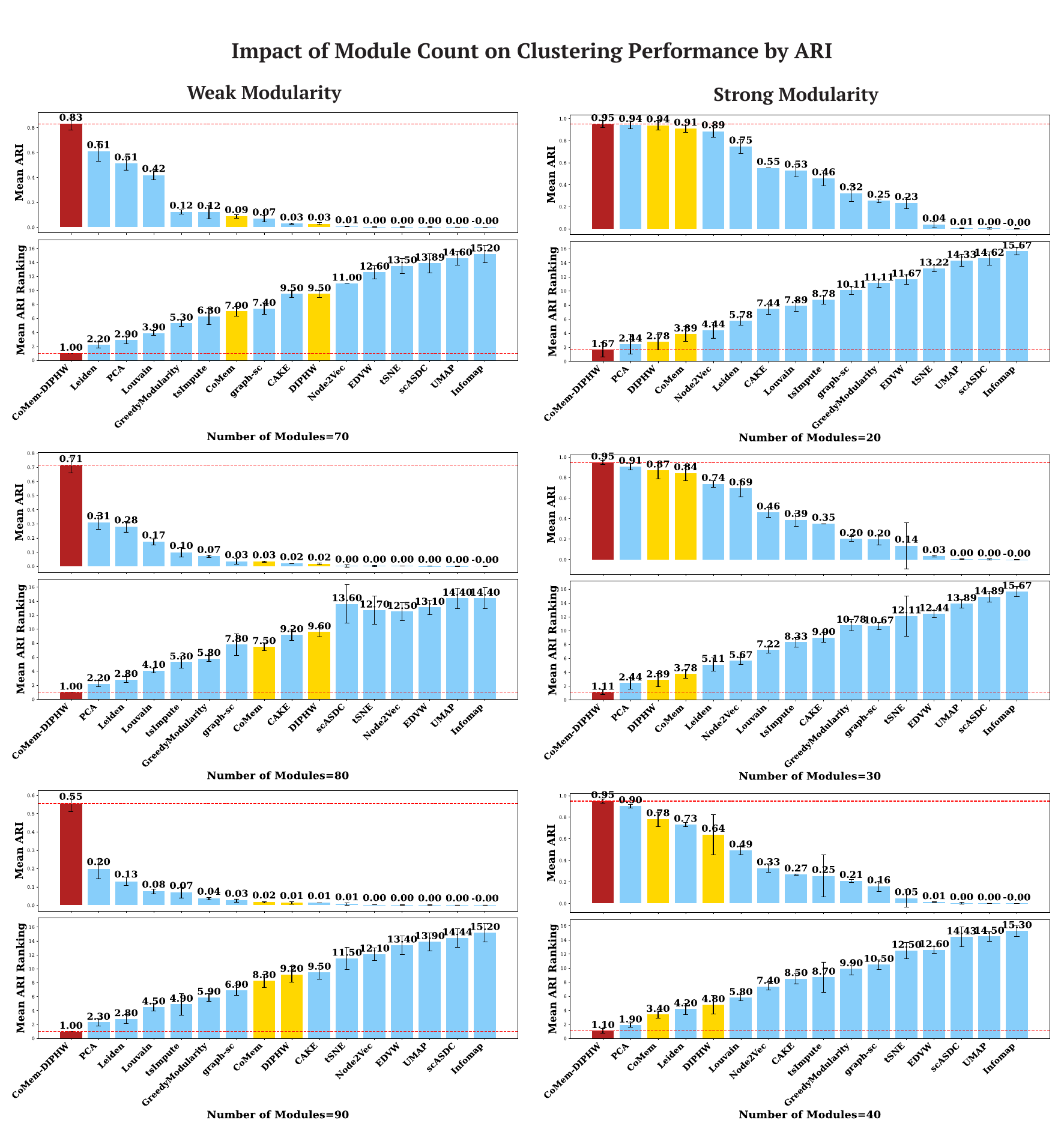}
\caption{Clustering Performance Comparison by ARI Across Varying Module Counts. Simulated scRNA-seq data were used for this evaluation. Here, we present the complementary results across intermediate parameter settings, which show consistent patterns: our proposed methods (highlighted in red and yellow) consistently ranked first by ARI in all scenarios, with a stronger advantage when data modularity is weak, i.e., when the number of embedded modules in the simulated scRNA-seq data is greater. Each experiment was repeated 10 times per parameter setting, with error bars representing the 95\% confidence interval. Red dashed lines indicate the highest ARI values or best ARI rankings. K-means was used to cluster the output of all embedding-based methods that do not directly assign cluster membership.}
\label{fig:Nmodules_ARI_supplement}
\end{figure}

\begin{figure}[H]
\centering
\includegraphics[width=0.85\textwidth]{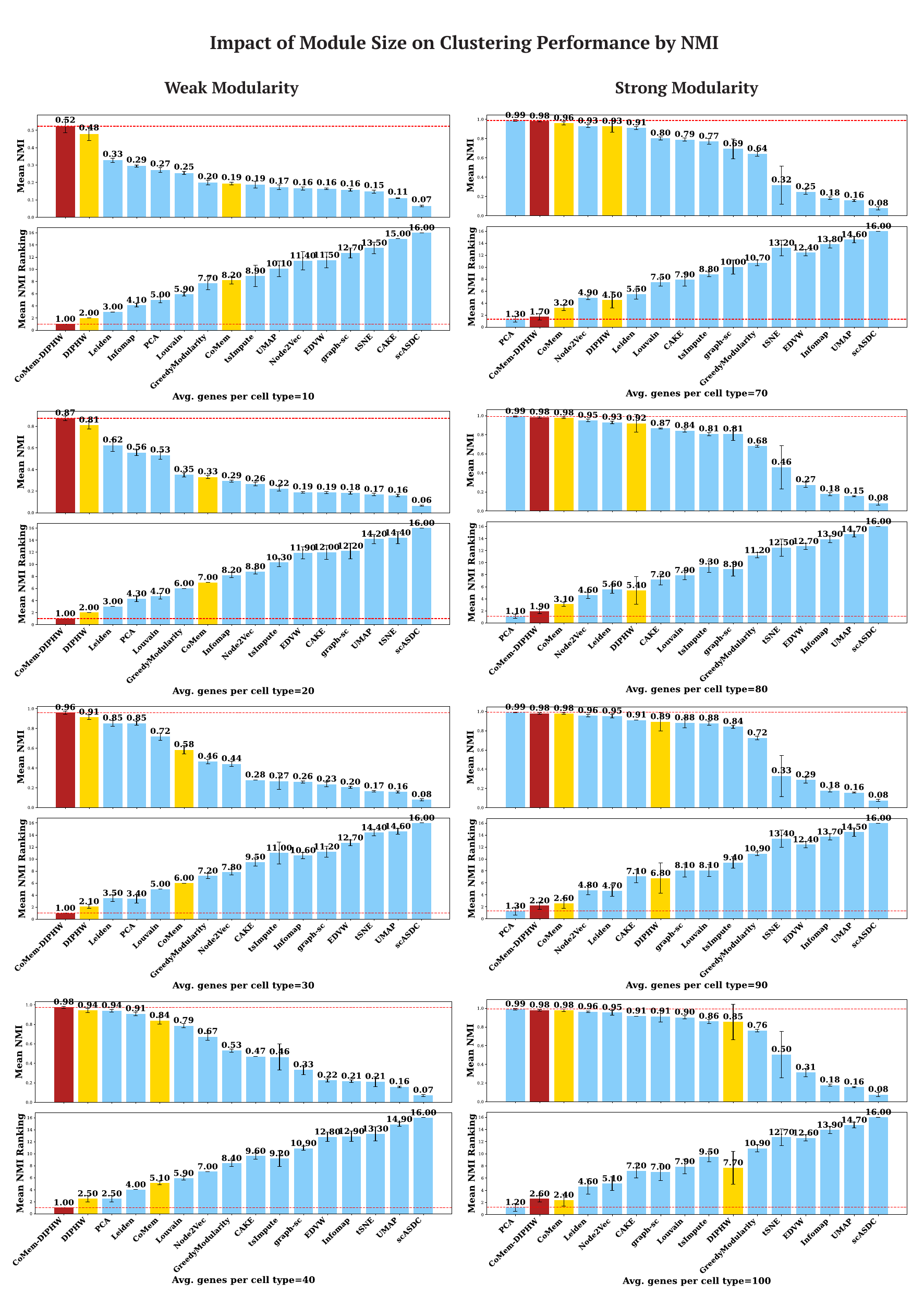}
\caption{Clustering Performance Comparison by NMI Across Varying Module Sizes. Simulated scRNA-seq data were used for this evaluation. The NMI results support the same conclusion: when modularity is weak (i.e., when the average number of co-expressed genes per module is small), the advantage of our proposed methods (highlighted in red and yellow) is more pronounced. Each experiment was repeated 10 times per parameter setting, with error bars representing the 95\% confidence interval. Red dashed lines indicate the highest NMI values or best NMI rankings. K-means was used to cluster the output of all embedding-based methods that do not directly assign cluster membership.}
\label{fig:Ngenes_NMI_supplement}
\end{figure}

\begin{figure}[H]
\centering
\includegraphics[width=0.85\textwidth]{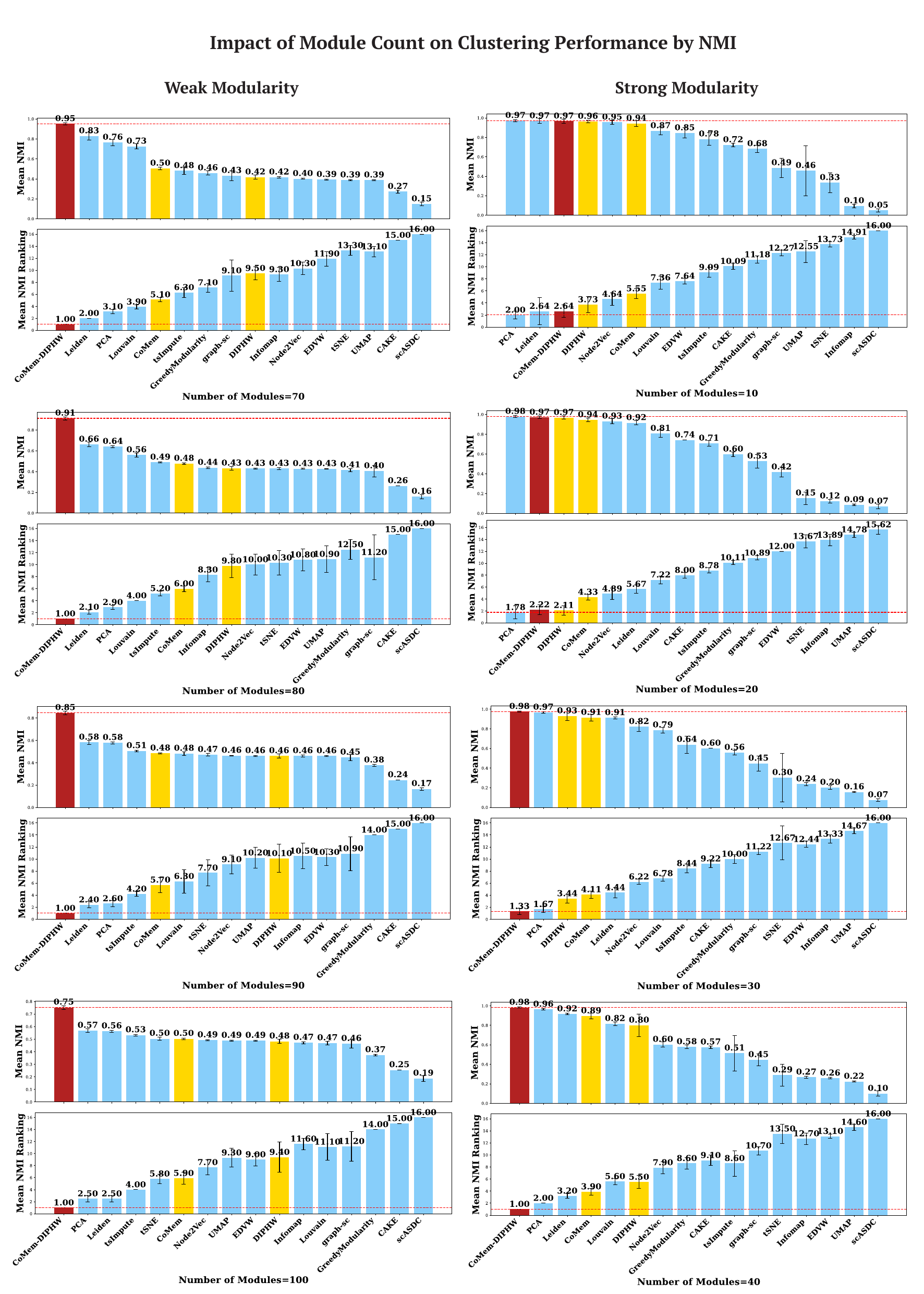}
\caption{Clustering Performance Comparison by NMI Across Varying Module Counts. Simulated scRNA-seq data were used for this evaluation. The results by NMI support the same conclusion: when modularity is weak (i.e., when the number of modules is greater), the advantage of our proposed methods (highlighted in red and yellow) is more pronounced. Each experiment was repeated 10 times per parameter setting, with error bars representing the 95\% confidence interval. Red dashed lines indicate the highest NMI values or best NMI rankings. K-means was used to cluster the output of all embedding-based methods that do not directly assign cluster membership.}
\label{fig:Nmodules_NMI_supplement}
\end{figure}

\begin{figure}[H]
\centering
\includegraphics[width=0.85\textwidth]{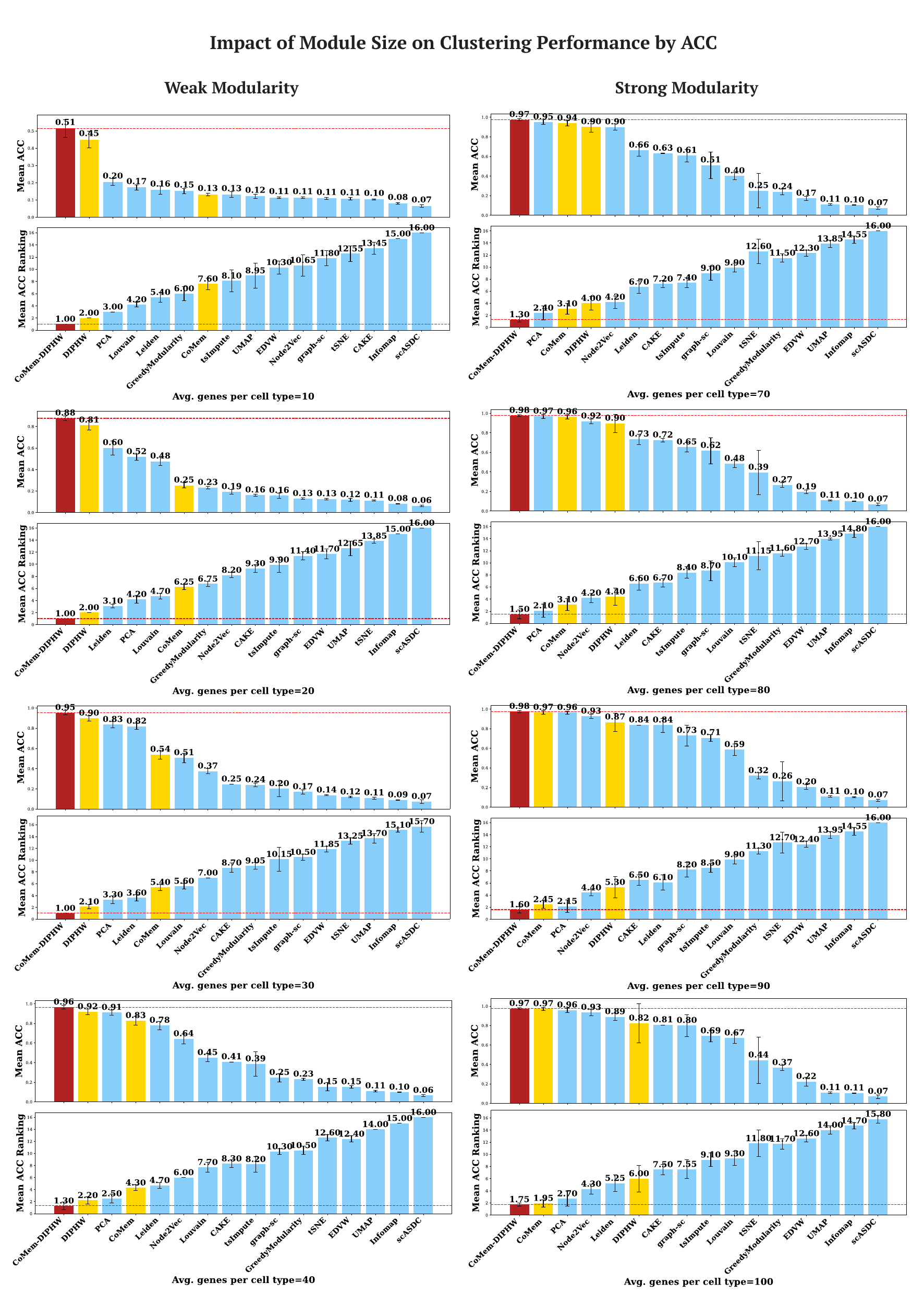}
\caption{Clustering Performance Comparison by ACC Across Varying Module Sizes. Simulated scRNA-seq data were used for this evaluation. The ACC results support the same conclusion: when modularity is weak (i.e., when the average number of co-expressed genes per module is small), the advantage of our proposed methods (highlighted in red and yellow) is more pronounced. Each experiment was repeated 10 times per parameter setting, with error bars representing the 95\% confidence interval. Red dashed lines indicate the highest ACC values or best ACC rankings. K-means was used to cluster the output of all embedding-based methods that do not directly assign cluster membership.}
\label{fig:Ngenes_ACC_supplement}
\end{figure}

\begin{figure}[H]
\centering
\includegraphics[width=0.85\textwidth]{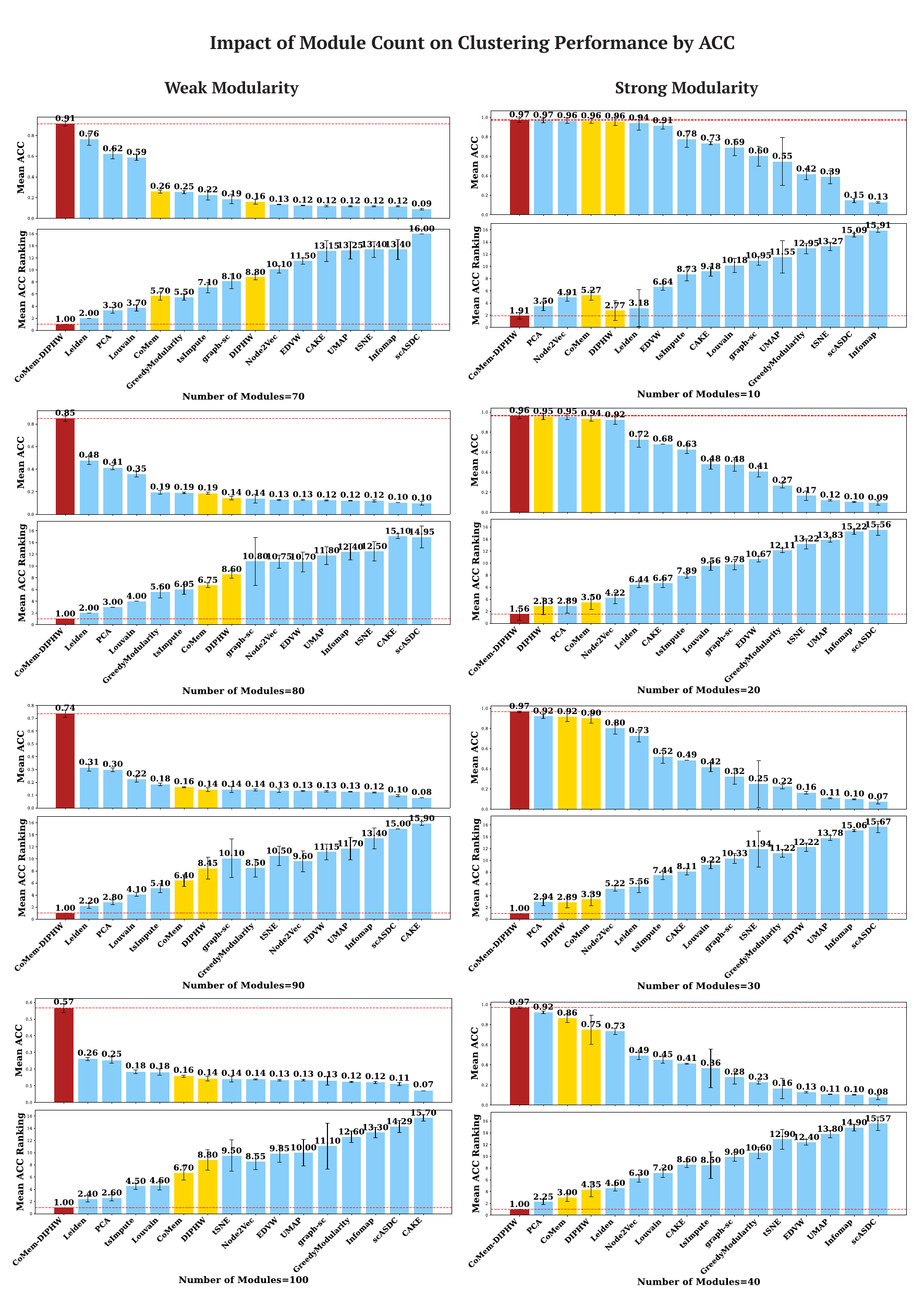}
\caption{Clustering Performance Comparison by ACC Across Varying Module Counts. Simulated scRNA-seq data were used for this evaluation. The results by ACC support the same conclusion: when modularity is weak (i.e., when the number of modules is greater), the advantage of our proposed methods (highlighted in red and yellow) is more pronounced. Each experiment was repeated 10 times per parameter setting, with error bars representing the 95\% confidence interval. Red dashed lines indicate the highest ACC values or best ACC rankings. K-means was used to cluster the output of all embedding-based methods that do not directly assign cluster membership.}
\label{fig:Nmodules_ACC_supplement}
\end{figure}

\begin{figure}[H]
\centering
\includegraphics[width=0.85\textwidth]{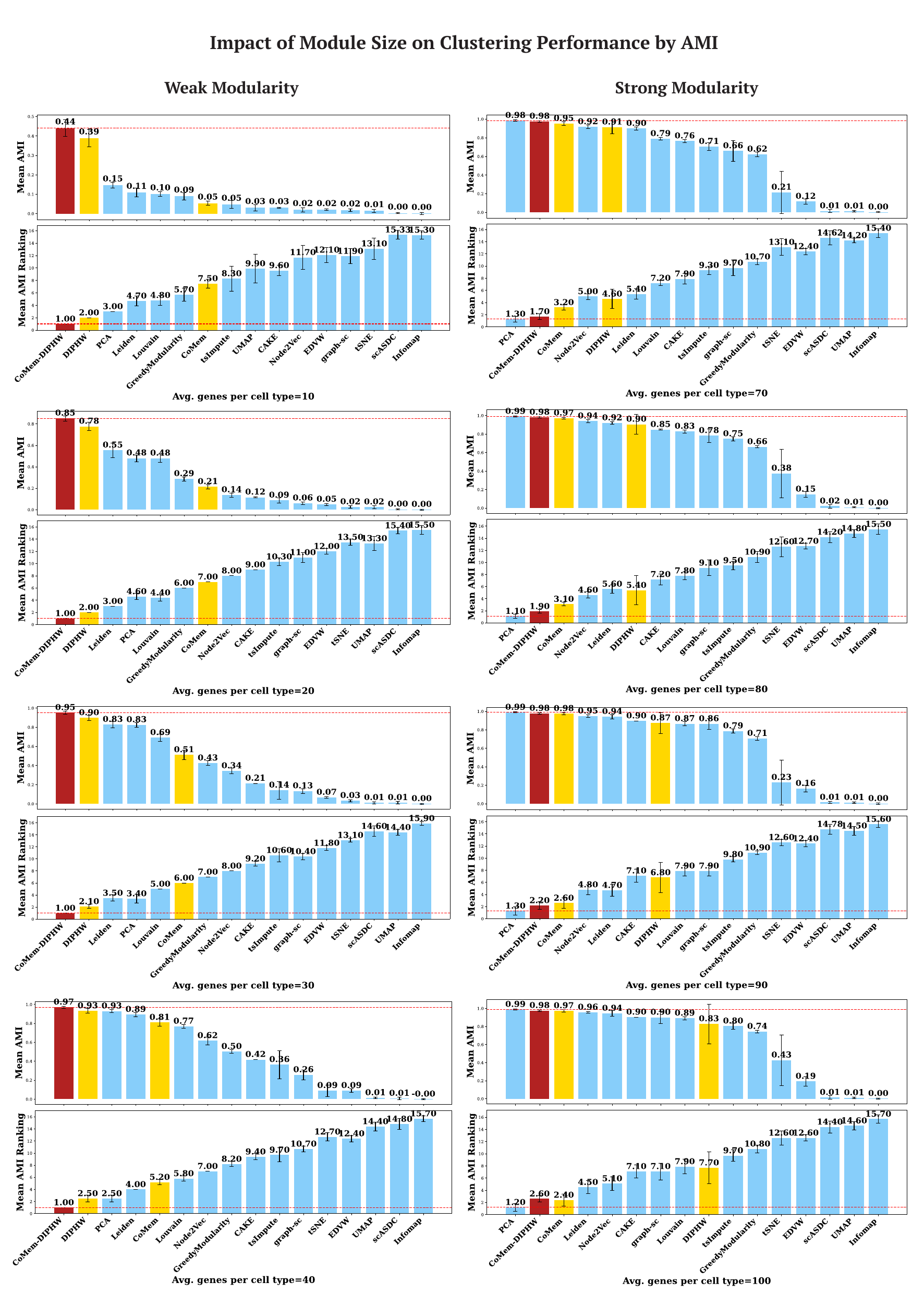}
\caption{Clustering Performance Comparison by AMI Across Varying Module Sizes. Simulated scRNA-seq data were used for this evaluation. The AMI results support the same conclusion: when modularity is weak (i.e., when the average number of co-expressed genes per module is small), the advantage of our proposed methods (highlighted in red and yellow) is more pronounced. Each experiment was repeated 10 times per parameter setting, with error bars representing the 95\% confidence interval. Red dashed lines indicate the highest AMI values or best AMI rankings. K-means was used to cluster the output of all embedding-based methods that do not directly assign cluster membership.}
\label{fig:Ngenes_AMI_supplement}
\end{figure}

\begin{figure}[H]
\centering
\includegraphics[width=0.85\textwidth]{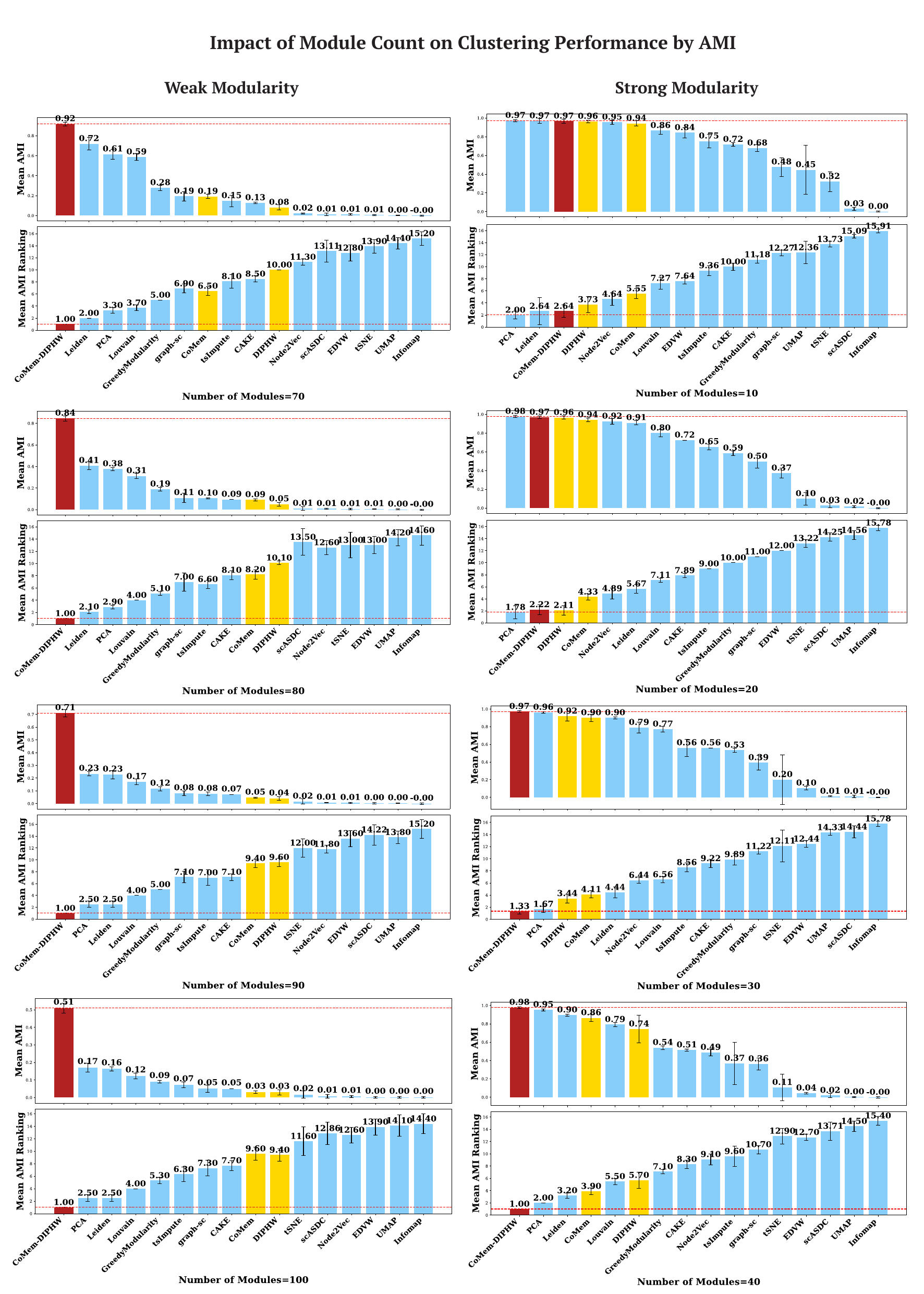}
\caption{Clustering Performance Comparison by AMI Across Varying Module Counts. Simulated scRNA-seq data were used for this evaluation. The results by AMI support the same conclusion: when modularity is weak (i.e., when the number of modules is greater), the advantage of our proposed methods (highlighted in red and yellow) is more pronounced. Each experiment was repeated 10 times per parameter setting, with error bars representing the 95\% confidence interval. Red dashed lines indicate the highest AMI values or best AMI rankings. K-means was used to cluster the output of all embedding-based methods that do not directly assign cluster membership.}
\label{fig:Nmodules_AMI_supplement}
\end{figure}

\begin{figure}[H]
\centering
\includegraphics[width=0.85\textwidth]{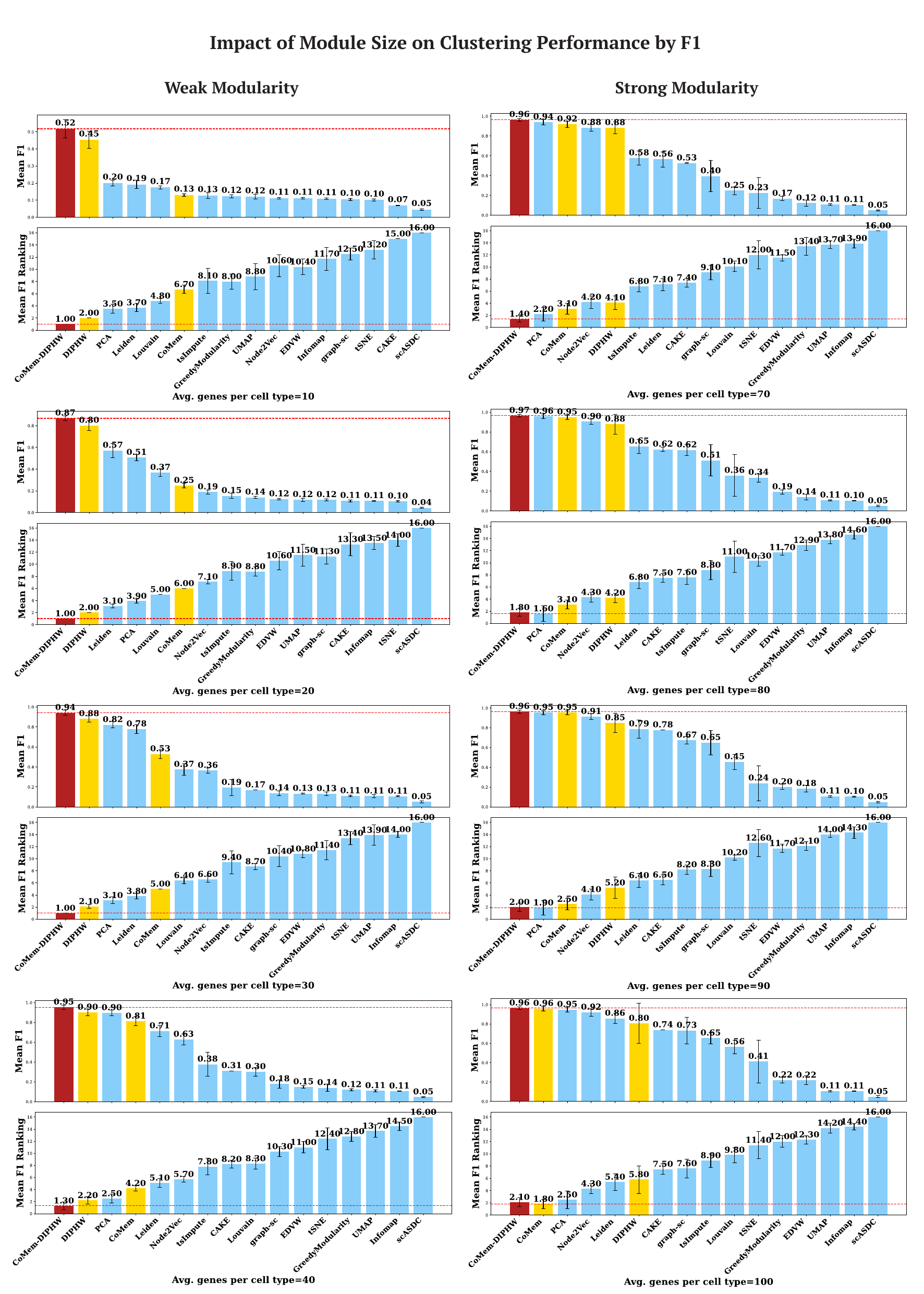}
\caption{Clustering Performance Comparison by F1 Across Varying Module Sizes. Simulated scRNA-seq data were used for this evaluation. The F1 results support the same conclusion: when modularity is weak (i.e., when the average number of co-expressed genes per module is small), the advantage of our proposed methods (highlighted in red and yellow) is more pronounced. Each experiment was repeated 10 times per parameter setting, with error bars representing the 95\% confidence interval. Red dashed lines indicate the highest F1 values or best F1 rankings. K-means was used to cluster the output of all embedding-based methods that do not directly assign cluster membership.}
\label{fig:Ngenes_F1_supplement}
\end{figure}

\begin{figure}[H]
\centering
\includegraphics[width=0.85\textwidth]{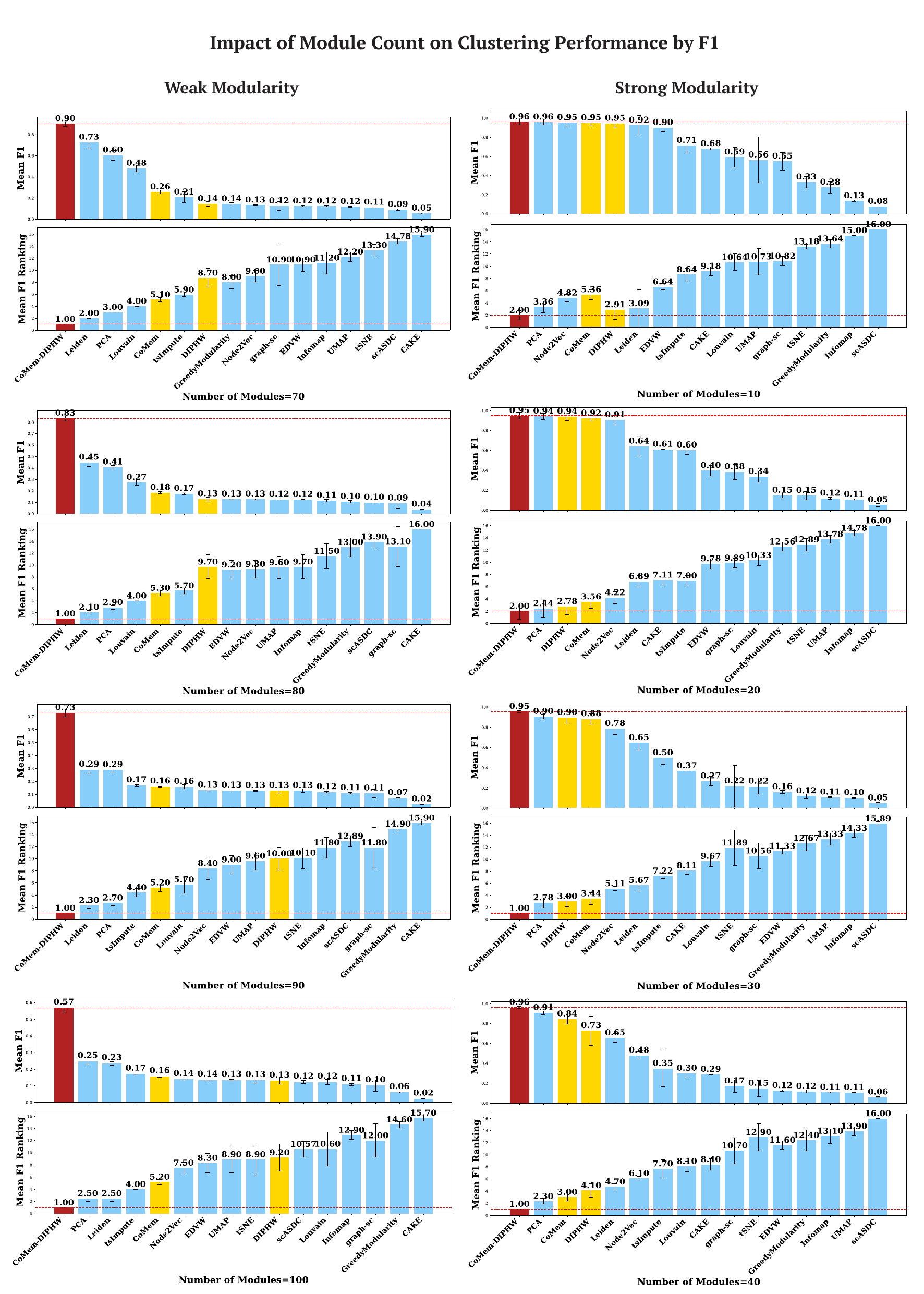}
\caption{Clustering Performance Comparison by F1 Across Varying Module Counts. Simulated scRNA-seq data were used for this evaluation. The results by F1 support the same conclusion: when modularity is weak (i.e., when the number of modules is greater), the advantage of our proposed methods (highlighted in red and yellow) is more pronounced. Each experiment was repeated 10 times per parameter setting, with error bars representing the 95\% confidence interval. Red dashed lines indicate the highest F1 values or best F1 rankings. K-means was used to cluster the output of all embedding-based methods that do not directly assign cluster membership.}
\label{fig:Nmodules_F1_supplement}
\end{figure}

\begin{figure}[H]
\centering
\includegraphics[width=\textwidth]{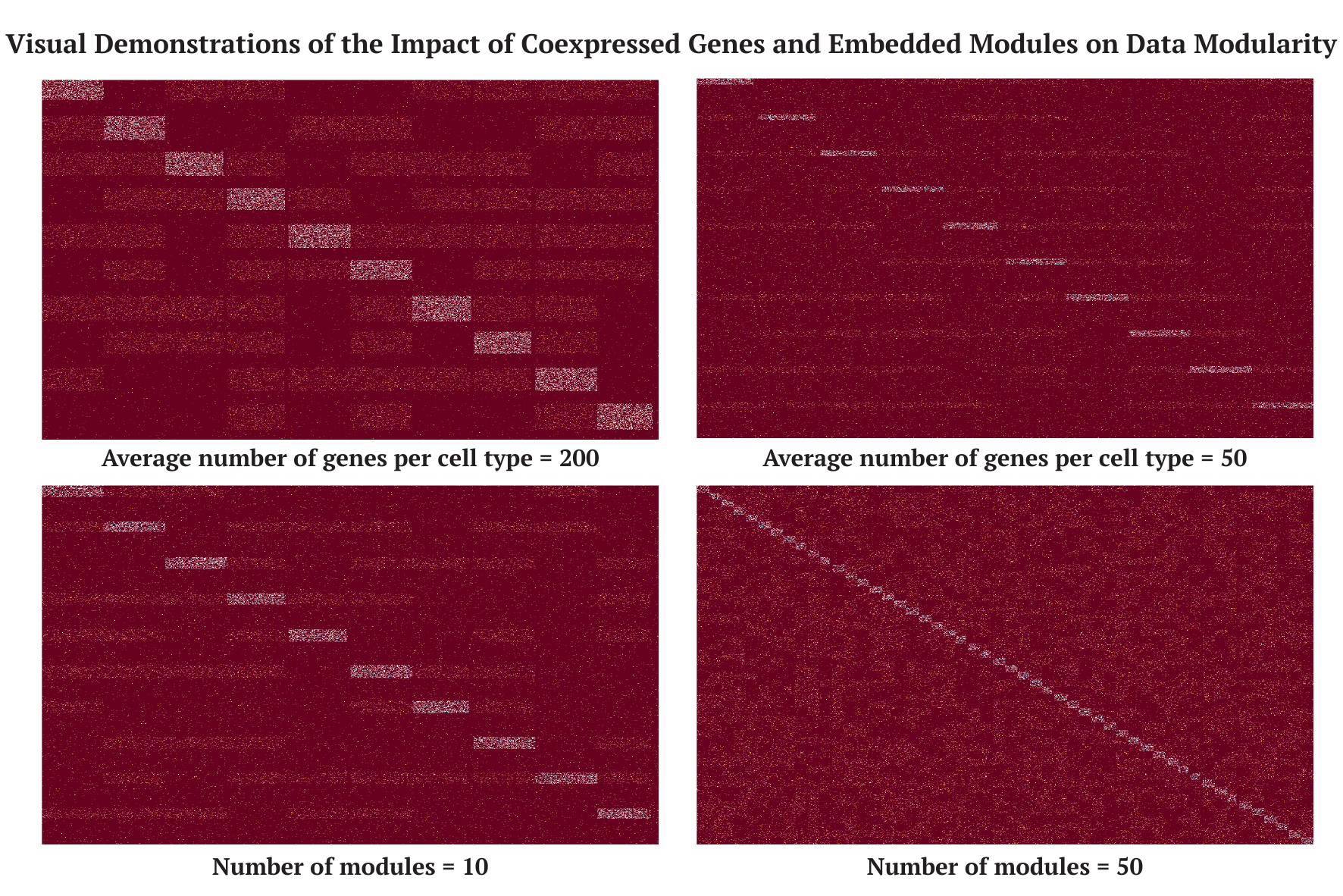}
\caption{Visualization of Modularity in Simulated scRNA-seq Data. Modularity increases with a higher average number of co-expressed genes per cell type and fewer modules. The number of modules models the number of cell types in the simulated data.}
\label{fig:ModularityVisual}
\end{figure}

Figures~\ref{fig:Ngenes_ARI_supplement} and~\ref{fig:Nmodules_ARI_supplement} show additional results on the impact of module size and module count on ARI clustering performance.
Figures~\ref{fig:Ngenes_NMI_supplement} and~\ref{fig:Nmodules_NMI_supplement} show the impact of module size and module count on NMI clustering performance.
Figures~\ref{fig:Ngenes_ACC_supplement} and~\ref{fig:Nmodules_ACC_supplement} show the impact of module size and module count on ACC clustering performance.
Figures~\ref{fig:Ngenes_AMI_supplement} and~\ref{fig:Nmodules_AMI_supplement} show the impact of module size and module count on AMI clustering performance.
Figures~\ref{fig:Ngenes_F1_supplement} and~\ref{fig:Nmodules_F1_supplement} show the impact of module size and module count on F1 clustering performance. Finally, Fig.~\ref{fig:ModularityVisual} shows a visualization of modularity in our imulated scRNA-seq Data.

\subsection{Results for the Human Brain Dataset}
\label{subsec:human_brain_results}

\begin{figure}[H]
  \centering
  \includegraphics[width=\textwidth]{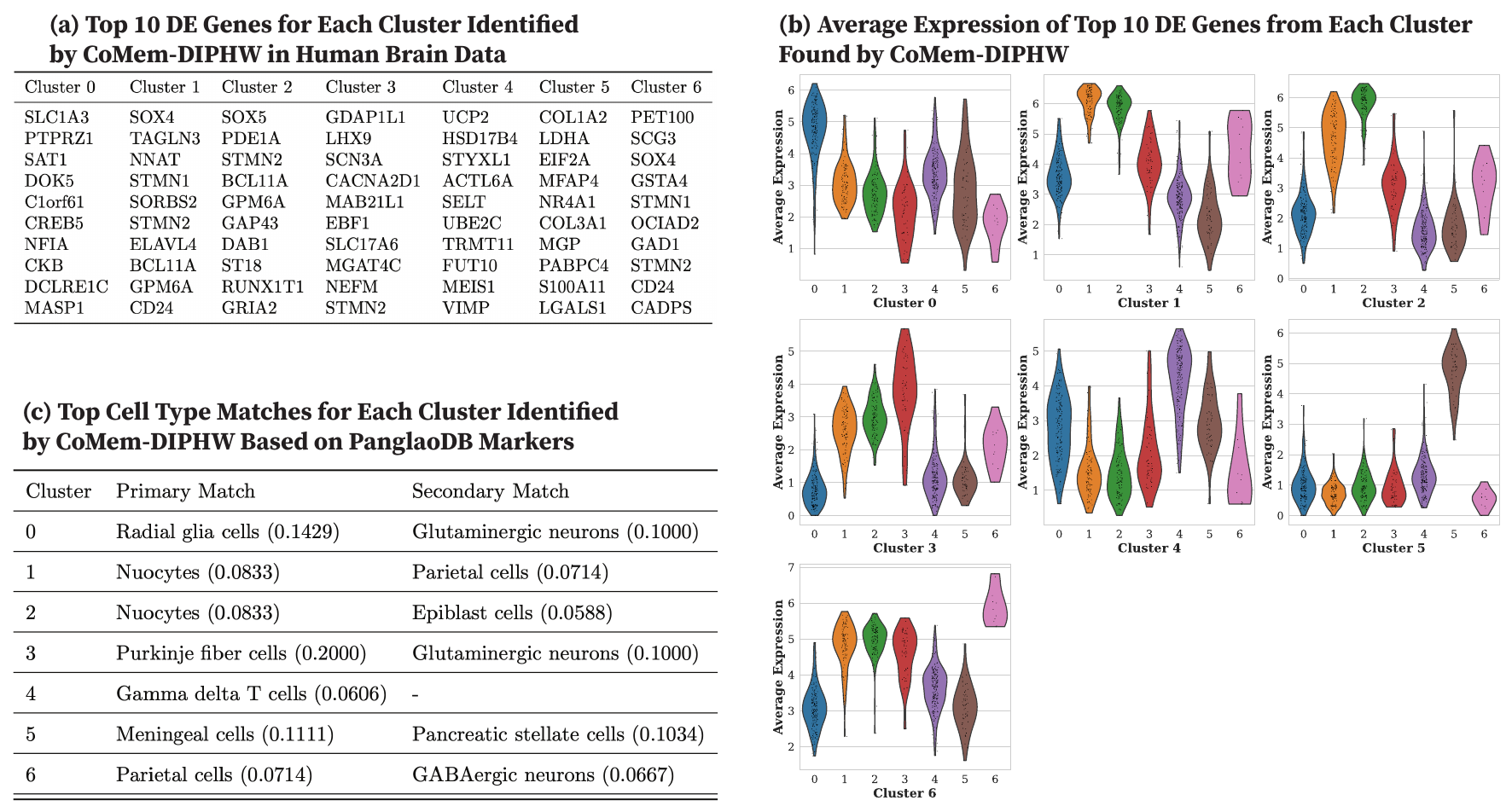} 
  \caption{Clustering Performance of CoMem-DIPHW on the Human Brain Dataset and Cell Type Annotation using DEGs and Canonical Markers. (a) Top 10 DEGs identified for each cluster by CoMem-DIPHW. (b) Across-cluster average expression of cluster-specific DEGs. Violin plots show the distribution of average expression levels of these DEGs across all clusters. Strong clustering performance is indicated by high expression of cluster-specific DEGs within their respective clusters and low expression in other clusters. (c) Cell type annotation using the PanglaoDB marker database. Cell types are determined by the overlap between each cluster's DEGs and cell type-specific markers in the PanglaoDB database, with match scores computed based on the proportion of matched markers.
}
\label{fig:CoMem_HumanBrain}
\end{figure}

\begin{figure}[H]
  \centering
  \includegraphics[width=\textwidth]{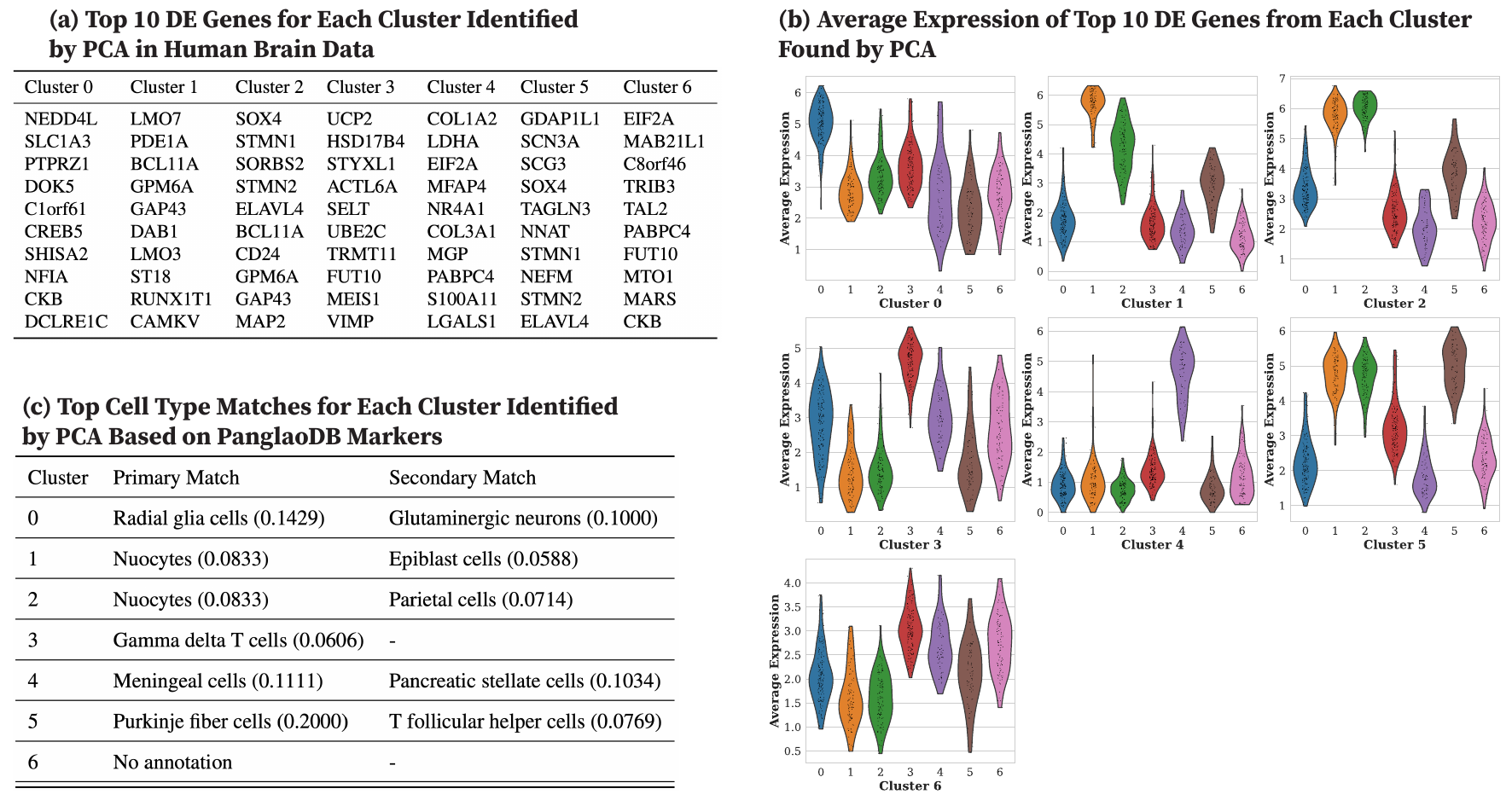} 
  \caption{Clustering Performance of PCA on the Human Brain Dataset and Cell Type Annotation using DEGs and Canonical Markers. (a) Top 10 DEGs identified for each cluster by PCA. (b) Across-cluster average expression of cluster-specific DEGs. Violin plots show the distribution of average expression levels of these DEGs across all clusters. Strong clustering performance is indicated by high expression of cluster-specific DEGs within their respective clusters and low expression in other clusters. (c) Cell type annotation using the PanglaoDB marker database. Cell types are determined by the overlap between each cluster's DEGs and cell type-specific markers in the PanglaoDB database, with match scores computed based on the proportion of matched markers.
}
\label{fig:PCA_HumanBrain}
\end{figure}

\begin{figure}[H]
  \centering
  \includegraphics[width=\textwidth]{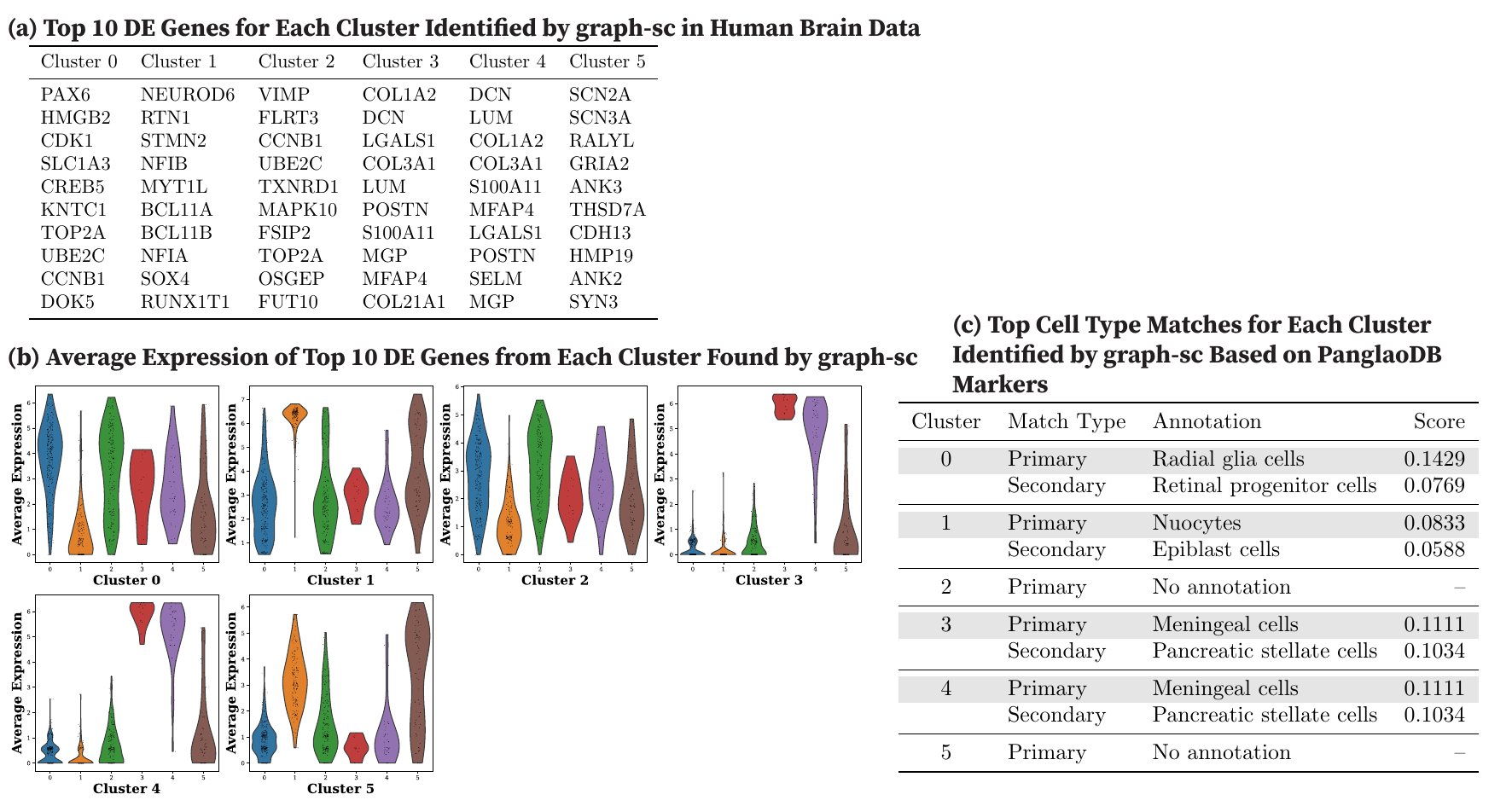} 
  \caption{Clustering Performance of graph-sc on the Human Brain Dataset and Cell Type Annotation using DEGs and Canonical Markers. (a) Top 10 DEGs identified for each cluster by graph-sc. (b) Across-cluster average expression of cluster-specific DEGs. Violin plots show the distribution of average expression levels of these DEGs across all clusters. Strong clustering performance is indicated by high expression of cluster-specific DEGs within their respective clusters and low expression in other clusters. (c) Cell type annotation using the PanglaoDB marker database. Cell types are determined by the overlap between each cluster's DEGs and cell type-specific markers in the PanglaoDB database, with match scores computed based on the proportion of matched markers. 
  Two of the graph-sc clusters had less than 5 cells, one was excluded from the expression plots in (b), and both failed to map to known cell types in (c).
}
\label{fig:graph-sc_HumanBrain}
\end{figure}

\begin{figure}[H]
  \centering
  \includegraphics[width=\textwidth]{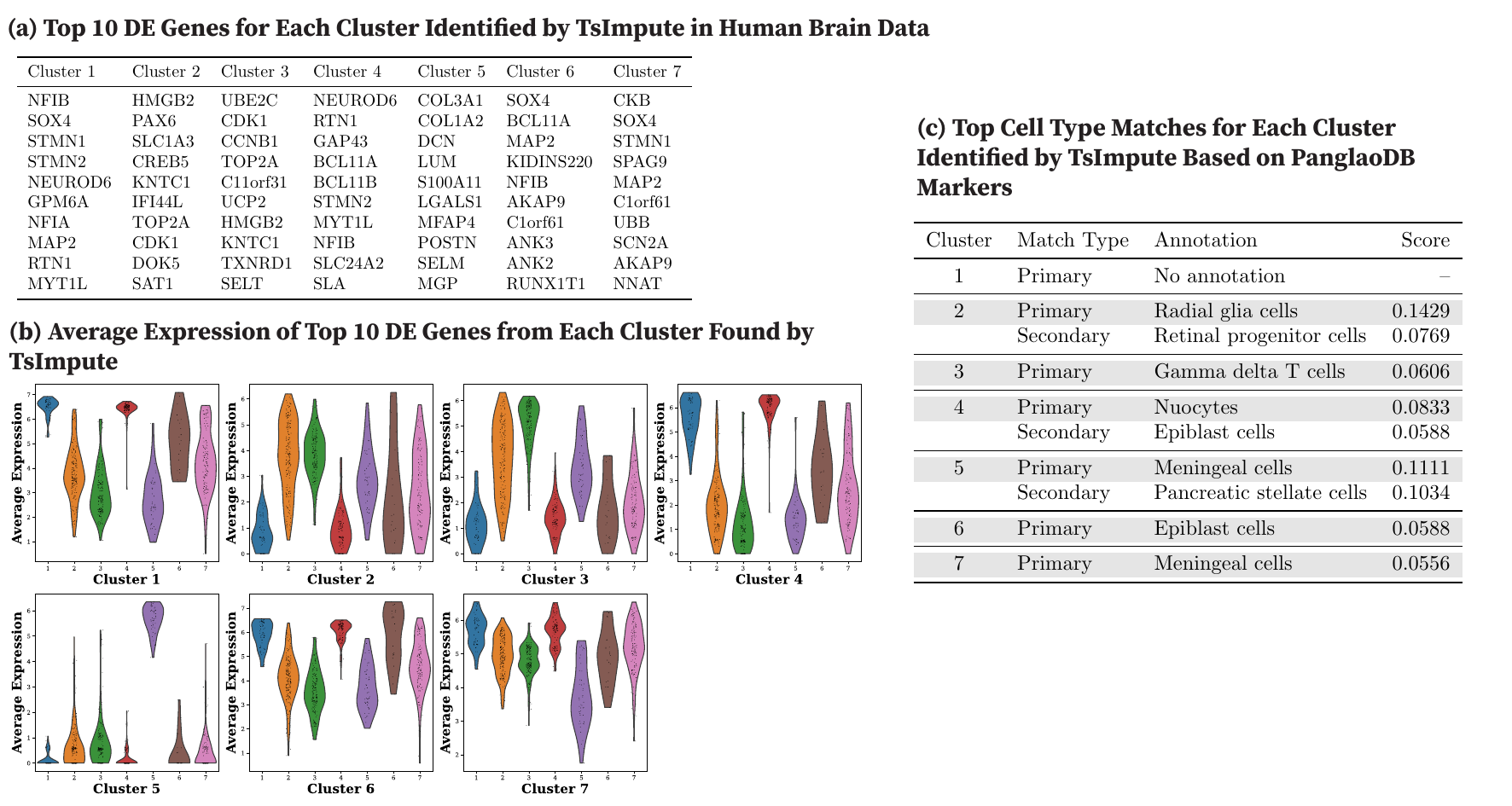} 
  \caption{Clustering Performance of tsImpute on the Human Brain Dataset and Cell Type Annotation using DEGs and Canonical Markers. (a) Top 10 DEGs identified for each cluster by tsImpute. (b) Across-cluster average expression of cluster-specific DEGs. Violin plots show the distribution of average expression levels of these DEGs across all clusters. Strong clustering performance is indicated by high expression of cluster-specific DEGs within their respective clusters and low expression in other clusters. (c) Cell type annotation using the PanglaoDB marker database. Cell types are determined by the overlap between each cluster's DEGs and cell type-specific markers in the PanglaoDB database, with match scores computed based on the proportion of matched markers.
}
\label{fig:tsImpute_HumanBrain}
\end{figure}

\begin{figure}[H]
  \centering
  \includegraphics[width=\textwidth]{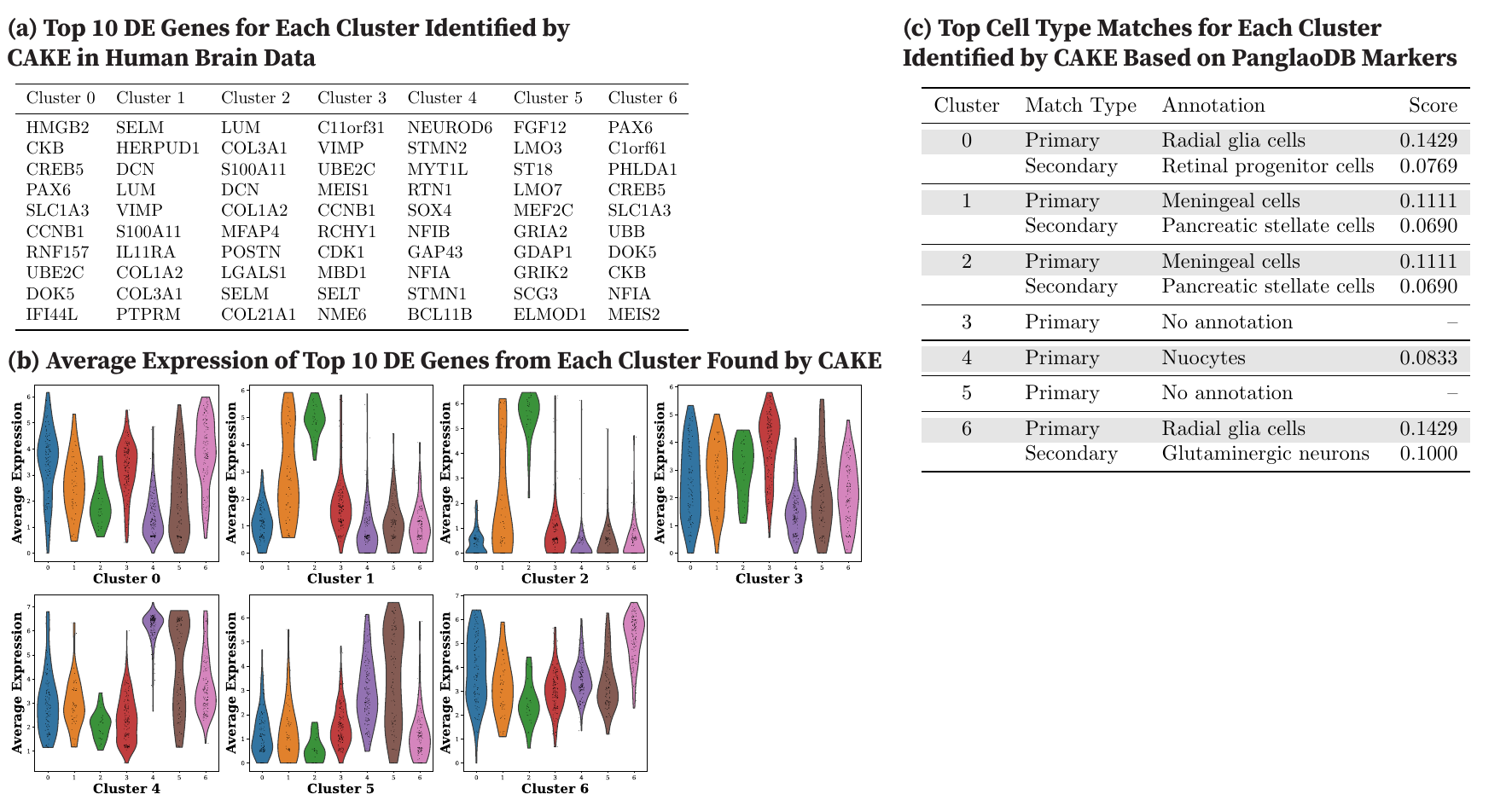} 
  \caption{Clustering Performance of CAKE on the Human Brain Dataset and Cell Type Annotation using DEGs and Canonical Markers. (a) Top 10 DEGs identified for each cluster by CAKE. (b) Across-cluster average expression of cluster-specific DEGs. Violin plots show the distribution of average expression levels of these DEGs across all clusters. Strong clustering performance is indicated by high expression of cluster-specific DEGs within their respective clusters and low expression in other clusters. (c) Cell type annotation using the PanglaoDB marker database. Cell types are determined by the overlap between each cluster's DEGs and cell type-specific markers in the PanglaoDB database, with match scores computed based on the proportion of matched markers.
}
\label{fig:CAKE_HumanBrain}
\end{figure}

\begin{figure}[H]
  \centering
  \includegraphics[width=\textwidth]{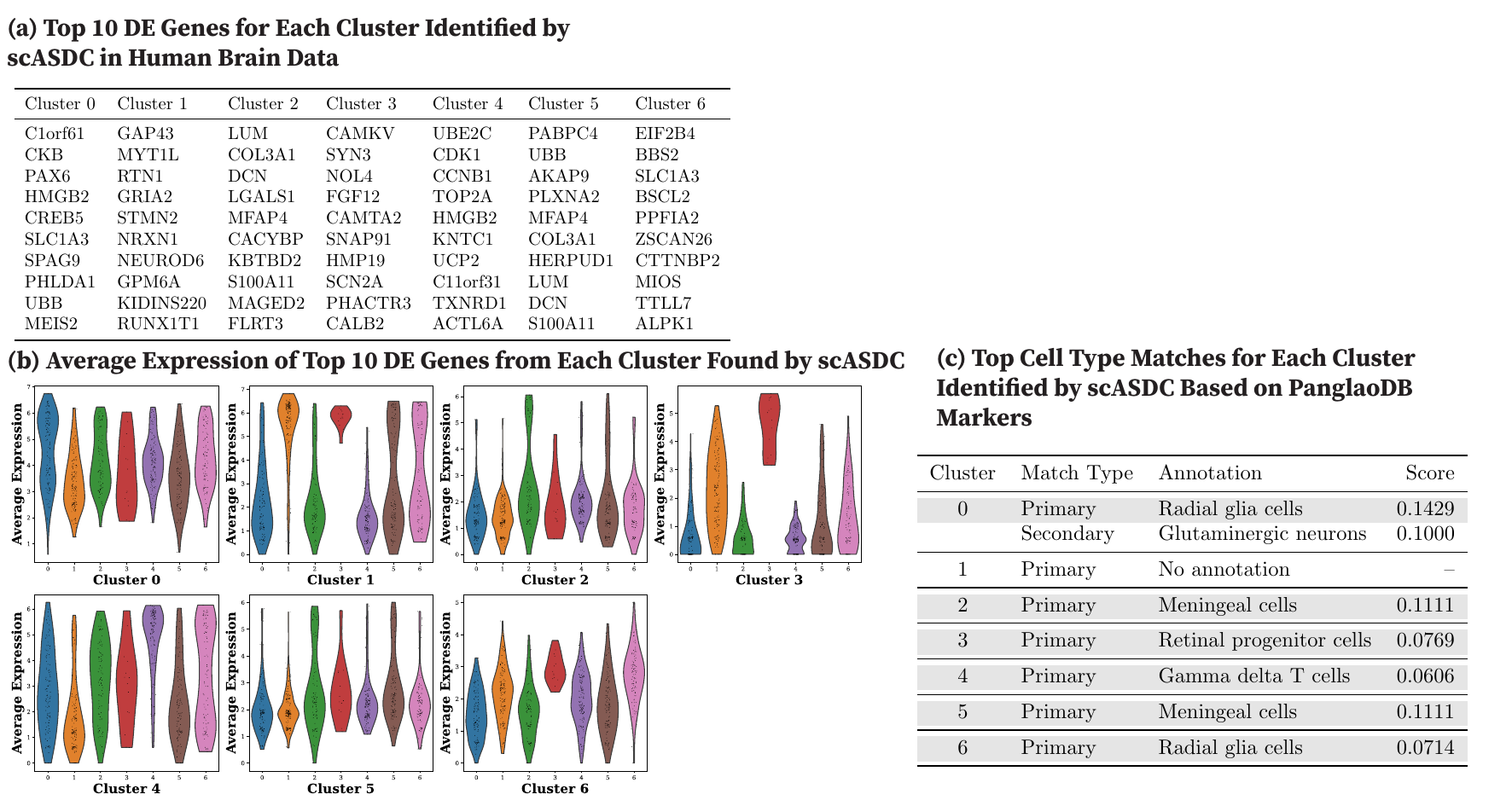} 
  \caption{Clustering Performance of scASDC on the Human Brain Dataset and Cell Type Annotation using DEGs and Canonical Markers. (a) Top 10 DEGs identified for each cluster by scASDC. (b) Across-cluster average expression of cluster-specific DEGs. Violin plots show the distribution of average expression levels of these DEGs across all clusters. Strong clustering performance is indicated by high expression of cluster-specific DEGs within their respective clusters and low expression in other clusters. (c) Cell type annotation using the PanglaoDB marker database. Cell types are determined by the overlap between each cluster's DEGs and cell type-specific markers in the PanglaoDB database, with match scores computed based on the proportion of matched markers.
}
\label{fig:scASDC_HumanBrain}
\end{figure}

Figures~\ref{fig:CoMem_HumanBrain},~\ref{fig:PCA_HumanBrain},~\ref{fig:graph-sc_HumanBrain},~\ref{fig:tsImpute_HumanBrain},~\ref{fig:CAKE_HumanBrain}, and~\ref{fig:scASDC_HumanBrain} show the results on the human brain dataset for CoMem-DIPHW, PCA, graph-sc, tsImpute, CAKE, and scASDC, respectively.

\subsection{Results for the Mouse Pancreas and the Mouse Brain Datasets}
\label{subsec:mouse_pancreas_brain_results}

\begin{figure}[H]
  \centering
  \includegraphics[width=\textwidth]{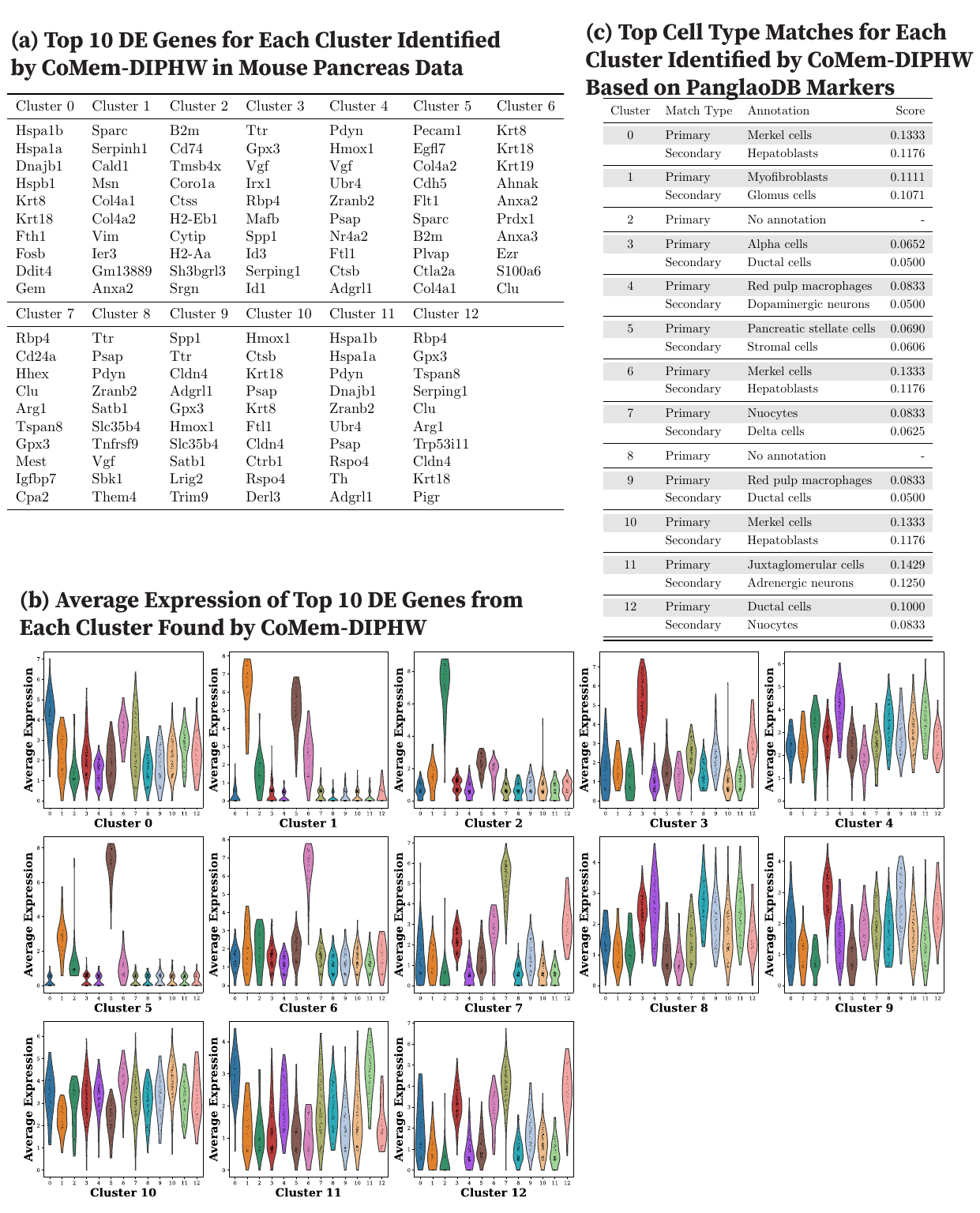} 
  \caption{Clustering Performance of CoMem-DIPHW on the Mouse Pancreas Dataset and Cell Type Annotation using DEGs and Canonical Markers. (a) Top 10 DEGs identified for each cluster by CoMem-DIPHW. (b) Across-cluster average expression of cluster-specific DEGs. Violin plots show the distribution of average expression levels of these DEGs across all clusters. Strong clustering performance is indicated by high expression of cluster-specific DEGs within their respective clusters and low expression in other clusters. (c) Cell type annotation using the PanglaoDB marker database. Cell types are determined by the overlap between each cluster's DEGs and cell type-specific markers in the PanglaoDB database, with match scores computed based on the proportion of matched markers.}
\label{fig:CoMem_MousePancreas}
\end{figure}

\begin{figure}[H]
  \centering
  \includegraphics[width=\textwidth]{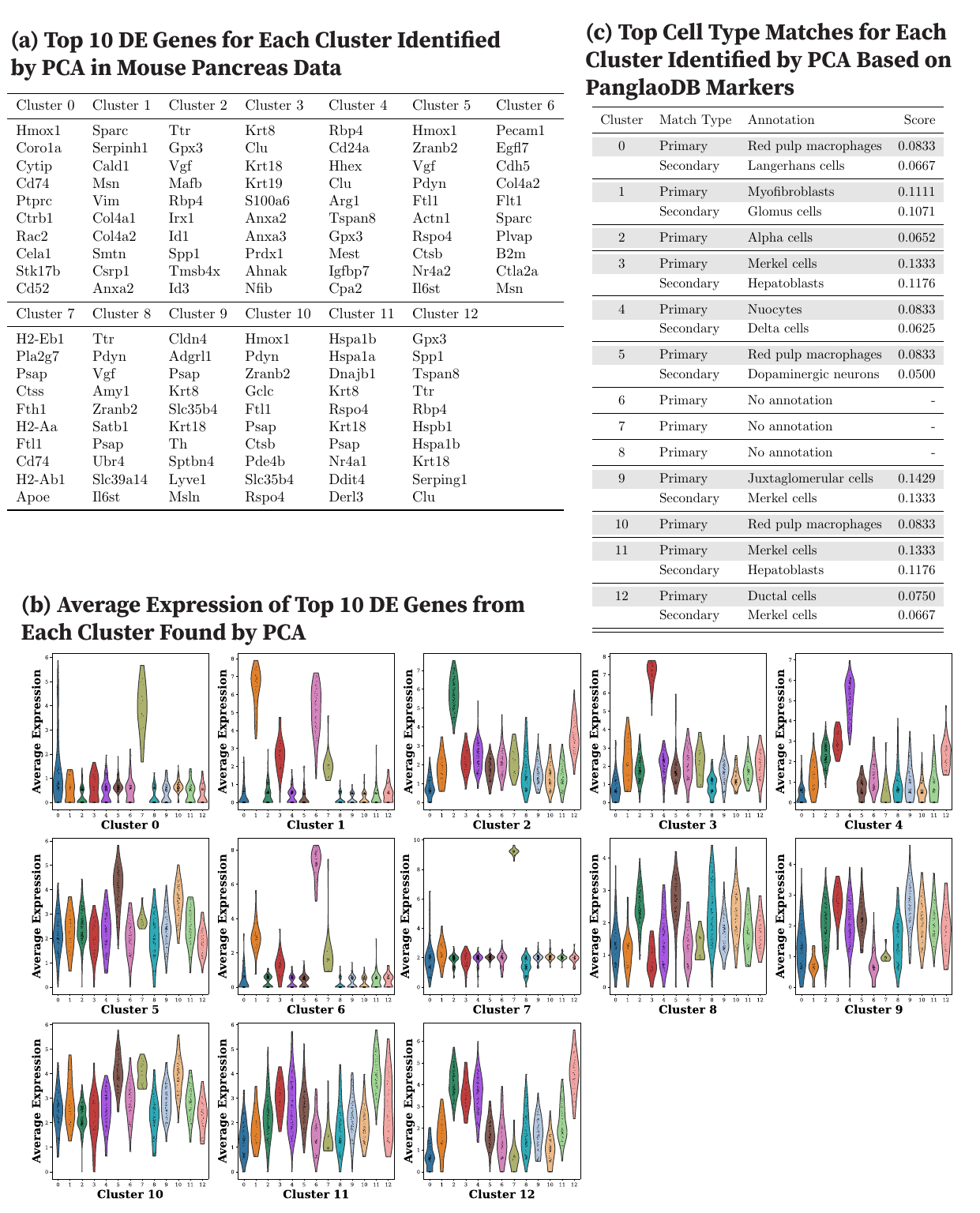} 
  \caption{Clustering Performance of PCA on the Mouse Pancreas Dataset and Cell Type Annotation using DEGs and Canonical Markers. (a) Top 10 DEGs identified for each cluster by PCA. (b) Across-cluster average expression of cluster-specific DEGs. Violin plots show the distribution of average expression levels of these DEGs across all clusters. Strong clustering performance is indicated by high expression of cluster-specific DEGs within their respective clusters and low expression in other clusters. (c) Cell type annotation using the PanglaoDB marker database. Cell types are determined by the overlap between each cluster's DEGs and cell type-specific markers in the PanglaoDB database, with match scores computed based on the proportion of matched markers.}
\label{fig:PCA_MousePancreas}
\end{figure}

\begin{figure}[H]
  \centering
  \includegraphics[width=\textwidth]{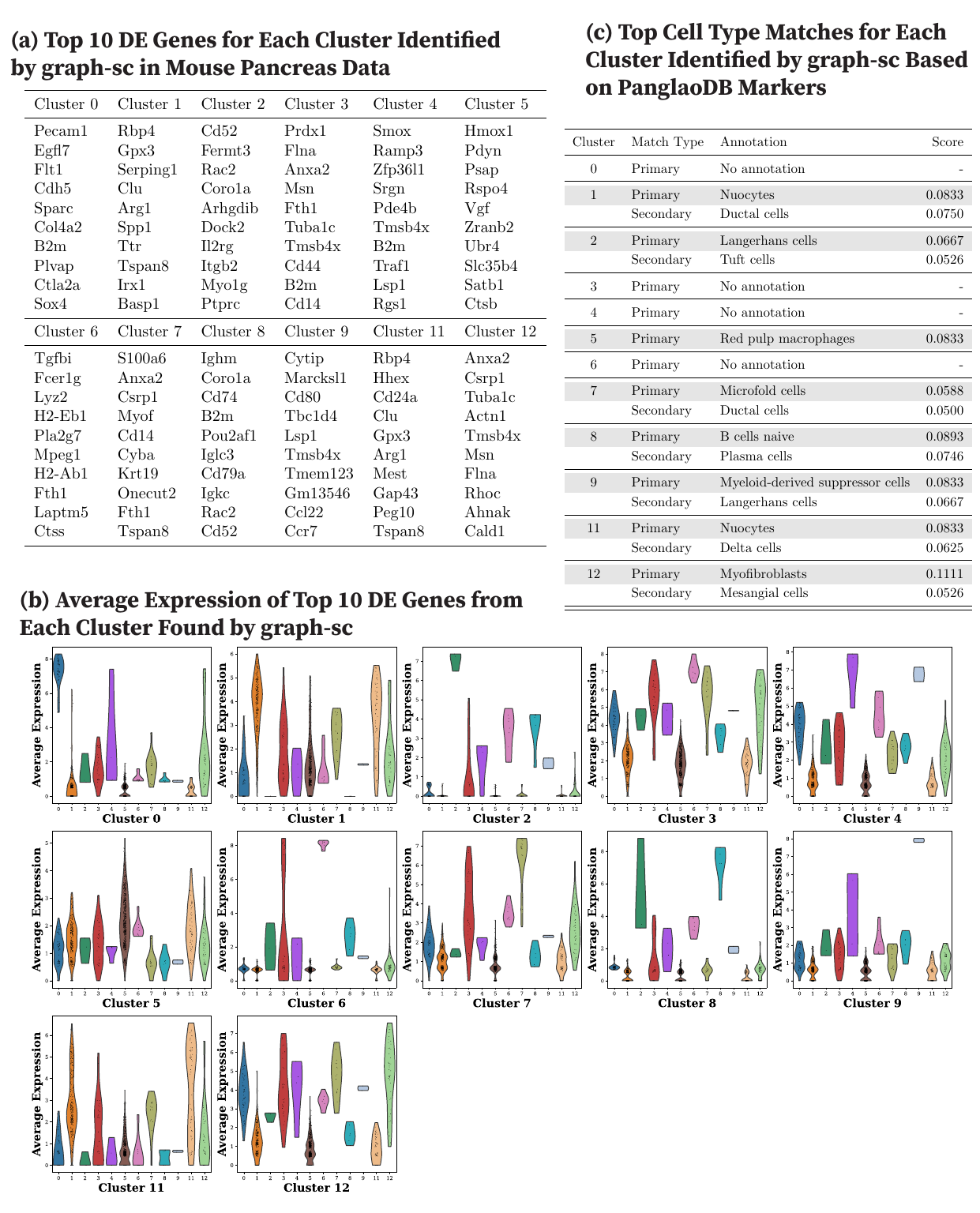} 
  \caption{Clustering Performance of graph-sc on the Mouse Pancreas Dataset and Cell Type Annotation using DEGs and Canonical Markers. (a) Top 10 DEGs identified for each cluster by graph-sc. (b) Across-cluster average expression of cluster-specific DEGs. Violin plots show the distribution of average expression levels of these DEGs across all clusters. Strong clustering performance is indicated by high expression of cluster-specific DEGs within their respective clusters and low expression in other clusters. (c) Cell type annotation using the PanglaoDB marker database. Cell types are determined by the overlap between each cluster's DEGs and cell type-specific markers in the PanglaoDB database, with match scores computed based on the proportion of matched markers.}
\label{fig:graph-sc_MousePancreas}
\end{figure}

\begin{figure}[H]
  \centering
  \includegraphics[width=\textwidth]{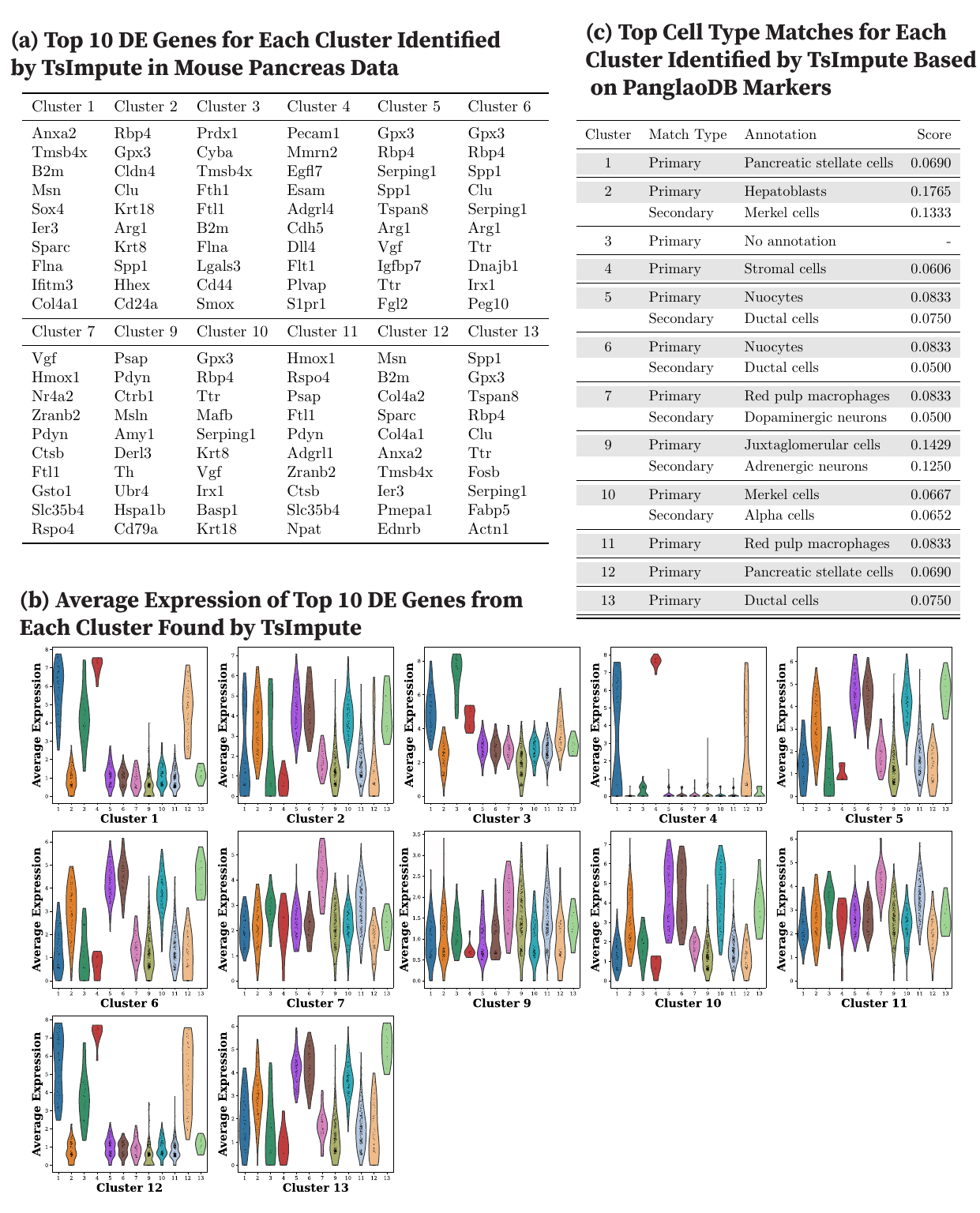} 
  \caption{Clustering Performance of tsImpute on the Mouse Pancreas Dataset and Cell Type Annotation using DEGs and Canonical Markers. (a) Top 10 DEGs identified for each cluster by tsImpute. (b) Across-cluster average expression of cluster-specific DEGs. Violin plots show the distribution of average expression levels of these DEGs across all clusters. Strong clustering performance is indicated by high expression of cluster-specific DEGs within their respective clusters and low expression in other clusters. (c) Cell type annotation using the PanglaoDB marker database. Cell types are determined by the overlap between each cluster's DEGs and cell type-specific markers in the PanglaoDB database, with match scores computed based on the proportion of matched markers.}
\label{fig:tsImpute_MousePancreas}
\end{figure}

\begin{figure}[H]
  \centering
  \includegraphics[width=\textwidth]{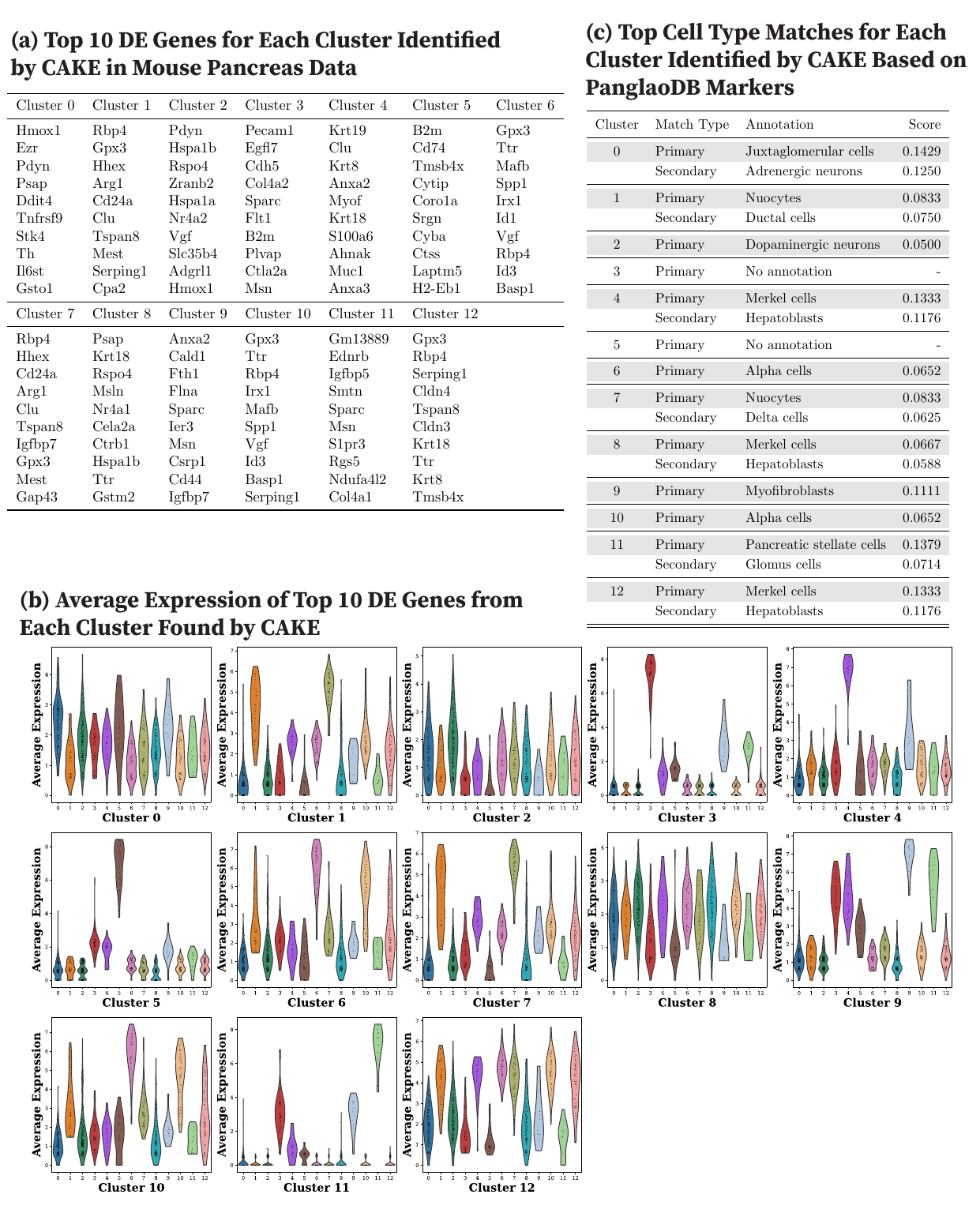} 
  \caption{Clustering Performance of CAKE on the Mouse Pancreas Dataset and Cell Type Annotation using DEGs and Canonical Markers. (a) Top 10 DEGs identified for each cluster by CAKE. (b) Across-cluster average expression of cluster-specific DEGs. Violin plots show the distribution of average expression levels of these DEGs across all clusters. Strong clustering performance is indicated by high expression of cluster-specific DEGs within their respective clusters and low expression in other clusters. (c) Cell type annotation using the PanglaoDB marker database. Cell types are determined by the overlap between each cluster's DEGs and cell type-specific markers in the PanglaoDB database, with match scores computed based on the proportion of matched markers.}
\label{fig:CAKE_MousePancreas}
\end{figure}

\begin{figure}[H]
  \centering
  \includegraphics[width=\textwidth]{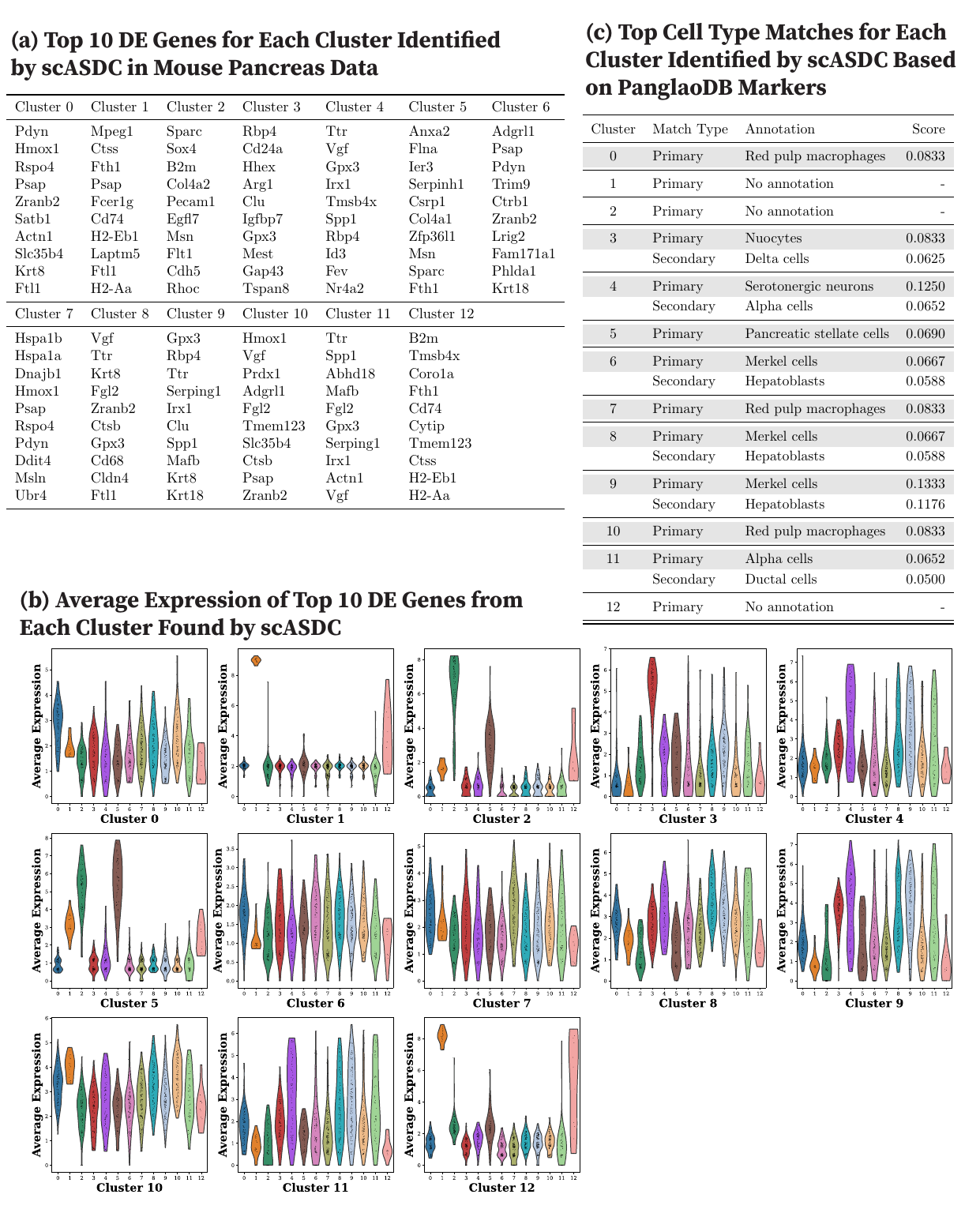} 
  \caption{Clustering Performance of scASDC on the Mouse Pancreas Dataset and Cell Type Annotation using DEGs and Canonical Markers. (a) Top 10 DEGs identified for each cluster by scASDC. (b) Across-cluster average expression of cluster-specific DEGs. Violin plots show the distribution of average expression levels of these DEGs across all clusters. Strong clustering performance is indicated by high expression of cluster-specific DEGs within their respective clusters and low expression in other clusters. (c) Cell type annotation using the PanglaoDB marker database. Cell types are determined by the overlap between each cluster's DEGs and cell type-specific markers in the PanglaoDB database, with match scores computed based on the proportion of matched markers.}
\label{fig:scASDC_MousePancreas}
\end{figure}

Figures~\ref{fig:CoMem_MousePancreas},~\ref{fig:PCA_MousePancreas},~\ref{fig:graph-sc_MousePancreas},~\ref{fig:tsImpute_MousePancreas},~\ref{fig:CAKE_MousePancreas}, and~\ref{fig:scASDC_MousePancreas} show the results on the mouse pancreas dataset for CoMem-DIPHW, PCA, graph-sc, tsImpute, CAKE, and scASDC, respectively.

\begin{figure}[H]
  \centering
  \includegraphics[width=0.98\textwidth]{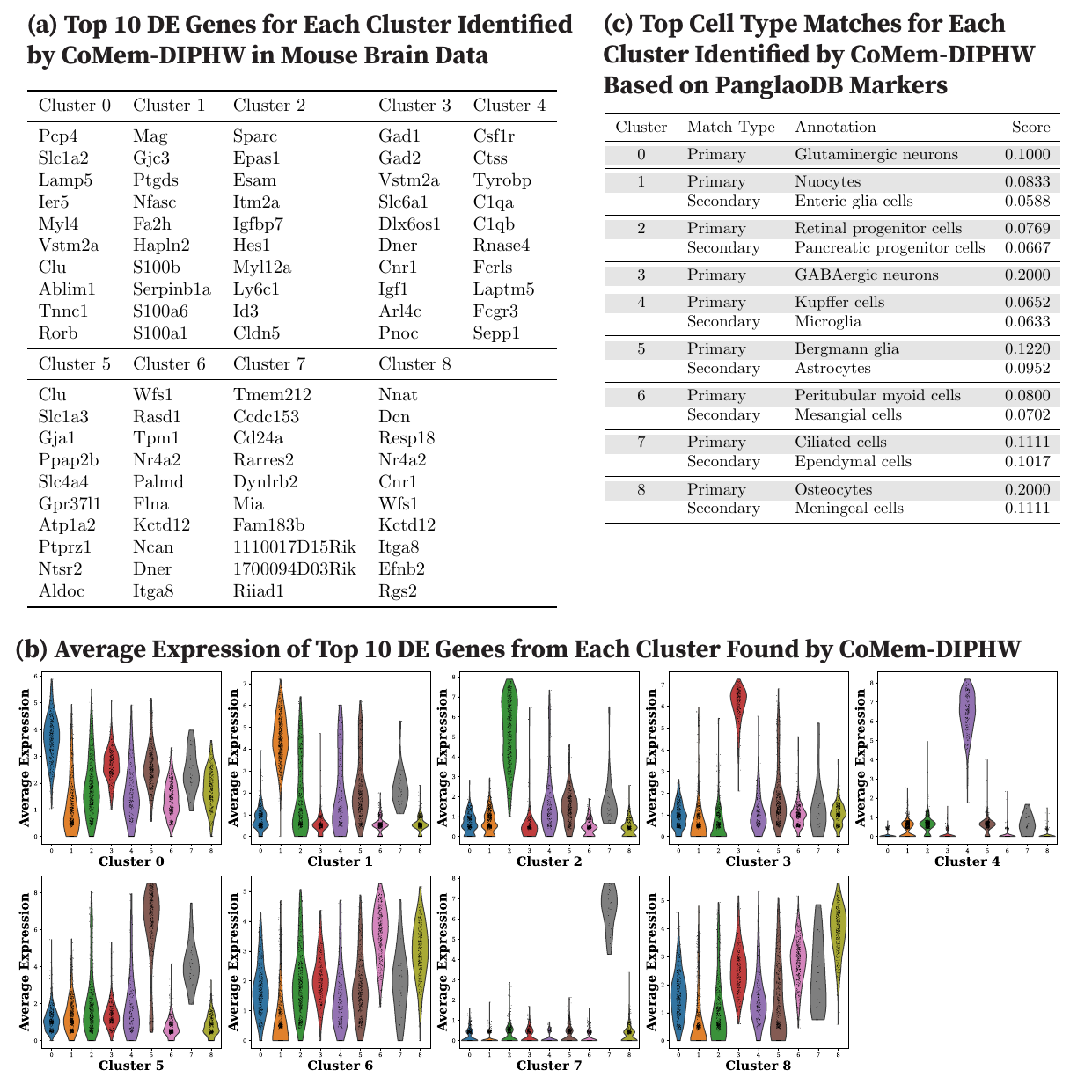} 
  \caption{Clustering Performance of CoMem-DIPHW on the Mouse Brain Dataset and Cell Type Annotation using DEGs and Canonical Markers. (a) Top 10 DEGs identified for each cluster by CoMem-DIPHW. (b) Across-cluster average expression of cluster-specific DEGs. Violin plots show the distribution of average expression levels of these DEGs across all clusters. Strong clustering performance is indicated by high expression of cluster-specific DEGs within their respective clusters and low expression in other clusters. (c) Cell type annotation using the PanglaoDB marker database.}
\label{fig:CoMem_MouseBrain}
\end{figure}

\begin{figure}[H]
  \centering
  \includegraphics[width=\textwidth]{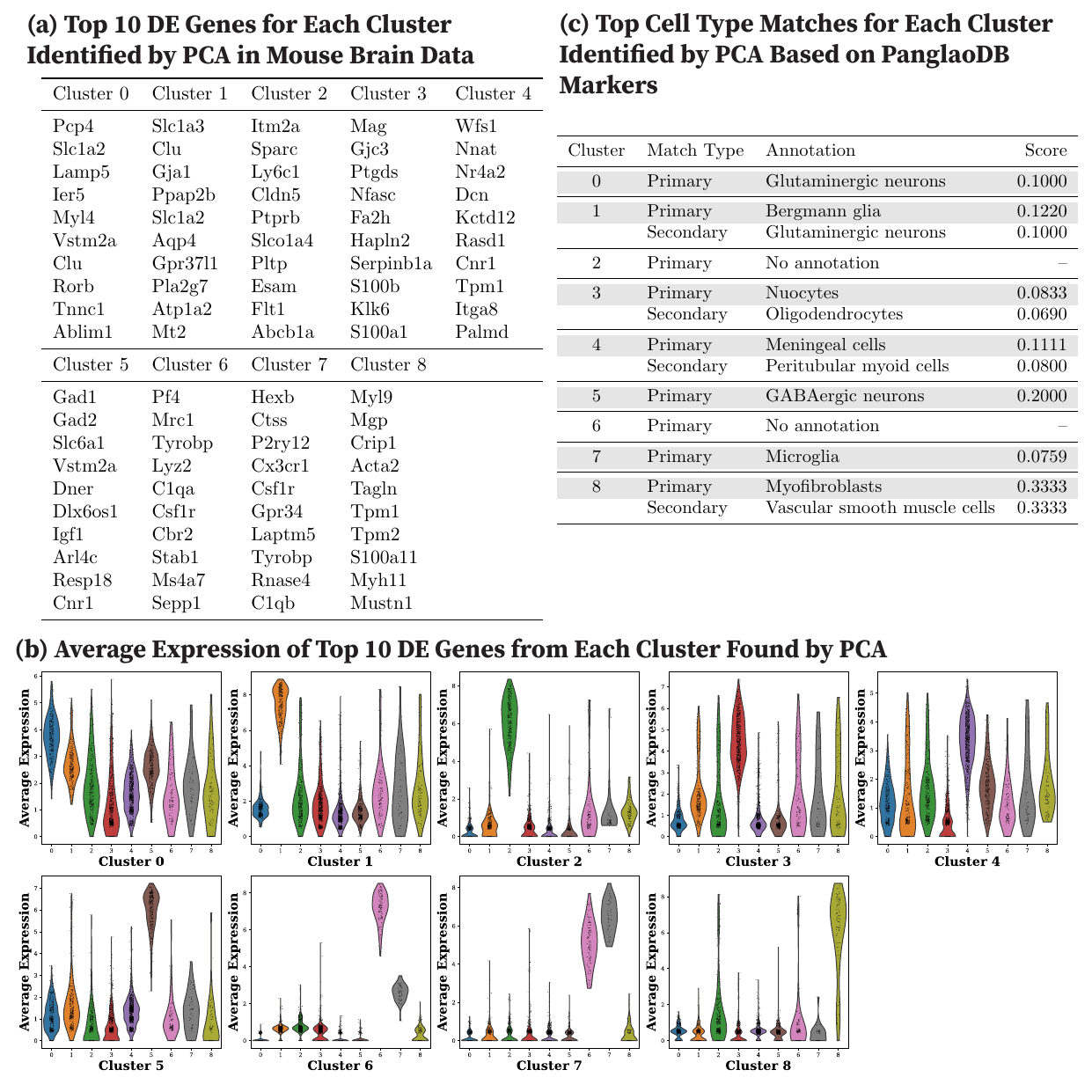} 
  \caption{Clustering Performance of PCA on the Mouse Brain Dataset and Cell Type Annotation using DEGs and Canonical Markers. (a) Top 10 DEGs identified for each cluster by PCA. (b) Across-cluster average expression of cluster-specific DEGs. Violin plots show the distribution of average expression levels of these DEGs across all clusters. (c) Cell type annotation using the PanglaoDB marker database.}
\label{fig:PCA_MouseBrain}
\end{figure}

\begin{figure}[H]
  \centering
  \includegraphics[width=\textwidth]{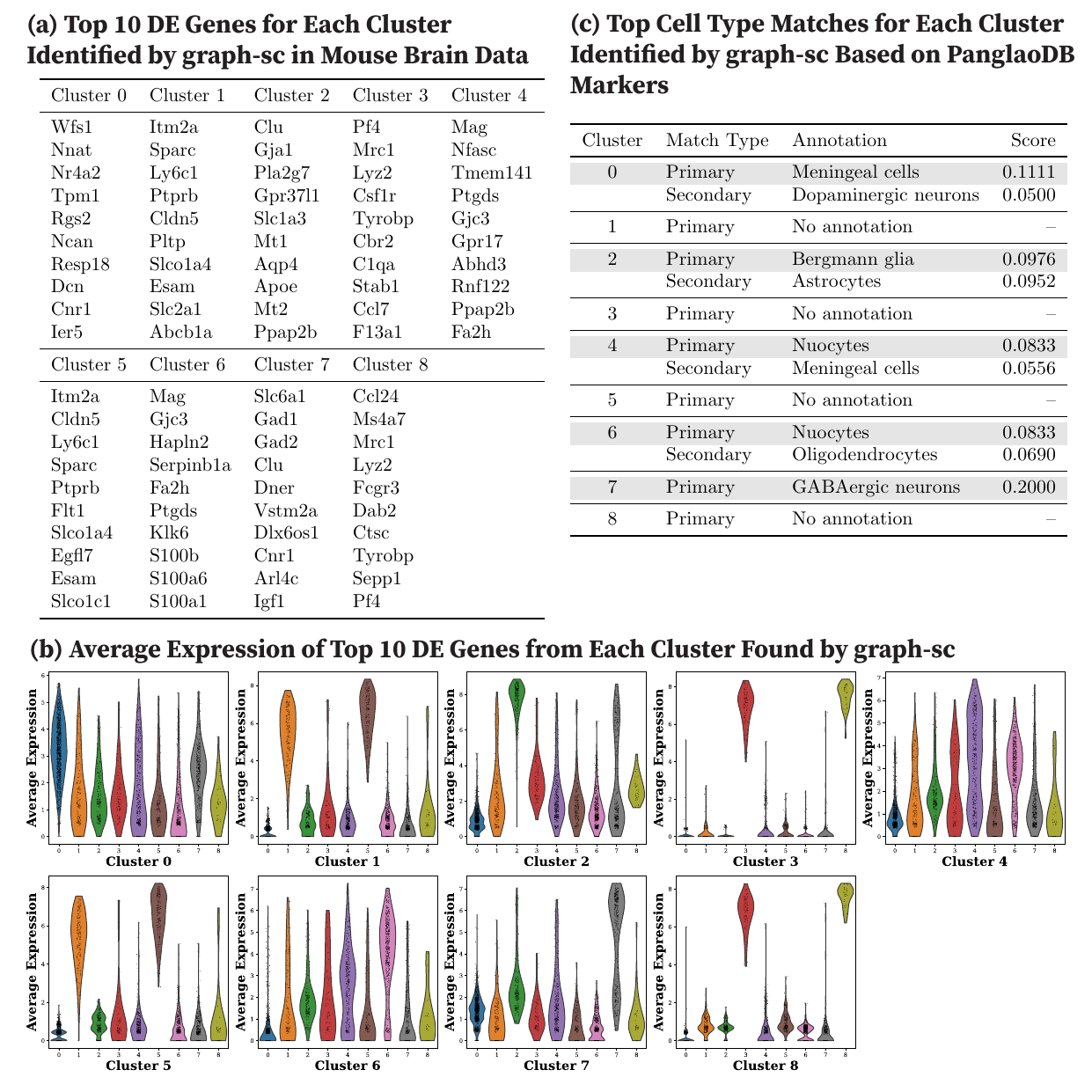} 
  \caption{Clustering Performance of graph-sc on the Mouse Brain Dataset and Cell Type Annotation using DEGs and Canonical Markers. (a) Top 10 DEGs identified for each cluster by graph-sc. (b) Across-cluster average expression of cluster-specific DEGs. Violin plots show the distribution of average expression levels of these DEGs across all clusters. (c) Cell type annotation using the PanglaoDB marker database.}
\label{fig:graph-sc_MouseBrain}
\end{figure}

\begin{figure}[H]
  \centering
  \includegraphics[width=\textwidth]{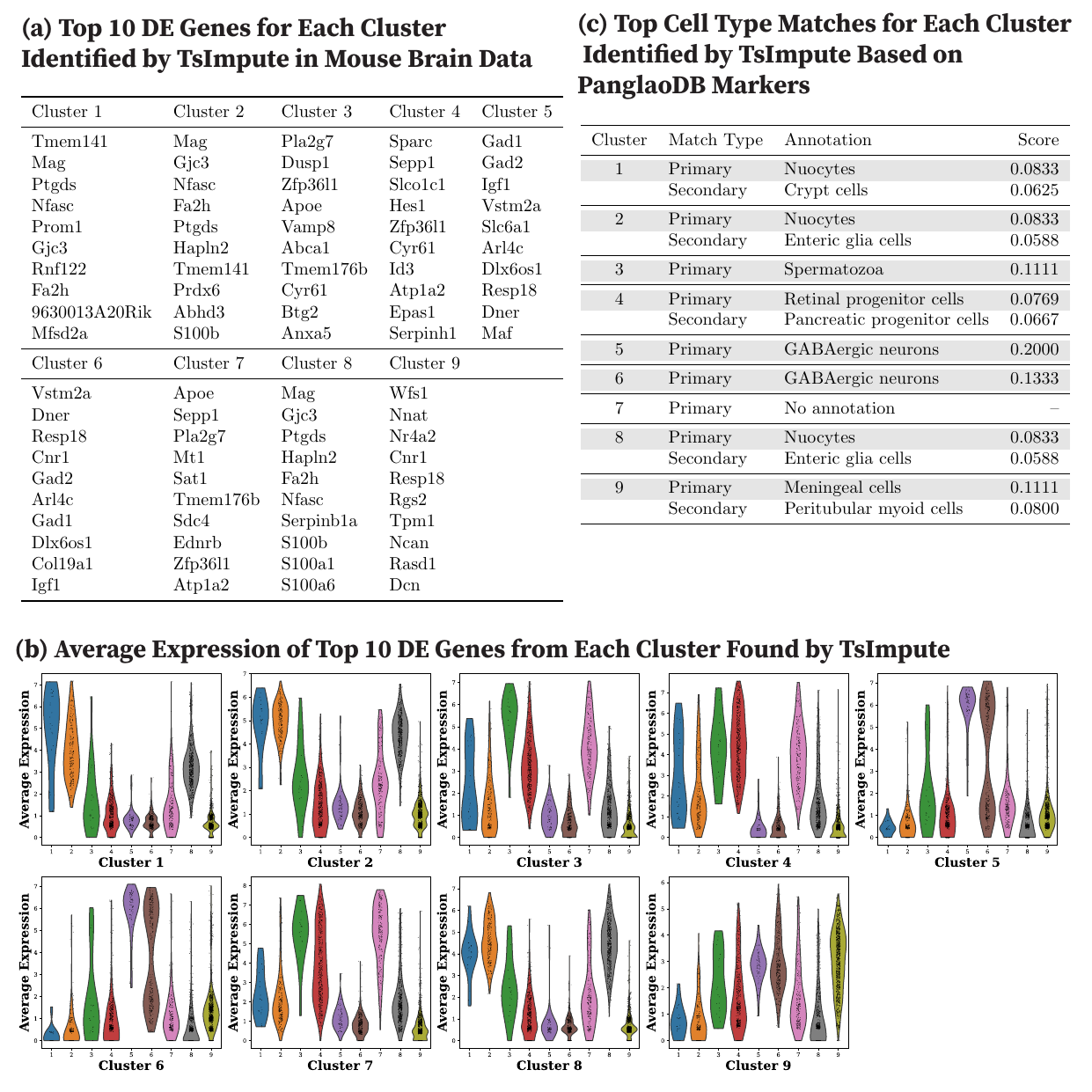} 
  \caption{Clustering Performance of tsImpute on the Mouse Brain Dataset and Cell Type Annotation using DEGs and Canonical Markers. (a) Top 10 DEGs identified for each cluster by tsImpute. (b) Across-cluster average expression of cluster-specific DEGs. Violin plots show the distribution of average expression levels of these DEGs across all clusters. (c) Cell type annotation using the PanglaoDB marker database.}
\label{fig:tsImpute_MouseBrain}
\end{figure}

\begin{figure}[H]
  \centering
  \includegraphics[width=\textwidth]{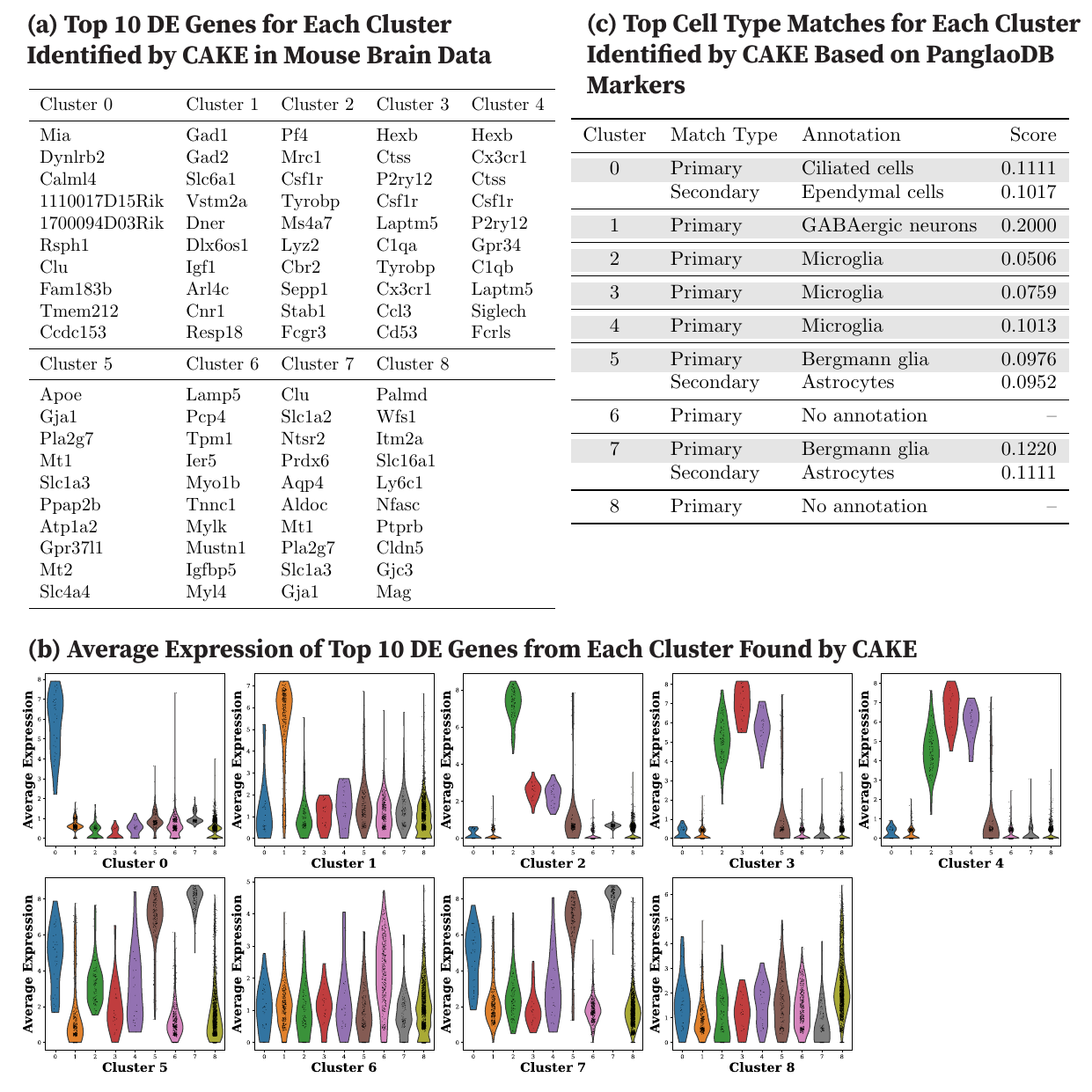} 
  \caption{Clustering Performance of CAKE on the Mouse Brain Dataset and Cell Type Annotation using DEGs and Canonical Markers. (a) Top 10 DEGs identified for each cluster by CAKE. (b) Across-cluster average expression of cluster-specific DEGs. Violin plots show the distribution of average expression levels of these DEGs across all clusters. (c) Cell type annotation using the PanglaoDB marker database.}
\label{fig:CAKE_MouseBrain}
\end{figure}

\begin{figure}[H]
  \centering
  \includegraphics[width=\textwidth]{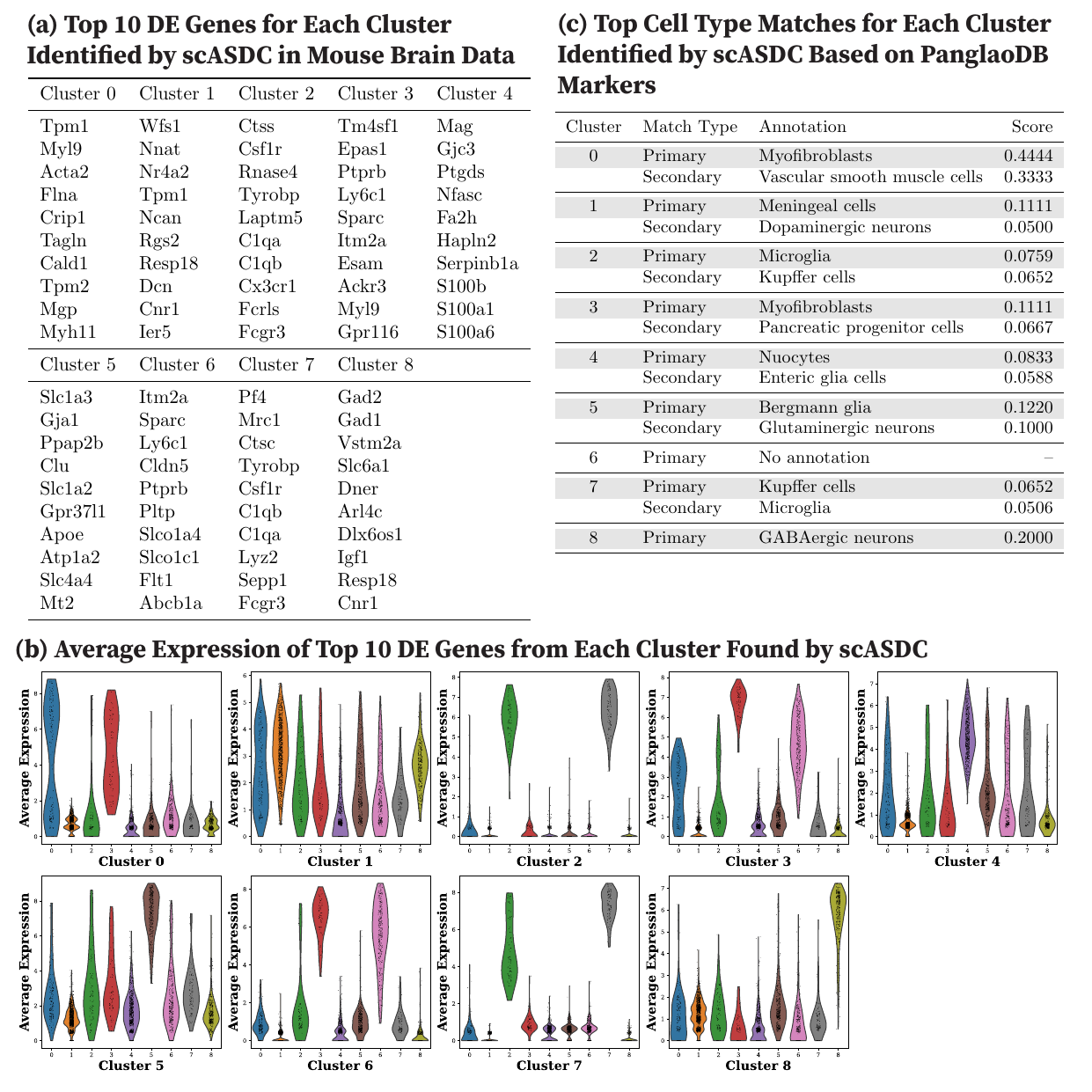} 
  \caption{Clustering Performance of scASDC on the Mouse Brain Dataset and Cell Type Annotation using DEGs and Canonical Markers. (a) Top 10 DEGs identified for each cluster by scASDC. (b) Across-cluster average expression of cluster-specific DEGs. Violin plots show the distribution of average expression levels of these DEGs across all clusters. (c) Cell type annotation using the PanglaoDB marker database.}
\label{fig:scASDC_MouseBrain}
\end{figure}

Figures~\ref{fig:CoMem_MouseBrain},~\ref{fig:PCA_MouseBrain},~\ref{fig:graph-sc_MouseBrain},~\ref{fig:tsImpute_MouseBrain},~\ref{fig:CAKE_MouseBrain}, and~\ref{fig:scASDC_MouseBrain} show the results on the mouse brain dataset for CoMem-DIPHW, PCA, graph-sc, tsImpute, CAKE, and scASDC, respectively.

\subsection{Hyperparameters}

Table~\ref{tab:hyperparameters} lists the hyperparameters used in each experiment.

\begin{table}[!h]
\centering
\fontsize{8pt}{9pt}\selectfont  
\begin{tabular}{|l|l|}
\hline
\textbf{Figure/Table/Section} & \textbf{Description and Value} \\
\hline
\multirow{16}{*}{\cell{0.3\textwidth}{Experiments on simulated scRNA-seq data across modularity levels (Fig.~\ref{fig:PerformanceByModularity} for average genes per cell type plots)}}
 & Number of cells = 3000  \\
 & Number of genes = 1000 \\
 & Number of embedded modules = 30  \\
 & Average number of cells per module = 38 \\
 & Target density of the gene expression matrix = 0.03 \\
 & Density within modules = 0.3  \\
 & Density between modules = 0.1 \\
 & Probability of inter-module connections = 0.6 \\
 & Average background expression = 10 \\
 & Average within-module expression = 20 \\
 & Average inter-module expression = 10 \\
 & Workers = 30 \\
 & Walk length = 30 \\
 & Embedding dimensions = 30 \\
 & Preference exponent = 50 \\
 & Average number of genes per module $\in$ \{10, 20, 30, 40, 70, 80, 90, 100\}\\
\hline
\multirow{4}{*}{\cell{0.3\textwidth}{Experiments on simulated scRNA-seq data across modularity levels (Fig.~\ref{fig:PerformanceByModularity} for number of modules plots)}}
 & Number of embedded modules $\in$ \{10, 20, 30, 40, 70, 80, 90, 100\}  \\
 & Average number of genes per module = 50 \\
 & Average number of cells per module = (\# genes / \# modules) +3 \\
 & Other parameters are the same as Fig.~\ref{fig:PerformanceByModularity} for average genes per cell type plots. \\
\hline
\cell{0.26\textwidth}{Experiments on the impact of module sizes on ARI clustering performance (Fig.~\ref{fig:Ngenes_ARI})} & 
Same as in Fig.~\ref{fig:PerformanceByModularity} for average genes per cell type plots. \\
\hline
\cell{0.26\textwidth}{Experiments on the impact of module counts on ARI clustering performance (Fig.~\ref{fig:Nmodules_ARI})} & 
Same as in Fig.~\ref{fig:PerformanceByModularity} for number of modules. \\
\hline
\multirow{5}{*}{\cell{0.3\textwidth}{Comparison of ARI and NMI across clustering methods for the ScMixology benchmark datasets (Table~\ref{tab:scMixology})}}
 & Number of highly expressed genes used = 500 \\
 & Preference exponent = 10 \\
 & Workers = 50 \\
 & Walk length = 10 \\
 & Embedding dimensions = 10 \\
\hline
\multirow{6}{*}{\cell{0.3\textwidth}{Experiments on human pancreas (Section~\ref{subsec:human_pancreas_results})}}
 & Number of highly expressed genes used = 1000 \\
 & Number of clusters = 9 \\
 & Preference exponent = 10 \\
 & Workers = 100 \\
 & Walk length = 50 \\
 & Embedding dimensions = 10 \\
\hline
\multirow{2}{*}{\cell{0.3\textwidth}{Experiments on mouse pancreas (Section~\ref{subsec:mouse_pancreas_brain_results})}}
 & Number of clusters = 13 \\
 & Other parameters are the same as for human pancreas analysis (Section~\ref{subsec:human_pancreas_results}). \\
\hline
\multirow{2}{*}{\cell{0.3\textwidth}{Experiments on human brain (Section~\ref{subsec:human_brain_results})}}
 & Number of clusters = 7 \\
 & Other parameters are the same as for human pancreas analysis (Section~\ref{subsec:human_pancreas_results}). \\
\hline
\multirow{2}{*}{\cell{0.3\textwidth}{Experiments on mouse brain (Section~\ref{subsec:mouse_pancreas_brain_results})}}
 & Number of clusters = 9 \\
 & Other parameters are the same as for human pancreas analysis (Section~\ref{subsec:human_pancreas_results}). \\
\hline
\end{tabular}
\caption{Hyperparameters Used in Each Experiment.}
\label{tab:hyperparameters}
\end{table}
\end{document}